\documentclass[12pt]{article}
\pdfoutput=1
\usepackage{amsmath,amssymb,amsthm,bm,graphicx,float,array,multirow,multicol,rotfloat,caption,subcaption,hyperref,cleveref,enumerate,geometry,mathdots,adjustbox,booktabs,parskip,mathtools,tikz,tikz-cd,pdflscape,csquotes,lscape,rotating,empheq,mathrsfs,longtable}
\usepackage[para]{threeparttable}
\usepackage[all]{xy}
\usepackage[normalem]{ulem}
\usepackage[numbers,sort&compress]{natbib}
\numberwithin{equation}{section}
\numberwithin{figure}{section}
\allowdisplaybreaks
\makeatletter
\usetikzlibrary{arrows,shapes.misc,positioning,decorations.pathmorphing,decorations.markings,decorations.pathreplacing,matrix,patterns,backgrounds}
\tikzset{
    rectangle/.style={draw,circle,inner sep=1pt}, 
    big arrow/.style={decoration={markings,mark=at position 1 with {\arrow[scale=1.5,#1]{>}}}, postaction={decorate}, shorten >=0.4pt}, 
    scale cd/.style={every label/.append style={scale=#1}, cells={nodes={scale=#1}}},
    brace/.style={decoration={brace, mirror},decorate}
}
\definecolor{lightmauve}{RGB}{255,187,255}
\definecolor{lightblue}{RGB}{238,238,255}
\definecolor{lightred}{RGB}{255,238,238}
\definecolor{midgreen}{rgb}{0, 0.5, 0}
\definecolor{ukyellow}{RGB}{235,155,0}
\newcommand\footscriptsize{\@setfontsize\footscriptsize{9.6}{9.6}}

\newcommand\notsoscript{\@setfontsize\notsoscript{9}{7}}

\newcommand{\email}[1]{\href{mailto:#1}{\vphantom{$\Bigl($}\tt #1}}

\theoremstyle{plain}
\newtheorem*{thm*}{Theorem}

\theoremstyle{definition}

\newtheorem*{defn*}{Definition}

\makeatother
\newcommand{\rankfour}{\operatorname{rank}_{(4)}}

\begin{document}

\begin{titlepage}
% Report number
\vspace*{-3cm} 
\begin{flushright}
{\tt CALT-TH-2022-044}\\
{\tt DESY-22-201}\\
{\tt UTWI 17-2022}\\
\end{flushright}
\begin{center}
\vspace{2cm}
{\LARGE\bfseries Isomorphisms of 4d $\mathcal{N}=2$ SCFTs from 6d} \\[1em]
%{\LARGE\bfseries 4d SCFTs in different guises: a 6d reveal}
\vspace{1.2cm}

{\large
Jacques Distler$^{1}$, Grant Elliot$^{1}$, Monica Jinwoo Kang$^{2}$, and Craig Lawrie$^{3}$\\}
\vspace{.7cm}
{ $^1$ Weinberg Institute for Theoretical Physics,\par Department of Physics, University of Texas at Austin}\par
{Austin, TX 78712, U.S.A.}\par
\vspace{.2cm}
{ $^2$ Walter Burke Institute for Theoretical Physics, California Institute of Technology}\par
{Pasadena, CA 91125, U.S.A.}\par
\vspace{.2cm}
{ $^3$ Deutsches Elektronen-Synchrotron DESY,}\par
{Notkestr.~85, 22607 Hamburg, Germany}\par
\vspace{.2cm}

\vspace{.3cm}

\scalebox{.75}{ \email{distler@golem.ph.utexas.edu}, \email{gelliot123@utexas.edu}, \email{monica@caltech.edu}, \email{craig.lawrie1729@gmail.com}
}\par
\vspace{1.2cm}
\textbf{Abstract}
\end{center}

There exist 4d $\mathcal{N}=2$ SCFTs in class $\mathcal{S}$ which have different constructions as punctured Riemann surfaces, but which nevertheless appear to describe the same physics. % -- at least as far as we can determine from the central charges, flavor symmetries and flavor central charges, Coulomb branch operator scaling dimensions, Schur and Hall--Littlewood indices, etc
Some of these class $\mathcal{S}$ theories have an alternative construction as torus-compactifications of 6d $(1,0)$ SCFTs. We demonstrate that the 6d SCFTs are isomorphic. Each 6d SCFT in question can be obtained from a parent 6d SCFT by Higgs branch renormalization group flow, and the parent theory possesses a discrete symmetry under which the relevant Higgs branch flows are exchanged. The existence of this discrete symmetry, which may be embedded in an enhanced continuous symmetry, proves that the original pair of class $\mathcal{S}$ theories are, in fact, isomorphic. 

\vfill 
\end{titlepage}
{\pagestyle{empty}
\tableofcontents
\newpage
}
\setcounter{page}{1}

\section{Introduction}\label{sec:intro}

A quantum field theory is characterized by its spectrum of local operators and their $n$-point correlation functions. If these differ for two QFTs, we can conclude that they are distinct theories. More subtly, even when the spectrum of local operators coincide, the theories may differ in their spectrum of line operators \cite{Aharony:2013hda} or other non-local observables. These more subtle differences may be detectable in the $n$-point correlation functions and/or by putting the theory on a curved $d$-dimensional manifold. For conformal field theories (CFTs), the infinite set of local data is determined by a much smaller (but, for $d>2$, necessarily still infinite) set of data: the scaling dimensions and the 3-point operator product expansion (OPE) coefficients of the conformal primary operators, namely, 
\begin{equation}
  \{\Delta_i, \,\lambda_{ijk}\} \,.
\end{equation} 
Since this set is still an infinite amount of data, it is not particularly computable and it is strongly believed to be highly redundant. Indeed, the goal of the conformal bootstrap program (see \cite{Poland:2018epd} and references therein) is to constrain these data via crossing symmetry and unitarity. Ideally, we would like to find a finite set of data from which the rest can be recovered. Then we could determine if two CFTs (or QFTs) are isomorphic, in finite time, by comparing the two finite sets of data.

Ideally, these finite sets of data should be computable from the presentation of the CFT, say, as a string theory construction. One  approach is to take the subset of the  CFT data that \emph{is} readily computable and ask if that subset is sufficient to distinguish between distinct CFTs.

For four-dimensional CFTs, the two central charges (Weyl-anomaly coefficients) $a$ and $c$, the flavor symmetry algebra $\mathfrak{f}$ (generated by conserved currents $J^a_\mu$), and the current algebra levels (the coefficient, in a certain normalization, of the identity operator in the OPE of two conserved currents) are readily computed. For 4d $\mathcal{N}=2$ superconformal field theories (SCFTs) of class $\mathcal{S}$ \cite{Gaiotto:2009hg,Gaiotto:2009we}, the global form of the flavor symmetry group $F$ is also readily computable \cite{Bhardwaj:2021ojs,Distler:2022nsn}. Moreover, every interacting 4d $\mathcal{N}=2$ SCFT has a Coulomb branch with a $\mathbb{C}^*$-action on it. The dimension of the Coulomb branch and the weights under this $\mathbb{C}^*$-action, \emph{i.e.}, the $U(1)_r$ charges of the generators of the Coulomb branch, are also readily computable; we call these the \emph{graded Coulomb branch dimensions}. Finally, the dimension of the Higgs branch (if the 4d $\mathcal{N}=2$ SCFT has one) is easily computable. Collectively\footnote{Sometimes, for reasons either historical or expository, we omit the global form of the flavor symmetry group from the list of conventional invariants. Hopefully, this will be clear from the context.}, we refer to these data as the ``conventional invariants'' of a 4d $\mathcal{N}=2$ SCFT.

In low-rank cases, these data (or even subsets thereof) suffices to characterize the 4d SCFT uniquely. For instance, if we encounter a rank one 4d $\mathcal{N}=2$ SCFT whose Coulomb branch generator has 
\begin{align}
    \Delta=6\quad\text{and}\quad (a,c)=\left(\frac{95}{24},\frac{31}{6}\right) , 
\end{align}
then it \emph{must} be the $(E_8)_{12}$ Minahan--Nemeschansky theory \cite{Argyres:2016xua}. Indeed, there are a multitude of distinct realizations of this SCFT in class $\mathcal{S}$ and they are all necessarily isomorphic.

For higher rank cases, however, these ``conventional invariants'' are known \emph{not} to suffice to characterize the 4d SCFT. There are distinct 4d SCFTs whose ``conventional invariants'' coincide \cite{Distler:2020tub,Distler:2022nsn}. Nevertheless, we do have examples where distinct class $\mathcal{S}$ constructions seem to lead to isomorphic SCFTs. One of the purposes of this paper is to show that, when that happens, the resulting isomorphism frequently has a 6d $(1,0)$ SCFT origin \cite{Baume:2021qho}.

The situation in six-dimensions is much better. The 6d $(1,0)$ SCFTs can be engineered in F-theory via Calabi--Yau threefolds that are elliptic fibrations over a non-compact complex surface $B$ \cite{Heckman:2013pva,Heckman:2015bfa}.\footnote{For the elliptically-fibered Calabi--Yau threefold compactifications via geometric-engineering process, see \cite{Morrison:2016djb,DelZotto:2018tcj,Heckman:2018pqx,Cabrera:2019izd,Apruzzi:2019vpe,Razamat:2019mdt,Apruzzi:2019opn,Hassler:2019eso,Apruzzi:2019enx,Baume:2020ure,Baume:2021qho,Esole:2017rgz,Esole:2017qeh,Esole:2017hlw,Esole:2018csl,Esole:2018mqb,Esole:2019asj,Esole:2014bka,Esole:2014hya}.} 
The configuration of exceptional divisors --- which are $\mathbb{P}^1$s with negative self-intersection numbers on $B$ --- and the elliptic fibers over them, which we will refer to as the ``curve configuration,'' is believed to uniquely characterize the interacting part of the 6d $(1,0)$ SCFT.\footnote{The 6d effective field theory consists of usual vector, tensor, and hypermultiplets, as well as a collection of tensionful BPS-strings. The curve configuration defines the effective field theory that exists at the generic point of the tensor branch of the interacting SCFT that lives at the origin. We reach the conformal fixed point by shrinking the two-cycles of $B$, where the two-cycles contribute tensionless strings and the two-cycles of the fiber contribute massless multiplets; \emph{i.e.}, going to the origin of the tensor branch is equivalent to taking the tension of all BPS-strings to zero simultaneously. Since the spectrum of BPS-strings is fixed by the curve configuration, the SCFT obtained by taking the tensionless limit is identical if two curve configurations are identical. It is shown in \cite{Distler:2022yse} that some specific curve configurations can have two different tensionful string spectra, which can be captured by including additional small data to the curve configuration. Such examples do not appear in this paper, and thus the tensionless string limit of identical curve configurations leads to the same SCFT at the origin.}

The cancellation of gauge anomalies tightly constrains the allowed curve configurations. Indeed, the flavor symmetry algebra and the anomaly 8-form $(\mathfrak{f}, I_8)$ determine the curve configuration almost but not completely. The anomaly polynomial $I_8$ of a given 6d $(1,0)$ SCFT is given by adding geometric contributions, each via characteristic classes of the inherent symmetry of the theory:
\begin{align}
    \begin{aligned}
    I_8 &=  \frac{\alpha}{24} c_2(R)^2+ \frac{\beta}{24}  c_2(R) p_1(T) + \frac{\gamma}{24}  p_1(T)^2 + \frac{\delta}{24} p_2(T) \cr &\quad + \sum_a \text{Tr}F_a^2 \left(\kappa_a p_1(T) + \nu_a c_2(R) + \sum_b \rho_{ab} \text{Tr}  F_b^2\right) + \sum_a \mu_a \text{Tr}F_a^4  \,.
  \end{aligned}
\end{align}

Upon toroidal compactification, the 6d $(1,0)$ SCFTs flow to 4d $\mathcal{N}=2$ SCFTs, whose conventional invariants (as we shall review below) are determined from a subset of the data in $(\mathfrak{f}, I_8)$, via the coefficients of the anomaly polynomial term: $(\mathfrak{f}, (\beta, \gamma, \delta, \{\kappa_a\}))$. The anomaly polynomial of the interacting SCFT associated to a curve configuration can completely be determined from the curve configuration itself \cite{Ohmori:2014kda,Intriligator:2014eaa,Baume:2021qho}. Subtracting the anomaly polynomial of the interacting sector defined via a curve configuration from anomaly polynomial of a given mixed SCFT yields the anomaly polynomial of a collection of free-hypermultiplets. Thus, we can determine the SCFT content by finding the interacting part of the SCFT and the number of free hypermultiplets, while the curve configuration provides a complete invariant of the interacting part of the 6d $(1,0)$ SCFT.

Our strategy, then, is very simple: to prove that two 4d $\mathcal{N}=2$ SCFTs are isomorphic, we show that they arise as the toroidal compactification of two 6d $(1,0)$ SCFTs which are isomorphic because they share the same curve configuration. Hence, we mostly consider 4d $\mathcal{N}=2$ SCFTs of class $\mathcal{S}$ that admits 6d $(1,0)$ SCFTs origins.\footnote{This does not exhaust the set of apparently-isomorphic 4d SCFTs, and we 
give some examples without a 6d $(1,0)$ origin in Section \ref{sec:oddballs}.}

The rest of the paper is organized as follows. In Section \ref{sec:4d}, we show how to find candidates of isomorphic SCFTs through constructing pairs of class $\mathcal{S}$ theories whose 4d conventional invariants coincide. We show that these theories also admit 6d $(1,0)$ SCFT origins via toroidal compactifications by a process of ungauging, followed by gauging. 
Using the 6d $(1,0)$ SCFT parents, we show in Section \ref{sec:6d} that each pair is isomorphic and find the renormalization group (RG) flows between the pairs. We take type $\mathfrak{e}_7$ SCFTs as the target example theories and study them explicitly. We revisit the 4d class $\mathcal{S}$ theories, that are from the 6d $(1,0)$ parents, in Section \ref{sec:same} and enumerate the isomorphic fixtures (\emph{i.e.}, 3-punctured spheres) in the class $\mathcal{S}$ theories of type $\mathfrak{e}_7$. In Section \ref{sec:different}, we relax some of the constraints in the algorithm of Section \ref{sec:4d} to construct more pairs of isomorphic theories. We even further relax the constraints in Section \ref{sec:isointer} and construct pairs of theories that differ in the number of free hypermultiplets, but whose  \emph{interacting} sectors are isomorphic SCFTs. While our analysis was on type $\mathfrak{e}_7$ theories, we consider some examples drawn from other ADE types in Section \ref{sec:other} to demonstrate that these are not specific to type $\mathfrak{e}_7$ theories. Finally, in Section \ref{sec:oddballs}, we discuss some examples which appear to be isomorphic, but which are not related to compactifications from 6d. For each example pair, we check that their Schur indices agree (up to the order to which we are able to compute them). In Section \ref{sec:discussion}, we discuss the source of the origins of these isomorphisms as $\mathbb{Z}_2$ outer-automorphisms of 6d $(1,0)$ SCFTs and how they differ for the ADE types.

\section{Isomorphisms of class \texorpdfstring{$\mathcal{S}$}{S} theories and the setup}
\label{sec:4d}

We consider 4d $\mathcal{N}=2$ SCFTs of class $\mathcal{S}$, the theories constructed as a compactification of 6d $(2,0)$ SCFTs of type $\mathfrak{g}$ on a genus $g$ $n$-punctured Riemann surface, with codimension-2 defect operators filling all of the 4d spacetime and situated at the $n$ marked points on the Riemann surface \cite{Gaiotto:2009we,Gaiotto:2009hg}. The defect operators located at the (``regular'') punctures are labeled by nilpotent orbits in $\mathfrak{g}$. In this way, a class $\mathcal{S}$ theory is encoded in the following data:\footnote{``Twisted'' class $\mathcal{S}$ theories can be constructed by incorporating outer-automorphism twists of $\mathfrak{g}$ around nontrivial cycles on the Riemann surface.  In this paper, we consider only untwisted class $\mathcal{S}$ theories.}
\begin{equation}\label{eqn:classstuple}
  \{ \,\, \mathfrak{g},\,\, g,\,\, O_1,\,\, \cdots,\,\, O_n\,\, \} \,,
\end{equation}
where the $O_i$ are nilpotent orbits in $\mathfrak{g}$. It is natural to ask: 
\begin{equation*}
\begin{gathered}
\textbf{When do two different tuples of data given by}\\
\textbf{equation \eqref{eqn:classstuple} lead to the same 4d $\mathcal{N}=2$ SCFTs?}
\end{gathered}
\end{equation*}
In this section, we will explore methods to generate such tuples which appear to correspond to isomorphic theories.

We focus on pairs of theories where the genus of the Riemann surface is the same, the 6d $(2,0)$ origin is the same, and all but two of the $n$ punctures are the same. That is, we wish to compare theories associated to the following data:
\begin{equation}\label{eqn:classspair}
    \{ \,\, \mathfrak{g},\,\, g,\,\, O_1,\,\,O_2,\,\,O_3,\,\, \cdots,\,\, O_{N+2}\,\, \} \qquad \text{and} \qquad \{ \,\, \mathfrak{g},\,\, g,\,\, O_1',\,\,O_2',\,\,O_3,\,\, \cdots,\,\, O_{N+2}\,\, \} \,.
\end{equation}
Then the question boils down to find under what circumstances are these two theories the same:
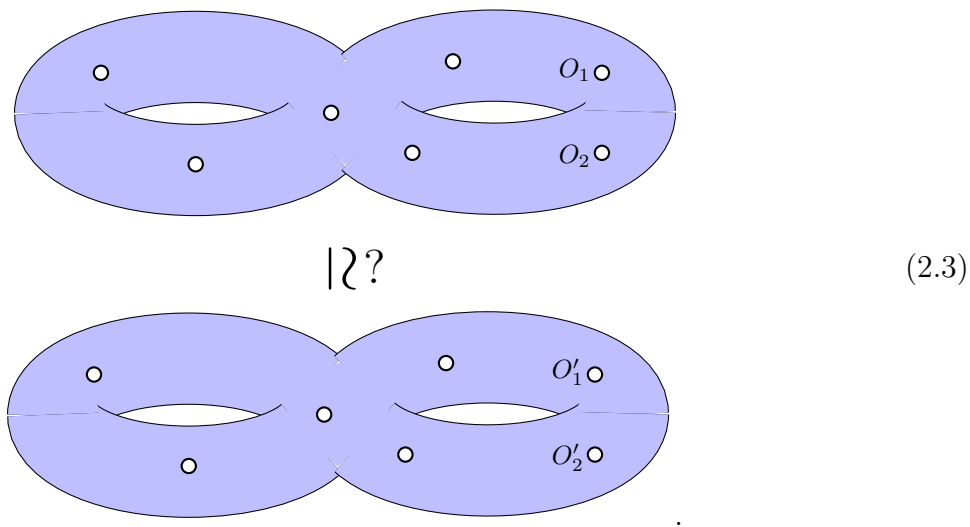
\begin{align}
\begin{gathered}
    \begin{tikzpicture}[xscale=1.8,yscale=1.8*cos(65)]
        \draw[double=blue!25,double distance=12mm,xshift=-.95mm,yshift=4mm] (0:1) arc (25:180:1);
        \draw[double=blue!25,double distance=12mm] (180:1) arc (180:335:1);
        \draw[double=blue!25,double distance=12mm,xshift=22.05mm] (0:1) arc (0:155:1);
        \draw[double=blue!25,double distance=12mm,xshift=23mm,yshift=-4mm] (180:1) arc (205:360:1);
        \draw[color=blue!25,fill=blue!25] (1.12,.9)--(1.5,0)--(1.1,-.9)--(.7,0)--(1.1,.9);
        \draw[thick,fill=white] (3,.7) ellipse (1.5pt and 3.55pt);
        \node[font=\footnotesize] (P1) at (2.8,.75) {$O_1$};
        \draw[thick,fill=white] (3,-.7) ellipse (1.5pt and 3.55pt);
        \node[font=\footnotesize] (P2) at (2.8,-.75) {$O_2$};
        \draw[thick,fill=white] (-.7,.7) ellipse (1.5pt and 3.55pt);
        \draw[thick,fill=white] (0,-.9) ellipse (1.5pt and 3.55pt);
        \draw[thick,fill=white] (1,0) ellipse (1.5pt and 3.55pt);
        \draw[thick,fill=white] (1.6,-.7) ellipse (1.5pt and 3.55pt);
        \draw[thick,fill=white] (1.9,.9) ellipse (1.5pt and 3.55pt);
    \end{tikzpicture}\\
    \raisebox{15pt}{\rotatebox{-90}{\scalebox{2}{\begin{tikzpicture}
        $\simeq$
    \end{tikzpicture}}}}\hspace{14pt} \scalebox{1.5}{\begin{tikzpicture}
        $?$
    \end{tikzpicture}}\\
    \begin{tikzpicture}[xscale=1.8,yscale=1.8*cos(65)]
        \draw[double=blue!25,double distance=12mm,xshift=-.95mm,yshift=4mm] (0:1) arc (25:180:1);
        \draw[double=blue!25,double distance=12mm] (180:1) arc (180:335:1);
        \draw[double=blue!25,double distance=12mm,xshift=22.05mm] (0:1) arc (0:155:1);
        \draw[double=blue!25,double distance=12mm,xshift=23mm,yshift=-4mm] (180:1) arc (205:360:1);
        \draw[color=blue!25,fill=blue!25] (1.12,.9)--(1.5,0)--(1.1,-.9)--(.7,0)--(1.1,.9);
        \draw[thick,fill=white] (3,.7) ellipse (1.5pt and 3.55pt);
        \node[font=\footnotesize] (P1) at (2.8,.75) {$O_1^\prime$};
        \draw[thick,fill=white] (3,-.7) ellipse (1.5pt and 3.55pt);
        \node[font=\footnotesize] (P2) at (2.8,-.75) {$O_2^\prime$};
        \draw[thick,fill=white] (-.7,.7) ellipse (1.5pt and 3.55pt);
        \draw[thick,fill=white] (0,-.9) ellipse (1.5pt and 3.55pt);
        \draw[thick,fill=white] (1,0) ellipse (1.5pt and 3.55pt);
        \draw[thick,fill=white] (1.6,-.7) ellipse (1.5pt and 3.55pt);
        \draw[thick,fill=white] (1.9,.9) ellipse (1.5pt and 3.55pt);
    \end{tikzpicture} \,.
\end{gathered}\label{genus2isom}
\end{align}
The two theories are evidently not isomorphic if they possess different conventional invariants. Thus, we would only like to consider pairs of the form in equation \eqref{eqn:classspair} such that the following quantities of the associated SCFTs are identical:
\begin{itemize}%
  \item the Weyl anomaly coefficients, $a$ and $c$,
  \item the flavor symmetry algebras and levels,
  \item the graded Coulomb branch dimensions,
  \item the Higgs branch dimension.
\end{itemize}
For any class $\mathcal{S}$ theory, the complex structure moduli of the punctured Riemann surface parametrize exactly-marginal deformations of the SCFT. Taking a degeneration limit, we can simplify the question in equation \eqref{genus2isom} to:
\begin{align}
\begin{gathered}
    \begin{tikzpicture}[xscale=1.8,yscale=1.8*cos(65)]
        \draw[double=blue!25,double distance=12mm,xshift=-.95mm,yshift=4mm] (0:1) arc (25:180:1);
        \draw[double=blue!25,double distance=12mm] (180:1) arc (180:335:1);
        \draw[double=blue!25,double distance=12mm,xshift=22.05mm] (0:1) arc (0:155:1);
        \draw[double=blue!25,double distance=12mm,xshift=23mm,yshift=-4mm] (180:1) arc (205:360:1);
        \draw[color=blue!25,fill=blue!25] (1.12,.9)--(1.5,0)--(1.1,-.9)--(.7,0)--(1.1,.9);
        \draw[thick,fill=white] (3.3,0) ellipse (1.5pt and 3.55pt);
        \node[font=\footnotesize] (PL) at (3.1,0) {$0$};
        \draw[thick,fill=white] (-.7,.7) ellipse (1.5pt and 3.55pt);
        \draw[thick,fill=white] (0,-.9) ellipse (1.5pt and 3.55pt);
        \draw[thick,fill=white] (1,0) ellipse (1.5pt and 3.55pt);
        \draw[thick,fill=white] (1.6,-.7) ellipse (1.5pt and 3.55pt);
        \draw[thick,fill=white] (1.9,.9) ellipse (1.5pt and 3.55pt);
        \draw[thick,fill=white] (5,.7) ellipse (1.5pt and 3.55pt);
        \draw[fill=blue!25] (4.5,0) ellipse (18pt and 42.6pt);
        \draw[thick,fill=white] (4.8,.7) ellipse (1.5pt and 3.55pt);
        \node[font=\footnotesize] (P1) at (4.6,.75) {$O_1$};
        \draw[thick,fill=white] (4.8,-.7) ellipse (1.5pt and 3.55pt);
        \node[font=\footnotesize] (P2) at (4.6,-.75) {$O_2$};
        \draw[thick,fill=white] (4.1,0) ellipse (1.5pt and 3.55pt);
        \node[font=\footnotesize] (PR) at (4.3,0) {$0$};
        \draw[thick] (3.35,0)--(4.05,0);
    \end{tikzpicture}\\
    \raisebox{15pt}{\rotatebox{-90}{\scalebox{2}{\begin{tikzpicture}
        $\simeq$
    \end{tikzpicture}}}}\hspace{14pt} \scalebox{1.5}{\begin{tikzpicture}
        $?$
    \end{tikzpicture}}\\[5pt]
    \begin{tikzpicture}[xscale=1.8,yscale=1.8*cos(65)]
        \draw[double=blue!25,double distance=12mm,xshift=-.95mm,yshift=4mm] (0:1) arc (25:180:1);
        \draw[double=blue!25,double distance=12mm] (180:1) arc (180:335:1);
        \draw[double=blue!25,double distance=12mm,xshift=22.05mm] (0:1) arc (0:155:1);
        \draw[double=blue!25,double distance=12mm,xshift=23mm,yshift=-4mm] (180:1) arc (205:360:1);
        \draw[color=blue!25,fill=blue!25] (1.12,.9)--(1.5,0)--(1.1,-.9)--(.7,0)--(1.1,.9);
        \draw[thick,fill=white] (3.3,0) ellipse (1.5pt and 3.55pt);
        \node[font=\footnotesize] (PL) at (3.1,0) {$0$};
        \draw[thick,fill=white] (-.7,.7) ellipse (1.5pt and 3.55pt);
        \draw[thick,fill=white] (0,-.9) ellipse (1.5pt and 3.55pt);
        \draw[thick,fill=white] (1,0) ellipse (1.5pt and 3.55pt);
        \draw[thick,fill=white] (1.6,-.7) ellipse (1.5pt and 3.55pt);
        \draw[thick,fill=white] (1.9,.9) ellipse (1.5pt and 3.55pt);
        \draw[thick,fill=white] (5,.7) ellipse (1.5pt and 3.55pt);
        \draw[fill=blue!25] (4.5,0) ellipse (18pt and 42.6pt);
        \draw[thick,fill=white] (4.8,.7) ellipse (1.5pt and 3.55pt);
        \node[font=\footnotesize] (P1) at (4.6,.75) {$O_1^\prime$};
        \draw[thick,fill=white] (4.8,-.7) ellipse (1.5pt and 3.55pt);
        \node[font=\footnotesize] (P2) at (4.6,-.75) {$O_2^\prime$};
        \draw[thick,fill=white] (4.1,0) ellipse (1.5pt and 3.55pt);
        \node[font=\footnotesize] (PR) at (4.3,0) {$0$};
        \draw[thick] (3.35,0)--(4.05,0);
    \end{tikzpicture} \,.
\end{gathered}\label{genus2degen}
\end{align}
The $(N+1)$-punctured genus $g$ Riemann surfaces on the left are identical in this degeneration limit, and if the cylinder connecting them to the three-punctured spheres on the right is unambiguous,\footnote{See Section \ref{sec:so12} for an example of the rare and special cases where there are inequivalent choices for the connecting cylinders in this degeneration limit.} then the two theories in equation \eqref{genus2isom} are isomorphic when the two three-punctured spheres describe isomorphic SCFTs:
\begin{align}\label{isom3p}
    \begin{tikzpicture}[xscale=1.8,yscale=1.8*cos(65)]
        \draw[fill=blue!25] (4.5,0) ellipse (18pt and 42.6pt);
        \draw[thick,fill=white] (4.8,.7) ellipse (1.5pt and 3.55pt);
        \node[font=\footnotesize] (P1) at (4.6,.75) {$O_1$};
        \draw[thick,fill=white] (4.8,-.7) ellipse (1.5pt and 3.55pt);
        \node[font=\footnotesize] (P2) at (4.6,-.75) {$O_2$};
        \draw[thick,fill=white] (4.1,0) ellipse (1.5pt and 3.55pt);
    \end{tikzpicture}
    \raisebox{24pt}{\begin{tikzpicture}
        \node[font=\Large] (SIM) at (0,0) {$\simeq$};
        \node[font=\large] (Q) at (0,.4) {?};
    \end{tikzpicture}}
    \begin{tikzpicture}[xscale=1.8,yscale=1.8*cos(65)]
        \draw[fill=blue!25] (4.5,0) ellipse (18pt and 42.6pt);
        \draw[thick,fill=white] (4.8,.7) ellipse (1.5pt and 3.55pt);
        \node[font=\footnotesize] (P1) at (4.6,.75) {$O_1^\prime$};
        \draw[thick,fill=white] (4.8,-.7) ellipse (1.5pt and 3.55pt);
        \node[font=\footnotesize] (P2) at (4.6,-.75) {$O_2^\prime$};
        \draw[thick,fill=white] (4.1,0) ellipse (1.5pt and 3.55pt);
    \end{tikzpicture} \,.
\end{align}

In equation \eqref{genus2degen}, we have depicted the cylinder connecting the genus $g$ surface and the three-punctured sphere as joining together two full punctures, labeled by $0$. However, this connecting puncture may be different in non-generic cases. If $O_1$ and $O_2$ are sufficiently low on the Hasse diagram, then the connecting puncture on the right may instead be forced to be an irregular puncture (in the sense of \cite{Chacaltana:2010ks}); these are labeled \cite{Chacaltana:2012zy} by pairs $(O,H)$, consisting of a nilpotent orbit $O$ and a subgroup $H$ of its flavor symmetry group. The puncture on the $(N+1)$-punctured surface on the left is the puncture $O$. In a sequence of works \cite{Chacaltana:2010ks,Chacaltana:2011ze,Chacaltana:2014jba,Chacaltana:2017boe,Chacaltana:2018vhp}, the three-punctured spheres with irregular punctures were catalogued, and we can simply look up the results for the pairs $(O_1, O_2)$ which yield a 3-punctured sphere with an irregular puncture.

Alternatively, if the collection of punctures on the genus $g$ surface on the left are sufficiently low down the Hasse diagram, and $g = 0$, then the full puncture $0$, which connects the $(N+1)$-punctured sphere on the left, is replaced by an irregular puncture, $(O,H)$ and the three-punctured sphere on the right has punctures $O_1$, $O_2$ and $O$. This will, in fact, be the generic situation for the examples we study. The three-punctured spheres that we find to be isomorphic SCFTs of class $\mathcal{S}$ will invariably have the 3rd puncture in equation \eqref{isom3p} being a less-than-full puncture.

Thus the general question about isomorphisms of class $\mathcal{S}$ theories of the form in equation \eqref{genus2isom} can be simplified to a question about isomorphisms of three-punctures spheres,
\begin{align}
    \begin{tikzpicture}[xscale=1.8,yscale=1.8*cos(65)]
        \draw[fill=blue!25] (4.5,0) ellipse (18pt and 42.6pt);
        \draw[thick,fill=white] (4.8,.7) ellipse (1.5pt and 3.55pt);
        \node[font=\footnotesize] (P1) at (4.6,.75) {$O_1$};
        \draw[thick,fill=white] (4.8,-.7) ellipse (1.5pt and 3.55pt);
        \node[font=\footnotesize] (P2) at (4.6,-.75) {$O_2$};
        \draw[thick,fill=white] (4.1,0) ellipse (1.5pt and 3.55pt);
        \node[font=\footnotesize] (P3) at (4.3,0) {$O$};
    \end{tikzpicture}
    \raisebox{24pt}{\begin{tikzpicture}
        \node[font=\Large] (SIM) at (0,0) {$\simeq$};
        \node[font=\large] (Q) at (0,.4) {?};
    \end{tikzpicture}}
    \begin{tikzpicture}[xscale=1.8,yscale=1.8*cos(65)]
        \draw[fill=blue!25] (4.5,0) ellipse (18pt and 42.6pt);
        \draw[thick,fill=white] (4.8,.7) ellipse (1.5pt and 3.55pt);
        \node[font=\footnotesize] (P1) at (4.6,.75) {$O_1^\prime$};
        \draw[thick,fill=white] (4.8,-.7) ellipse (1.5pt and 3.55pt);
        \node[font=\footnotesize] (P2) at (4.6,-.75) {$O_2^\prime$};
        \draw[thick,fill=white] (4.1,0) ellipse (1.5pt and 3.55pt);
        \node[font=\footnotesize] (P3) at (4.3,0) {$O$};   
    \end{tikzpicture} \,,
    \label{eqn:generalfixtures}
\end{align}
for some choice of third puncture $O$.

It remains to construct suitable pairs $(O_1,O_2)$ and $(O'_1,O'_2)$, such that the resulting 4d $\mathcal{N}=2$ SCFTs have all the same conventional invariants. As in \cite{Distler:2022nsn}, a mechanism for doing so can be constructed via a suitable minimal nilpotent Higgsings, \emph{i.e.} the Higgs branch renormalization group flow associated to turning on the highest-root moment map of some simple factor in the manifest flavor symmetry of one of the punctures. The effect is to replace the puncture $O$ with the puncture $O'$ such that $O$ is the minimal degeneration of $O'$.

Suppose that there exists a pair of nilpotent Higgsings by the same simple Lie algebra $\mathfrak{f}$ at the same level $k$ such that
\begin{equation}
\begin{split}
    O_1\xrightarrow{\mathfrak{f}_k} O'_1\,,\\
    O'_2\xrightarrow{\mathfrak{f}_k} O_2 \,.
\end{split}
\label{eqn:higgspair}
\end{equation}
In the class $\mathcal{S}$ construction, each puncture gives rise to a flavor algebra, referred to as the manifest flavor algebra $\mathfrak{f}(O)$, which is a subalgebra of the full flavor algebra of the SCFT.
It is shown in \cite{Distler:2022nsn} that matching the manifest flavor symmetries, \emph{i.e.}~imposing
\begin{equation}
    \mathfrak{f}(O_1)\oplus\mathfrak{f}(O_2) = \mathfrak{f}(O'_1)\oplus\mathfrak{f}(O'_2)
    \label{eqn:noenhanced}
\end{equation}
automatically leads to candidate pairs $(O_1,O_2)$ and $(O'_1,O'_2)$ such that the theories associated to the three-punctured spheres in equation \eqref{eqn:generalfixtures} have the same conventional invariants.\footnote{It is necessary to choose the third puncture $O$ to be sufficiently high up on the Hasse diagram of nilpotent orbits of $\mathfrak{g}$ to guarantee that the theories are not bad, in the sense of Gaiotto--Witten \cite{Gaiotto:2008ak}.} Later in this paper we will see that \eqref{eqn:noenhanced} is a sufficient, but not necessary, condition for obtaining isomorphic pairs. But, for now, let us impose it.

Even if we relax \eqref{eqn:noenhanced}, we still require that the flavor symmetry algebras and levels coincide. Indeed, we should go further and demand that the global forms of the flavour symmetry \emph{groups} coincide. When these differ, the theories are clearly not isomorphic. But imposing this stronger condition still does not suffice for the theories to be isomorphic.

In \cite{Distler:2022nsn}, a family of examples was constructed to illustrate this point. Consider the class $\mathcal{S}$ theory of type $\mathfrak{e}_7$ with the punctures\footnote{We denote the nilpotent orbits of exceptional Lie algebras using Bala--Carter notation \cite{MR417306,MR417307}. See the standard reference \cite{MR1251060} or the paper \cite{Chacaltana:2012zy} for a review.} 
\begin{equation}\label{eqn:firstE7example}
(O_1,O_2)=\bigl(A_3,D_6(a_1)\bigr)\,,\qquad(O'_1,O'_2)=\bigl((A_3+A_1)'',D_5\bigr) \,.
\end{equation}
Then different choices of the third puncture $O$ lead to various possibilities for the underlying pairs of SCFTs whose conventional invariants agree. We have depicted the Hasse diagram of $\mathfrak{e}_7$ nilpotent orbits in Figure \ref{fig:e7hasseO}, where we have color coded the choice of third picture $O$ according to three different cases based on whether the flavor symmetry algebra and the global form of the flavor symmetry coincides or not as the following:
\begin{itemize}
\item (black): the flavor symmetry algebras agree, but the global form of the flavor symmetry groups differ,
\item ({\color{ukyellow}yellow}): the global form of the flavor symmetry groups agree, but the theories are not isomorphic,
\item ({\color{midgreen}green}): the global form of the flavor symmetry groups agree and the theories appear to be isomorphic,
\item ({\color{red}red}): ``bad" three-punctured spheres, that are not associated to nontrivial 4d SCFTs.
\end{itemize}

\begin{figure}[H]
    \centering
    \begin{center}
    \scalebox{.62}{
    \begin{tikzpicture}
        \node (0) at (0,0) {$0$};
        \node[below=.75cm of 0] (A1) {$A_1$};
        \node[below=.75cm of A1] (A1A1) {$2A_1$};
        \node[below left=.25cm and 3cm of A1A1] (3A1pp) {${\color{ukyellow}(3A_1)''}$};
        \node[below right=.25cm and 3cm of A1A1] (3A1p) {$(3A_1)'$};
        \node[below=1cm of 3A1pp] (A1A1A1A1) {${\color{ukyellow}4A_1}$};
        \node[below=1cm of 3A1p] (A2) {$A_2$};
        \node[below=3.25cm of A1A1] (A2A1) {$A_2+A_1$};
        \node[below=.75cm of A2A1] (A2A1A1) {${A_2+2A_1}$};
        \node[below=3cm of A1A1A1A1] (A2A2) {$2A_2$};
        \node[below=3cm of A2] (A2A1A1A1) {${\color{ukyellow}A_2+3A_1}$};
        \node[below=2.7cm of A2A1A1] (A3) {$A_3$};
        \node[below=1cm of A2A1A1A1] (A2A2A1) {${\color{ukyellow}2A_2+A_1}$};
        \node[below=1cm of A3] (A3A1pp) {${\color{ukyellow}(A_3+A_1)''}$};
        \node[below=1cm of A2A2A1] (A3A1p) {${(A_3+A_1)'}$};
        \node[below=1cm of A3A1pp] (A3A1A1) {${A_3+2A_1}$};
        \node[below=1cm of A3A1p] (D4a1) {$D_4(a_1)$};
        \node[below=1cm of A3A1A1] (D4a1A1) {${\color{ukyellow}D_4(a_1)+A_1}$};
        \node[below=1cm of D4a1A1] (A3A2poo) {};
        \node[below=3.5cm of D4a1] (D4poo) {};
        \node[right=14cm of 0] (D4a1A1poo) {};
        \node[right=18.5cm of 0] (D4a1poo) {};
        \node[below=1.5cm of D4a1A1poo] (A3A2) {${\color{ukyellow}A_3+A_2}$};
        \node[below=1cm of D4a1poo] (D4) {$D_4$};
        \node[below left=3.125cm and 4cm of D4a1A1poo] (A4) {${\color{ukyellow}A_4}$};
        \node[below=1cm of A3A2] (A3A2A1) {${\color{ukyellow}A_3+A_2+A_1}$};
        \node[below=1cm of A3A2A1] (A4A1) {$A_4+A_1$};
        \node[below=2.8cm of D4] (D4A1) {${\color{ukyellow}D_4+A_1}$};
        \node[below=2.7cm of A4] (A5pp) {${\color{midgreen}(A_5)''}$};
        \node[below=1cm of A4A1] (A4A2) {${A_4+A_2}$};
        \node[below=1cm of D4A1] (D5a1) {$D_5(a_1)$};
        \node[below=1cm of A5pp] (A5A1) {${\color{midgreen}A_5+A_1}$};
        \node[below=1cm of A4A2] (A5p) {$(A_5)'$};
        \node[below=1cm of D5a1] (D5a1A1) {${\color{ukyellow}D_5(a_1)+A_1}$};
        \node[below=1cm of A5p] (D6a2) {${\color{midgreen}D_6(a_2)}$};
        \node[below=1cm of D5a1A1] (E6a3) {$E_6(a_3)$};
        \node[below=1cm of D6a2] (E7a5) {${\color{midgreen}E_7(a_5)}$};
        \node[below=1cm of E6a3] (D5) {${\color{red}D_5}$};
        \node[below=4.4cm of A5A1] (A6) {$A_6$};
        \node[below=1cm of E7a5] (D6a1) {${\color{red}D_6(a_1)}$};
        \node[below=1cm of D5] (D5A1) {${\color{red}D_5+A_1}$};
        \node[below=1cm of D6a1] (E7a4) {${\color{red}E_7(a_4)}$};
        \node[below=2.25cm of A6] (D6) {${\color{red}D_6}$};
        \node[below=2.25cm of D5A1] (E6a1) {${\color{red}E_6(a_1)}$};
        \node[below=1cm of D6] (E7a3) {${\color{red}E_7(a_3)}$};
        \node[below=1cm of E6a1] (E6) {${\color{red}E_6}$};
        \node[below=3cm of E7a4] (E7a2) {${\color{red}E_7(a_2)}$};
        \node[below=.5cm of E7a2] (E7a1) {${\color{red}E_7(a_1)}$};
        \path[->] (0) edge node[left] {$(\mathfrak{e}_{7})_{36}$} (A1)
        (A1) edge node[left] {$(\mathfrak{so}_{12})_{28}$} (A1A1)
        (A1A1) edge node[above left=-.125cm and -.125cm] {$(\mathfrak{su}_2)_{20}$} (3A1pp)
        (A1A1) edge node[above right=-.125cm and -.125cm] {$(\mathfrak{so}_9)_{24}$} (3A1p)
        (3A1pp) edge node[left] {$(\mathfrak{f}_4)_{24}$} (A1A1A1A1)
        (3A1p) edge node[below right=-.125cm and -.125cm] {$(\mathfrak{sp}_3)_{20}$} (A1A1A1A1)
        (3A1p) edge node[right] {$(\mathfrak{su}_2)_{19}$} (A2)
        (A1A1A1A1) edge node[below left=-.125cm and -.125cm] {$(\mathfrak{sp}_3)_{19}$} (A2A1)
        (A2) edge node[below right=-.125cm and -.125cm] {$(\mathfrak{su}_6)_{20}$} (A2A1)
        (A2A1) edge node[left] {$(\mathfrak{su}_4)_{18}$} (A2A1A1)
        (A2A1A1) edge node[above left=-.125cm and -.125cm] {$(\mathfrak{su}_2)_{28}$} (A2A2)
        (A2A1A1) edge node[above right=-.125cm and -.125cm] {$(\mathfrak{su}_2)_{16}$} (A2A1A1A1)
        (A2A1A1) edge[dashed] (A3)
        (A2A2) edge node[below left=-.125cm and -.125cm] {$(\mathfrak{g}_2)_{16}$} (A2A2A1)
        (A2A2) edge node[left] {$(\mathfrak{su}_2)_{36}$} (A3A1pp)
        (A2A1A1A1) edge node[right] {$(\mathfrak{g}_2)_{28}$} (A2A2A1)
        (A3) edge node[right] {$(\mathfrak{su}_2)_{12}$} (A3A1pp)
        (A3) edge node[above right=-.125cm and -.125cm] {$(\mathfrak{so}_{7})_{16}$} (A3A1p)
        (A2A2A1) edge[blue] node[right,black] {$(\mathfrak{su}_2)_{38}$} (A3A1p)
        (A3A1pp) edge node[left] {$(\mathfrak{so}_{7})_{16}$} (A3A1A1)
        (A3A1p) edge node[below right=-.125cm and -.125cm] {$(\mathfrak{su}_2)_{12}$} (A3A1A1)
        (A3A1p) edge node[right] {$(\mathfrak{su}_2)_{13}$} (D4a1)
        (A3A1A1) edge node[left] {$(\mathfrak{su}_2)_{13}$} (D4a1A1)
        (D4a1) edge node[below right=-.125cm and -.125cm] {$(\mathfrak{su}_2)_{12}$} (D4a1A1)
        (D4a1) edge[dashed] (D4poo)
        (D4a1A1) edge node[left] {$(\mathfrak{su}_2)_{12}$} (A3A2poo)
        (D4a1poo) edge[dashed] (D4)
        (D4a1A1poo) edge node[left] {$(\mathfrak{su}_2)_{12}$} (A3A2)
        (A3A2) edge[dashed] (A4)
        (A3A2) edge node[right] {$(\mathfrak{su}_2)_{12}$} (A3A2A1)
        (D4) edge node[right] {$(\mathfrak{sp}_3)_{12}$} (D4A1)
        (A4) edge[dashed] (A5pp)
        (A4) edge node[above right=-.125cm and -.125cm] {$(\mathfrak{su}_{3})_{12}$} (A4A1)
        (A3A2A1) edge[dashed] (A4A1)
        (A3A2A1) edge[dashed] (D4A1)
        (A4A1) edge[dashed] (A4A2)
        (A4A1) edge[dashed] (D5a1)
        (D4A1) edge node[right] {$(\mathfrak{sp}_{2})_{11}$} (D5a1)
        (A5pp) edge node[left] {$(\mathfrak{g}_2)_{12}$} (A5A1)
        (A4A2) edge[dashed] (A5A1)
        (A4A2) edge[dashed] (A5p)
        (A4A2) edge[dashed] (D5a1A1)
        (D5a1) edge node[right] {$(\mathfrak{su}_2)_{10}$} (D5a1A1)
        (A5A1) edge[blue] node[below left=-.125cm and -.125cm,black] {$(\mathfrak{su}_2)_{26}$} (D6a2)
        (A5p) edge node[left] {$(\mathfrak{su}_2)_{20}$} (D6a2)
        (A5p) edge node[above right=.125cm and -1cm] {$(\mathfrak{su}_2)_9$} (E6a3)
        (D5a1A1) edge[dashed] (D6a2)
        (D5a1A1) edge[dashed] (E6a3)
        (D6a2) edge node[left] {$(\mathfrak{su}_2)_{9}$} (E7a5)
        (E6a3) edge node[above left=-.125cm and -.125cm] {$(\mathfrak{su}_2)_{20}$} (E7a5)
        (E6a3) edge[dashed] (D5)
        (E7a5) edge[dashed] (A6)
        (E7a5) edge[dashed,red] (D6a1)
        (E7a5) edge[dashed,red] (D5A1)
        (D5) edge[red] node[above left=.25cm and -1cm] {$(\mathfrak{su}_2)_{12}$} (D6a1)
        (D5) edge[red] node[right] {$(\mathfrak{su}_2)_{8}$} (D5A1)
        (A6) edge[red] node[below left=-.125cm and -.125cm] {$(\mathfrak{su}_2)_{36}$} (E7a4)
        (D6a1) edge[red] node[left] {$(\mathfrak{su}_2)_{8}$} (E7a4)
        (D5A1) edge[red] node[below right=-.125cm and -.125cm] {$(\mathfrak{su}_2)_{12}$} (E7a4)
        (E7a4) edge[dashed,red] (D6)
        (E7a4) edge[dashed,red] (E6a1)
        (D6) edge[red] node[left] {$(\mathfrak{su}_2)_{7}$} (E7a3)
        (E6a1) edge[dashed,red] (E7a3)
        (E6a1) edge[dashed,red] (E6)
        (E7a3) edge[dashed,red] (E7a2)
        (E6) edge[red] node[below right=-.125cm and -.125cm] {$(\mathfrak{su}_2)_{12}$} (E7a2)
        (E7a2) edge[dashed, red] (E7a1)
        ;
    \end{tikzpicture}}
    \end{center}
    \caption{The possible choices of the third puncture $O$ in equation \eqref{eqn:generalfixtures} for the pairs of punctures in equation \eqref{eqn:firstE7example}.} %The black ones refer to the cases where flavor symmetry algebras agree, but the global form of the flavor symmetry groups differ; the {\color{ukyellow}yellow} ones correspond to the cases where the global form of the flavor symmetry groups agree, but the theories are not isomorphic; the {\color{midgreen}green} ones are when the global form of the flavor symmetry groups agree and the theories appear to be isomorphic. We depict in {\color{red}red} for the ``bad" three-punctured spheres that are not associated to nontrivial 4d SCFTs.}
    \label{fig:e7hasseO}
\end{figure}

In this family of examples, there are four choices for $O$ that appear to lead to isomorphic pairs of theories: 
\begin{align}
    (A_5)''\,,\quad A_5+A_1\,, \quad D_6(a_2)\,, \quad E_7(a_5)\,. 
\end{align}
The Schur indices up to $O(\tau^{12})$ were computed in \cite{Distler:2022nsn} and shown to coincide. While persuasive, this is far from sufficient to prove that the pairs of theories are isomorphic.

On the other hand, once we have shown that the pair of theories with third puncture $O=(A_5)''$ are isomorphic, then the isomorphism follows for the other three choices of third puncture. In particular, it was shown in \cite{Distler:2022nsn}, that any (not necessarily minimal) nilpotent Higgsing of the third puncture of a pair of isomorphic theories leads to a new pair of isomorphic theories. Consider the following sequence of local Higgsings: 
\begin{equation}\label{eqn:swiper}
\begin{tikzpicture}
\node[color=midgreen] (A5pp) at (0,0) {$(A_5)''$};
\node[color=midgreen] (A5A1) at (2,0) {$A_5+A_1$};
\node[color=midgreen] (D6a2) at (5,0) {$D_6(a_2)$};
\node[color=midgreen] (E7a5) at (7.5,0) {$E_7(a_5)$};
\path[thick, ->,draw,color=olive] 
(A5pp) edge[color=black] node[color=black,below] {$\scriptstyle (\mathfrak{g}_2)_{12}$} (A5A1)
(A5A1) edge[color=blue] node[color=black,below] {$\scriptstyle \mathfrak{su}(2)_{26}$} (D6a2)
(D6a2) edge[color=black] node[color=black,below] {$\scriptstyle \mathfrak{su}(2)_{9}$} (E7a5)
(0.3,0.3) arc[color=black,radius=4.2, start angle=120, end angle=60] node[color=black,above right=.5cm and -2.5cm] {$\scriptstyle (\mathfrak{g}_2)_{12}$} (D6a2)
;
\end{tikzpicture}
 \,.
\end{equation}
Black arrows correspond to minimal nilpotent Higgsings associated to giving a VEV to the highest-root moment map operator of the non-Abelian flavor symmetry which decorates the arrow. The olive arrow is a non-minimal nilpotent Higgsing corresponding to giving a VEV to a $(\mathfrak{g}_2)_{12}$ moment map in the next-to-minimal nilpotent orbit. The Higgsing from $A_5+A_1\to D_6(a_2)$ is not a nilpotent Higgsing, as it does not involve solely giving a VEV to a moment map operator, but nevertheless such a Higgs branch renormalization group flow exists, as shown in \cite{DEE8}. Non-nilpotent Higgsings are in general noteworthy and will be explored in more detail in a future work \cite{DKL}. But we do not need to discuss them here; all of the punctures corresponding to additional isomorphic pairs are obtainable from $(A_5)''$ by a sequence of nilpotent Higgsings.

What remains then is to prove the isomorphism for $O = (A_5)''$. The proof involves a detour via a similar question about isomorphisms between 6d $(1,0)$ SCFTs known as Higgsed rank $N$ $(\mathfrak{g}, \mathfrak{g})$ conformal matter. Each of these 6d SCFTs corresponds to a pair of nilpotent orbits, $O_1$ and $O_2$, as we explain in detail in Section \ref{sec:6d}. The compactification of such SCFTs on a torus are dual to class $\mathcal{S}$ of type $\mathfrak{g}$ on a sphere with $N+2$ punctures, $N$ of which are simple punctures, and the remaining two are associated to the nilpotent orbits $O_1$ and $O_2$.\footnote{More precisely, it is a certain codimension $(N-2)$ sublocus of the conformal manifold of the class $\mathcal{S}$ theory that is dual to the $T^2$-compactification of the 6d $(1,0)$ SCFT.} This is precisely the kind of class $\mathcal{S}$ theories discussed around equation \eqref{genus2isom}.

Specializing to the case of $\mathfrak{g} = \mathfrak{e}_7$, where the simple puncture is denoted by $E_7(a_1)$, and picking the two punctures $(O_a,O_b)=(O_1,O_2)$ or $(O'_1,O'_2)$ we can further degenerate the genus $g=0$ Riemann surface on the left in equation \eqref{genus2degen} to write the $(N+2)$-punctured sphere as 
\begin{align}
\begin{aligned}
    \scalebox{.61}{
    \begin{tikzpicture}
        \draw[radius=40pt,fill=lightred] circle;
        \draw[radius=2pt,fill=white]  (-.5,.9) circle node[right=2pt] {$E_7(a_1)$};
        \draw[radius=2pt,fill=white]  (-.5,-.9) circle node[right=2pt] {$E_7(a_1)$};
        \draw[radius=2pt,fill=white]  (1,0) circle node[left=2pt] (D6su2) {$\bigl(D_6,SU(2)\bigr)$};
        \node at (0,-2) {$\tfrac{1}{2}(2,1,1)$};
        \draw[radius=40pt,fill=lightred] (4,0) circle;
        \draw[radius=2pt,fill=white]  (4,1) circle node[below=2pt] {$E_7(a_1)$};
        \draw[radius=2pt,fill=white]  (3,0) circle node[right=2pt] (D6) {$D_6$};
        \draw[radius=2pt,fill=white]  (5,0) circle node[below left=2pt and -5pt] (A5ppG2){$\bigl({(A_5)}'',G_2\bigr)$};
        \path (1.1,0) edge node[above] {$SU(2)$}  (2.9,0);
        \node at (4,-2) {$\tfrac{1}{2}(2,7,1)$};
        \draw[radius=40pt,fill=lightblue] (8,0) circle;
        \draw[radius=2pt,fill=white]  (8,1) circle node[below=2pt] {$E_7(a_1)$};
        \draw[radius=2pt,fill=white]  (7,0) circle node[right=2pt] (A5pp) {${(A_5)}''$};
        \draw[radius=2pt,fill=white]  (9,0) circle node[below left=2pt and -5pt] (A5ppG2){$\bigl({(3A_1)}'',F_4\bigr)$};
        \path (5.1,0) edge node[above] {$G_2$}  (6.9,0);
        \node at (8,-2) {$[{(E_8)}_{12}\;\text{SCFT}]$};
        \draw[radius=40pt,fill=lightblue] (12,0) circle;
        \draw[radius=2pt,fill=white]  (12,1) circle node[below=2pt] {$E_7(a_1)$};
        \draw[radius=2pt,fill=white]  (11,0) circle node[below right=2pt and -5pt] (A1A1A1pp) {${(3A_1)}''$};
        \draw[radius=2pt,fill=white]  (13,0) circle node[below left=2pt and -5pt] {$0$};
        \path (9.1,0) edge node[above] {$F_4$}  (10.9,0);
        \node at (11.6,-2) {$[{(F_4)}_{24}\times {(E_7)}_{36}\;\text{SCFT}]$};
        \draw[radius=40pt,fill=lightblue] (16,0) circle;
        \draw[radius=2pt,fill=white]  (16,1) circle node[below=2pt] {$E_7(a_1)$};
        \draw[radius=2pt,fill=white]  (15,0) circle node[below right=2pt and -5pt] (A1A1A1pp) {$0$};
        \draw[radius=2pt,fill=white]  (17,0) circle node[below left=2pt and -5pt] {$0$};
        \path (13.1,0) edge node[above] {$E_7$}  (14.9,0);
        \node at (16,-2) {$[{(E_7)}_{36}\times {(E_7)}_{36}\;\text{SCFT}]$};
        \node at (18.5,0) (dots) {$\dots$};
        \path (17.1,0) edge node[above right=0pt and -5pt] {$E_7$}  (dots);
        \draw[radius=40pt,fill=lightblue] (21,0) circle;
        \draw[radius=2pt,fill=white]  (20,0) circle node[right=2pt] {$0$};
        \draw[radius=2pt,fill=white]  (21.5,.9) circle node[left=2pt] {$O_a$};
        \draw[radius=2pt,fill=white]  (21.5,-.9) circle node[left=2pt] {$O_b$};
        \path (dots) edge node[above left=0pt and -5pt] {$E_7$}  (19.9,0);
    \end{tikzpicture}}
\end{aligned} \,. \label{eqn:chainofP1}
\end{align}
When $N\geq5$, the three-punctured sphere on the right has the full puncture, $0$, along with $O_a$ and $O_b$. For $N=4,3,2$, the full puncture is replaced by $(3A_1)''$, $(A_5)''$ or $D_6$, respectively. Thus, for low values of $N$, we are probing the physics of (an $F_4$, $G_2$ or $SU(2)$ gauging of) the three-punctured sphere with $(3A_1)''$, $(A_5)''$ or $D_6$ as the third puncture. If the two 6d $(1,0)$ SCFTs that yields these theories with $(O_a, O_b)$ are isomorphic, then the three-punctured spheres on the right are also isomorphic. For the pairs of punctures in equation \eqref{eqn:firstE7example}, the 6d $(1,0)$ theories are manifestly non-isomorphic for $N>3$. Hence their $T^2$ compactification to 4d does not yield isomorphic theories. However, for $N=3$, we find that the theories are manifestly isomorphic as 6d $(1,0)$ SCFTs. Hence so are their compactifications to 4d.

We note that in equation \eqref{eqn:chainofP1}, we made an assumption that $O_a$ and $O_b$ are high-enough up on the Hasse diagram such that the theory
\begin{align}
\begin{aligned}
    \begin{tikzpicture}
        \draw[radius=40pt,fill=lightblue] (0,0) circle;
        \draw[radius=2pt,fill=white]  (-1,0) circle node[right=2pt] {$0$};
        \draw[radius=2pt,fill=white]  (.5,.9) circle node[left=2pt] {$O_a$};
        \draw[radius=2pt,fill=white]  (.5,-.9) circle node[left=2pt] {$O_b$};
    \end{tikzpicture}  
\end{aligned} \quad\,, 
\end{align}
is ``good'', \emph{i.e.}, that the compactification of the $(2,0)$ theory from 6d yields a nontrivial 4d SCFT. Of the 990 pairs, $(O_a,O_b)$ in the $\mathfrak{e}_7$ theory, 49 of them are ``bad''.

\section{Nilpotent Higgsing and identical 6d \texorpdfstring{{\boldmath{$(1,0)$}}}{(1,0)} SCFTs}\label{sec:6d}

In Section \ref{sec:4d}, we showed how to construct pairs of class $\mathcal{S}$ theories whose conventional invariants coincide. These are are candidates for being isomorphic $\mathcal{N}=2$ SCFTs. With considerable additional effort, we could narrow down the list of candidate isomorphic SCFTs by computing their spectra of Schur operators up to some high order. But this was still far from sufficient to show that the theories are, in fact, isomorphic.

In this section, we utilize a different, and rather simple, approach to determining whether or not two of the 4d $\mathcal{N}=2$ SCFTs of interest are truly identical. First, we ask a similar question in the context of a class of 6d $(1,0)$ SCFTs that are each associated to a pair of nilpotent orbits of an ADE Lie algebra; on torus-compactification, such SCFTs are known to be dual to class $\mathcal{S}$ theories on spheres where two of the punctures are the same as the nilpotent orbits in the 6d theory, and the rest are simple punctures. At first, it appears that we have merely uplifted the same problem of determining when two SCFTs are the same to six dimensions; however, the landscape of consistent 6d $(1,0)$ SCFTs is highly constrained from string theory \cite{Heckman:2013pva,Heckman:2015bfa}.\footnote{For recent reviews of the power of string theory in constraining (6d) SCFTs, see \cite{Heckman:2018jxk,Argyres:2022mnu}.} In this way, knowledge of the conventional invariants of the 6d $(1,0)$ SCFT is often enough to fully determine the SCFT, or else leaving only a small number of possibilities. We can determine for what pairs of nilpotent orbits two such 6d SCFTs are the same, and this leads to the same conclusion for the 4d $\mathcal{N}=2$ class $\mathcal{S}$ theories obtained via torus-compactification, and related SCFTs obtained from degenerations and partial puncture closure.

The rank $N$ $(\mathfrak{g}, \mathfrak{g})$ conformal matter is the 6d $(1,0)$ SCFT that lives on the worldvolume of $N$ M5-branes probing a $\mathbb{C}^2/\Gamma_\mathfrak{g}$ singularity \cite{DelZotto:2014hpa}. Here $\mathfrak{g}$ can be any ADE Lie algebra, and $\Gamma_\mathfrak{g}$ is the finite subgroup of $SU(2)$ associated to $\mathfrak{g}$ via the McKay correspondence \cite{MR604577}. Each conformal matter theory has a non-Abelian flavor algebra which is
\begin{equation}
    \mathfrak{f} = \mathfrak{g} \oplus \mathfrak{g} \,,
\end{equation}
though this can be enhanced for small values of $N$. From this starting point, new 6d $(1,0)$ SCFTs can be obtained via Higgs branch renormalization group flows triggered by giving nilpotent vacuum expectation values (VEVs) to the moment map operators associated to these two flavor symmetries. In particular, we can consider the family of theories
\begin{equation}
  \mathcal{T}_{\mathfrak{g}, N}(O_L, O_R) \,,
\end{equation}
where $O_L$ and $O_R$ are nilpotent orbits in $\mathfrak{g}$. When $O_L = O_R = 0$, \emph{i.e.}, the trivial nilpotent orbit, then we recover the original conformal matter theory, which is often referred to as simply $\mathcal{T}_{\mathfrak{g}, N}$. Such families of 6d $(1,0)$ SCFTs related via a nilpotent hierarchy have been studied in great detail, see, for example, \cite{Heckman:2015ola,Heckman:2016ssk,Mekareeya:2016yal,Heckman:2018pqx,DeLuca:2018zbi,Hassler:2019eso,Baume:2021qho,Bourget:2022tmw}. When $\mathcal{T}_{\mathfrak{g}, N}$ is compactified on a torus, the 4d $\mathcal{N}=2$ SCFT that is obtained is known to have a dual description in terms of the class $\mathcal{S}$ construction \cite{Ohmori:2015pua,DelZotto:2015rca,Ohmori:2015pia}. In particular, the $T^2$ compactification is dual to the compactification of the 6d $(2,0)$ SCFT of type $\mathfrak{g}$ on a sphere with two full punctures and $N$ simple punctures. Higgsing of the 6d $(1,0)$ SCFT by giving a VEV valued in a nilpotent orbit of $\mathfrak{g}$ to the moment map operators then corresponds to the partial closure of the full punctures in the dual class $\mathcal{S}$ description; this proposal has been tested extensively in \cite{Mekareeya:2017sqh,Baume:2021qho,Distler:2022yse}.

In this section, the question we will attempt to answer is: 
\begin{gather*}
\textbf{When do the interacting sectors of two 6d $(1,0)$}\\ 
\textbf{SCFTs $\mathcal{T}_{\mathfrak{g}, N}(O_L, O_R)$ and $\mathcal{T}_{\mathfrak{g}, N'}(O_L', O_R')$ match?}
\end{gather*}
It is straightforward to see that $\mathcal{T}_{\mathfrak{g}, N}(O_L, O_R)$ and $\mathcal{T}_{\mathfrak{g}, N'}(O_L', O_R')$ can only have identical interacting parts if $N = N'$. Rank $N$ conformal matter, for any $\mathfrak{g}$, possesses a sequence of Higgs branch renormalization group flows which eventually ends at the 6d $(2,0)$ SCFT of type $A_{N-1}$ \cite{Ohmori:2015pia}. In particular, after going to the superconformal point at the origin of the tensor branch, the Type IIB geometry is an elliptic fibration over a base $\mathbb{C}^2/\mathbb{Z}_N$. The defect group of a 6d SCFT depends only on the information of the base \cite{DelZotto:2015isa}; for $\mathbb{C}^2/\mathbb{Z}_N$, it is simply $\mathbb{Z}_N$. Hence, two interacting SCFTs with different defect groups cannot be identical. A priori, we could consider a more general version of this question where $\mathfrak{g}' \neq \mathfrak{g}$, as there are known examples where such different Higgsed conformal matter theories lead to the same 6d $(1,0)$ SCFTs. However, we will not consider these cases in this paper and will restrict to the setting where $\mathfrak{g}$ remains the same.\footnote{In fact, in this paper, we focus on the case where $\mathfrak{g} = \mathfrak{e}_7$; however, we include some examples for other gauge algebras in Section \ref{sec:other}.} In special cases, this question has been answered previously, for example, all the theories $\mathcal{T}_{\mathfrak{g}, 1}(O_L, O_R)$ and $\mathcal{T}_{\mathfrak{g}, 1}(O_L', O_R')$, where $\mathfrak{g}$ is an exceptional Lie algebra, with isomorphic interacting sectors have been tabulated in \cite{Baume:2021qho}.

The analysis in this paper makes much use of the atomic construction of 6d $(1,0)$ SCFTs, which we briefly review here. We consider Type IIB string theory compactified to six dimensions on a complex K\"ahler surface $B$, and with a non-trivial axio-dilaton profile turned on along $B$ so as to preserve eight supercharges. The consistency of the axio-dilaton profile can be rephrased as an elliptic fibration over $B$, such that the total space of the fibration is a Calabi--Yau threefold. Compactifications on non-compact elliptically-fibered Calabi--Yau threefolds, satisfying some conditions such as the absence of compact complex curves in $B$, lead to 6d $(1,0)$ SCFTs. Such threefolds have non-minimal singular fibers and may also have orbifold-like singularities in the base: $B = \mathbb{C}^2/\Gamma$, where $\Gamma$ is a finite subgroup of $U(2)$. Performing a sequence of K\"ahler deformations leads to a new non-compact elliptically-fibered Calabi--Yau threefold for which $B$ is smooth and all singular fibers are minimal. Physically, this procedure involves giving vacuum expectation values to the scalar primaries inside all of the tensor multiplets of the SCFT; thus, the new Calabi--Yau threefold describes the tensor branch effective field theory associated to the SCFT.

It turns out that the Calabi--Yau geometries that can correspond to a tensor branch configuration are highly constrained. The only compact curves that the base can contain are rational curves, and those must intersection in an intersection matrix $A^{ij}$ which is negative-definite; furthermore, the self-intersection number of each rational curve is constrained to be $\geq -12$, and adjacent curves can only intersect with intersection number one. The detailed conditions have recently been reviewed in \cite{Heckman:2018jxk} and were summarized in recent works of (subsets of) the current authors \cite{Baume:2021qho,Distler:2022yse}. Enumerating tensor branch geometry then reduces to a problem of combining rational curves and singular fibers/algebras in such a way that the necessary conditions are satisfied. In the end, we utilize a common shorthand notation, which we explain via an example. Consider
\begin{equation}
    \overset{\mathfrak{su}_3}{3}1\overset{\mathfrak{su}_3}{3} \,.
\end{equation}
This represents a non-compact elliptically-fibered Calabi--Yau threefold where the base contains three $\mathbb{P}^1$s, which intersect in the following intersection matrix:
\begin{equation}
    A_{ij} = \begin{pmatrix}
        -3 & 1 & 0 \\
        1 & -1 & 1 \\
        0 & 1 & -3 
    \end{pmatrix}_{ij} \,.
\end{equation}
The two $(-3)$-curves are written with an $\mathfrak{su}_3$ above them, this indicates that the singular fiber over those curves is of type IV; physically each singular fiber is associated to an algebra, and that algebra provides a gauge algebra of the effective field theory on the tensor branch. The tensor branch field theory also contains hypermultiplets transforming in representations of the gauge algebra, however it is redundant to write them in the shorthand notation, as the number and representation is entirely fixed by gauge-anomaly cancellation, after specifying the self-intersection number and the gauge algebra: $\overset{\mathfrak{g}}{n}$.\footnote{There are a small number of cases where the hypermultiplet spectrum is ambiguous even after specifying the self-intersection number and the algebra, however these situations will not arise in this paper.} We use this concise notation to refer to a tensor branch effective field theory for a 6d $(1,0)$ SCFT throughout this paper; we refer the reader to the review \cite{Heckman:2018jxk} for a comprehensive explanation.

Now that we have introduced a construction for 6d $(1,0)$ SCFTs from string theory, we would like to know what physical properties of the resulting SCFTs can be determined from knowledge of the tensor branch description. When a 6d $(1,0)$ SCFT is compactified on a $T^2$, without any additional bells and whistles, then the dimension of the Coulomb branch of the resulting 4d $\mathcal{N}=2$ SCFT is equal to the sum of number of tensor multiplets plus the sum of the ranks of the gauge algebras of the 6d theory. To avoid confusion, we will call this number $\rankfour$, even though it is an intrinsic property of the 6d SCFT.

The hallmark of 6d $(1,0)$ SCFTs is the anomaly polynomial. The anomaly polynomial of a 6d $(1,0)$ SCFT is a formal eight-form written in terms of the characteristic classes of the R-symmetry, Lorentz symmetry, and the flavor symmetry. It takes the form
\begin{equation}\label{eqn:apgeneral}
  \begin{aligned}
    I_8 &=  \frac{\alpha}{24} c_2(R)^2+ \frac{\beta}{24}  c_2(R) p_1(T) + \frac{\gamma}{24}  p_1(T)^2 + \frac{\delta}{24} p_2(T) \cr &\quad + \sum_a \text{Tr}F_a^2 \left(\kappa_a p_1(T) + \nu_a c_2(R) + \sum_b \rho_{ab} \text{Tr}  F_b^2\right) + \sum_a \mu_a \text{Tr}F_a^4  \,,
  \end{aligned}
\end{equation}
where each summation over $a$ or $b$ runs over the simple non-Abelian flavor symmetries of the theory. The coefficients in the anomaly polynomial are referred to as the 't Hooft coefficients. The anomaly polynomial can be determined from the effective tensor branch description of the SCFT following \cite{Ohmori:2014kda,Intriligator:2014eaa,Cordova:2020tij,DelZotto:2018tcj,Baume:2021qho}.

As we have discussed, in four dimensions the $\mathcal{N}=2$ conventional invariants are the central charges, the flavor algebras, and the flavor central charges;\footnote{Generally, we include more information in the 4d conventional invariants, such as the graded Coulomb branch scaling dimensions.} and these quantities are neatly summarized in the 4d anomaly polynomial 
\begin{equation}\label{eqn:I6}
    I_6 = 24(a-c)\left(\frac{1}{3}c_1(R)^3 - \frac{1}{12}c_1(R)p_1(T_4) \right) - 4(2a - c)c_1(R)c_2(R) + \sum_a k_a c_1(R) c_2(F_a) \,,
\end{equation}
where $c_1(R)$ is the first Chern class of the $U(1)$ R-symmetry, $c_2(R)$ is the second Chern class of the $SU(2)$ R-symmetry, $p_1(T_4)$ is the first Pontryagin class of the tangent bundle to the 4d spacetime, and $c_2(F_a)$ are the curvatures of the simple non-Abelian flavor symmetry factors.\footnote{For ease of explanation we do not write the Abelian flavor symmetries in the anomaly polynomial in equation \ref{eqn:I6}, however they are of course included in spirit.} A natural generalization of this notion of conventional invariants to the 6d $(1,0)$ context is to again to take the anomaly polynomial $I_8$. This quantity, $I_8$, satisfies the necessary condition to form an invariant: SCFTs with different anomalies polynomials are necessarily different CFTs. Interestingly, if a 6d $(1,0)$ SCFT $\mathcal{T}_{\mathfrak{g},N}(O_L, O_R)$ is compactified on a torus, then the resulting 4d $\mathcal{N}=2$ anomaly polynomial depends only on a subset of the anomaly coefficients appearing in equation \eqref{eqn:apgeneral}. In particular, we can consider the ``$\mathcal{N}=2$ subsector of the 6d conventional invariants'', which includes only the coefficients
\begin{equation}
    \beta\,,\quad \gamma\,,\quad \delta\,,\quad \{\kappa_a\} \,.
\end{equation}
It is noteworthy then that there can exist 6d $(1,0)$ SCFTs which have different conventional invariants, but which compactify on a $T^2$ to 4d $\mathcal{N}=2$ SCFTs with the same conventional invariants. Generally these two 4d SCFTs will be different, as the differences in six dimensions should be reflected in the torus-compactification; in particular, we would like to understand how the 6d $(1,0)$ anomaly coefficient $\alpha$ affects the 4d physics. We leave a detailed answer to this question for future work.

While the 6d $(1,0)$ anomaly polynomial is a powerful invariant of the Higgsed conformal matter theories that we consider in this paper, it is not complete. There exist a small number of interacting 6d $(1,0)$ SCFTs with the same anomaly polynomial, and yet which are distinct SCFTs. Modulo the subtleties\footnote{\label{fn:pm}The usual tensor branch description, involving a collection of exceptional curves and the elliptic fibers over them may need to be supplemented, as in \cite{Distler:2022yse}, by a choice of $\pm$ chiral projection for each $(-1)$ curve. As in \cite{Distler:2022yse}, these sign choices are well-defined modulo outer-automorphisms of the gauge and flavor algebras. In the examples discussed in this paper, they are completely removable.} explored in \cite{Distler:2022yse}, the effective field theory on the tensor branch does provide a complete invariant. To determine whether two theories $\mathcal{T}_{\mathfrak{g}, N}(O_L, O_R)$ and $\mathcal{T}_{\mathfrak{g}, N}(O_L', O_R')$ possess the same interacting sector, it suffices to determine the curve configuration/tensor branch description of each theory. If the curve configurations are the same, then the SCFTs that live at the origin of the tensor branch are the same.

When a Higgs branch renormalization group flow is triggered by giving a nilpotent vacuum expectation value, associated to a nilpotent orbit $O$, to the moment map of a $\mathfrak{g}$ flavor symmetry, the resulting SCFT typically has a reduced flavor symmetry. The nilpotent orbit $O$ is associated to an embedding $\rho_O : \mathfrak{su}_2 \rightarrow \mathfrak{g}$, and the residual flavor symmetry is defined to be the commutant in $\mathfrak{g}$ of this embedded $\mathfrak{su}_2$; we call this $\mathfrak{f}(O)$. Then, the manifest non-Abelian flavor algebra of $\mathcal{T}_{\mathfrak{g}, N}(O_L, O_R)$ is 
\begin{equation}
    \mathfrak{f}^\text{manifest} = \mathfrak{f}(O_L) \oplus \mathfrak{f}(O_R) \,.
\end{equation}
However, the flavor symmetry of the interacting sector of $\mathcal{T}_{\mathfrak{g}, N}(O_L, O_R)$ may differ from this manifest symmetry. That is, the flavor symmetry and the flavor central charges are conventional invariants of the 6d $(1,0)$ SCFT which cannot be read off from the pair of nilpotent orbits directly, but one must first go through the intermediate step of constructing the tensor branch description, and then use the procedure described in \cite{Baume:2021qho} to read off the correct non-Abelian flavor algebras and their flavor central charges. 

There is no guarantee that $\mathcal{T}_{\mathfrak{g}, N}(O_L, O_R)$ is an interacting SCFTs with no free sector and a single stress-energy tensor. Therefore, it is not only important to know the effective tensor branch description of the interacting sector, but also to be able to determine the number of free hypermultiplets in the spectrum. The quaternionic dimension of the Higgs branch of $\mathcal{T}_{\mathfrak{g}, N}(O_L, O_R)$ is
\begin{equation}\label{eqn:HBfull}
    \operatorname{dim}(\mathcal{H}) = N + \operatorname{dim}(\mathfrak{g}) - \operatorname{dim}(O_L) - \operatorname{dim}(O_R) \,.
\end{equation}
Here the dimension of the nilpotent orbit is as defined in \cite{MR1251060}. In contrast, if we have a tensor branch configuration corresponding to a single interacting SCFT that is Higgsable to a 6d $(2,0)$ SCFT of rank $N-1$, then the dimension of the Higgs branch of this interacting SCFT can be obtained from the anomaly polynomial \cite{Ohmori:2015pua,Ohmori:2015pia}. It is 
\begin{equation}\label{eqn:HBinter}
    \operatorname{dim}(\mathcal{H}) = -60\delta - 29(N-1) \,,
\end{equation}
where $\delta$ is the coefficient of the $p_2(T)$ term appearing in the anomaly polynomial in equation \eqref{eqn:apgeneral}. The difference between the dimension of the Higgs branch in equations \eqref{eqn:HBfull} and \eqref{eqn:HBinter} is the number of free hypermultiplets in $\mathcal{T}_{\mathfrak{g}, N}(O_L, O_R)$. A free hypermultiplet transforming in a representation $\bm{R}$ of a flavor algebra $\mathfrak{f}$ contributes to the anomaly polynomial as
\begin{equation}
    I_8^\text{free hyper}(\mathfrak{f}, \bm{R}, F) = \frac{1}{24} \text{Tr}_{\bm{R}} F^4 + \frac{1}{48}p_1(T)  \text{Tr}_{\bm{R}} F^2 + \frac{\text{dim}(\bm{R})}{5760}\left( 7p_1(T)^2 - 4p_2(T) \right) \,,
\end{equation}
where $F$ is the curvature associated to $\mathfrak{f}$. Thus, we can see that adding free hypermultiplets does not modify the anomaly coefficients $\alpha$ and $\beta$ in the combined theory. As these coefficients are insensitive to the inclusion of free sectors, they are conventional invariants which are well-suited for searching for $\mathcal{T}_{\mathfrak{g}, N}(O_L, O_R)$ SCFTs with the same interacting sector.

For the remainder of the section, we specialize to studying isomorphic pairs of the 6d $(1,0)$ SCFTs from rank $N$ $(\mathfrak{e}_7, \mathfrak{e}_7)$ conformal matter that leads to class $\mathcal{S}$ theories via toroidal compactifications. It is a curious quirk that the isomorphism of 6d $(1,0)$ SCFTs is more obvious when considering $\mathfrak{g} = \mathfrak{e}_n$ than when considering $\mathfrak{g}$ a classical Lie algebra. We first highlight the power of the curve configuration in determining isomorphisms by analyzing in detail for two examples in Section \ref{sec:6degg}, and then provide all the isomorphic pairs with $\mathfrak{g}=\mathfrak{e}_7$ in Section \ref{sec:isoe7}.

\subsection{Examples of two pairs of isomorphic SCFTs with \texorpdfstring{$\mathfrak{g} = \mathfrak{e}_7$}{g = e7}}\label{sec:6degg}

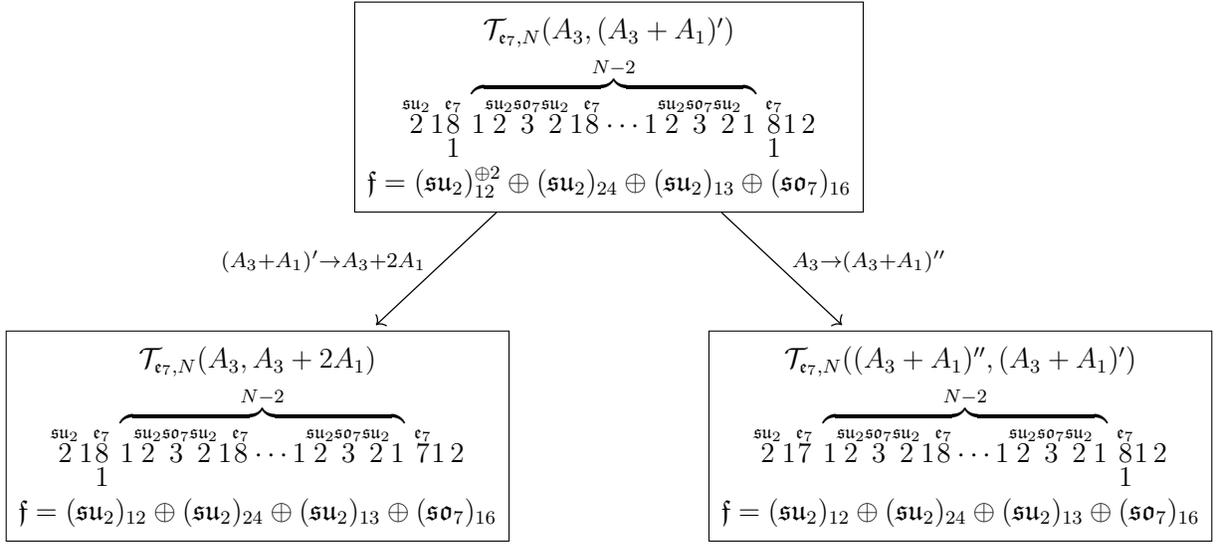
\begin{figure}[H]
\centering
    \begin{tikzcd}[cells={nodes={draw=black}},column sep=-5em,row sep=3.8em,scale cd=0.9]
        & \begin{tabular}{@{}c@{}}
            $\mathcal{T}_{\mathfrak{e}_7, N}(A_3, (A_3 + A_1)')$ \\[3pt]
            $\overset{\mathfrak{su}_2}{2}1\underset{\displaystyle 1}{\overset{\mathfrak{e}_7}{8}} \,
            \overbrace{1\overset{\mathfrak{su}_2}{2}\overset{\mathfrak{so}_7}{3}\overset{\mathfrak{su}_2}{2}1\overset{\mathfrak{e}_7}{8}\cdots1\overset{\mathfrak{su}_2}{2}\overset{\mathfrak{so}_7}{3}\overset{\mathfrak{su}_2}{2}1}^{N-2} \, \underset{\displaystyle 1}{\overset{\mathfrak{e}_7}{8}}1\,2$ \\[8pt]
            $\mathfrak{f} = (\mathfrak{su}_2)_{12}^{\oplus 2} \oplus (\mathfrak{su}_2)_{24} \oplus  (\mathfrak{su}_2)_{13} \oplus (\mathfrak{so}_7)_{16}$
        \end{tabular} \arrow[shorten >=3pt]{ld}[swap,yshift=-3pt]{(A_3 + A_1)' \rightarrow A_3 + 2A_1}\arrow[shorten >=3pt]{rd}[yshift=-3pt]{A_3 \rightarrow (A_3 + A_1)''} & \\
        \begin{tabular}{@{}c@{}}
            $\mathcal{T}_{\mathfrak{e}_7, N}(A_3, A_3 + 2A_1)$ \\[3pt]
            $\overset{\mathfrak{su}_2}{2}1\underset{\displaystyle 1}{\overset{\mathfrak{e}_7}{8}} \,
            \overbrace{1\overset{\mathfrak{su}_2}{2}\overset{\mathfrak{so}_7}{3}\overset{\mathfrak{su}_2}{2}1\overset{\mathfrak{e}_7}{8}\cdots1\overset{\mathfrak{su}_2}{2}\overset{\mathfrak{so}_7}{3}\overset{\mathfrak{su}_2}{2}1}^{N-2} \,
            \overset{\mathfrak{e}_7}{7}1\,2$ \\[8pt]
            $\mathfrak{f} = (\mathfrak{su}_2)_{12} \oplus (\mathfrak{su}_2)_{24} \oplus  (\mathfrak{su}_2)_{13} \oplus (\mathfrak{so}_7)_{16}$
        \end{tabular} & & \begin{tabular}{@{}c@{}}
            $\mathcal{T}_{\mathfrak{e}_7, N}((A_3 + A_1)'', (A_3 + A_1)')$ \\[3pt]
            $\overset{\mathfrak{su}_2}{2}1\overset{\mathfrak{e}_7}{7} \,
            \overbrace{1\overset{\mathfrak{su}_2}{2}\overset{\mathfrak{so}_7}{3}\overset{\mathfrak{su}_2}{2}1\overset{\mathfrak{e}_7}{8}\cdots1\overset{\mathfrak{su}_2}{2}\overset{\mathfrak{so}_7}{3}\overset{\mathfrak{su}_2}{2}1}^{N-2} \, \underset{\displaystyle 1}{\overset{\mathfrak{e}_7}{8}}1\,2$ \\[8pt]
            $\mathfrak{f} = (\mathfrak{su}_2)_{12} \oplus (\mathfrak{su}_2)_{24} \oplus  (\mathfrak{su}_2)_{13} \oplus (\mathfrak{so}_7)_{16}$
        \end{tabular} 
    \end{tikzcd}
	\caption{The Hasse diagram for the two nilpotent Higgsings of $\mathcal{T}_{\mathfrak{e}_7, N}(A_3, (A_3 + A_1)')$ that lead to $\mathcal{T}_{\mathfrak{e}_7, N}(A_3, A_3 + 2A_1)$ and $\mathcal{T}_{\mathfrak{e}_7, N}((A_3 + A_1)'', (A_3 + A_1)')$. In both cases the Higgsing involves giving a vacuum expectation value to the highest-root moment map of an $(\mathfrak{su}_2)_{12}$ flavor symmetry factor.}
    \label{fig:hasse1}
\end{figure}

Let us now present a couple of detailed examples to illustrate  pairs of theories with isomorphic interacting sectors. 
We first consider the theory
\begin{equation}\label{eqn:eg1}
  \mathcal{T}_{\mathfrak{e}_7, N}(A_3, A_3 + 2A_1) \,,
\end{equation}
which has
\begin{equation}
  \begin{gathered}
    \rankfour = 18N-22 \,,\quad \alpha = 2304N^3 - 13438N + 12586 - \frac{48}{N}  \,,\quad \beta = 269 -191N \,, \\[2mm] \operatorname{dim}\mathcal{H} = N + 44 \,, \quad \mathfrak{f}_\text{manifest} = (\mathfrak{su}_2)_{12} \oplus (\mathfrak{su}_2)_{24} \oplus  (\mathfrak{su}_2)_{13} \oplus (\mathfrak{so}_7)_{16} \,.
  \end{gathered}
\end{equation}
We compare this theory with a different Higgsed $(\mathfrak{e}_7, \mathfrak{e}_7)$ conformal matter,
\begin{equation}\label{eqn:eg2}
    \mathcal{T}_{\mathfrak{e}_7, N}\left((A_3 + A_1)'', (A_3 + A_1)'\right) \,.
\end{equation}
The physical properties of this latter theory are
\begin{equation}
  \begin{gathered}
  \rankfour = 18N - 22 \,,\quad \alpha = 2304N^3 -13438N + 12562 \,,\quad \beta = 269 - 191N \,, \\[2mm] \operatorname{dim}\mathcal{H} = N + 44 \,, \quad \mathfrak{f}_\text{manifest} = (\mathfrak{su}_2)_{12} \oplus (\mathfrak{su}_2)_{24} \oplus  (\mathfrak{su}_2)_{13} \oplus (\mathfrak{so}_7)_{16} \,.
  \end{gathered}
\end{equation}
It is clear that, for generic values of $N$, the theories in equations \eqref{eqn:eg1} and \eqref{eqn:eg2} are different, in particular, they have different values of the 't Hooft anomaly coefficient $\alpha$. This is also clear to see from the tensor branch descriptions of each theory. We have two theories
\begin{subequations}
\begin{align}
    \mathcal{T}_{\mathfrak{e}_7, N}(A_3, A_3 + 2A_1) \,: & \qquad \overset{\mathfrak{su}_2}{2}1\underset{\displaystyle 1}{\overset{\mathfrak{e}_7}{8}} \,
    \overbrace{1\overset{\mathfrak{su}_2}{2}\overset{\mathfrak{so}_7}{3}\overset{\mathfrak{su}_2}{2}1\overset{\mathfrak{e}_7}{8}\cdots1\overset{\mathfrak{su}_2}{2}\overset{\mathfrak{so}_7}{3}\overset{\mathfrak{su}_2}{2}1}^{N-2} \,
    \overset{\mathfrak{e}_7}{7}1\,2 \,,\\
    \mathcal{T}_{\mathfrak{e}_7, N}((A_3 + A_1)'', (A_3 + A_1)') \,: & \qquad \overset{\mathfrak{su}_2}{2}1\overset{\mathfrak{e}_7}{7} \,
    \overbrace{1\overset{\mathfrak{su}_2}{2}\overset{\mathfrak{so}_7}{3}\overset{\mathfrak{su}_2}{2}1\overset{\mathfrak{e}_7}{8}\cdots1\overset{\mathfrak{su}_2}{2}\overset{\mathfrak{so}_7}{3}\overset{\mathfrak{su}_2}{2}1}^{N-2} \, \underset{\displaystyle 1}{\overset{\mathfrak{e}_7}{8}}1\,2 \,.
\end{align}
\end{subequations}
Both of these theories can be obtained from a nilpotent Higgsing of a parent theory, in this case $\mathcal{T}_{\mathfrak{e}_7, N}(A_3, (A_3 + A_1)')$, as we can see from the Hasse diagram in Figure \ref{fig:hasse1}. The parent theory has two $(\mathfrak{su}_2)_{12}$ flavor symmetries, and giving a nilpotent vacuum expectation value to the associated moment map operators triggers a Higgs branch renormalization flow from 
\begin{align}
    A_3 \rightarrow (A_3+A_1)''\quad\text{and}\quad (A_3 + A_1)' \rightarrow A_3 + 2A_1\,, 
\end{align}
respectively. On the tensor branch, these Higgsings can be though of as shrinking either of the dangling $(-1)$-curves and deforming the geometry to remove the resulting singularity.

\begin{figure}[H]
    \centering
    \begin{tikzcd}[cells={nodes={draw=black}},column sep=-5em,row sep=3.8em,scale cd=0.9]
        & \begin{tabular}{@{}c@{}}
            $\mathcal{T}_{\mathfrak{e}_7, 2}(A_3, (A_3 + A_1)')$ \\[3pt]
            $\overset{\mathfrak{su}_2}{2}1\overset{\displaystyle 1}{\underset{\displaystyle 1}{\overset{\mathfrak{e}_7}{8}}}1\,2$ \\[8pt]
            $\mathfrak{f} = (\mathfrak{su}_2)_{12}^{\oplus 2} \oplus (\mathfrak{su}_2)_{24} \oplus  (\mathfrak{su}_2)_{13} \oplus (\mathfrak{so}_7)_{16}$
        \end{tabular} \arrow[shorten >=3pt]{ld}[swap,yshift=-3pt]{(A_3 + A_1)' \rightarrow A_3 + 2A_1}\arrow[shorten >=3pt]{rd}[yshift=-3pt]{A_3 \rightarrow (A_3 + A_1)''} & \\
        \begin{tabular}{@{}c@{}}
            $\mathcal{T}_{\mathfrak{e}_7, 2}(A_3, A_3 + 2A_1)$ \\[3pt]
            $\overset{\mathfrak{su}_2}{2}1\underset{\displaystyle 1}{\overset{\mathfrak{e}_7}{7}}1\,2$ \\[8pt]
            $\mathfrak{f} = (\mathfrak{su}_2)_{12} \oplus (\mathfrak{su}_2)_{24} \oplus  (\mathfrak{su}_2)_{13} \oplus (\mathfrak{so}_7)_{16}$
        \end{tabular}\arrow[equal, shorten=8mm]{rr} & & \begin{tabular}{@{}c@{}}
            $\mathcal{T}_{\mathfrak{e}_7, 2}((A_3 + A_1)'', (A_3 + A_1)')$ \\[3pt]
            $\overset{\mathfrak{su}_2}{2}1\underset{\displaystyle 1}{\overset{\mathfrak{e}_7}{7}}1\,2$ \\[8pt]
            $\mathfrak{f} = (\mathfrak{su}_2)_{12} \oplus (\mathfrak{su}_2)_{24} \oplus  (\mathfrak{su}_2)_{13} \oplus (\mathfrak{so}_7)_{16}$
        \end{tabular} 
    \end{tikzcd}
    \caption{The Hasse diagram in Figure \ref{fig:hasse1} when $N=2$. The parent theory, $\mathcal{T}_{\mathfrak{e}_7, 2}(A_3, (A_3 + A_1)')$, has a $\mathbb{Z}_2$ automorphism that interchanges the two $(\mathfrak{su}_2)_{12}$ flavor symmetry factors, and thus nilpotent Higgsing by either factor leads to the \emph{same} infrared 6d $(1,0)$ SCFT.}
    \label{fig:hasse2}
\end{figure}
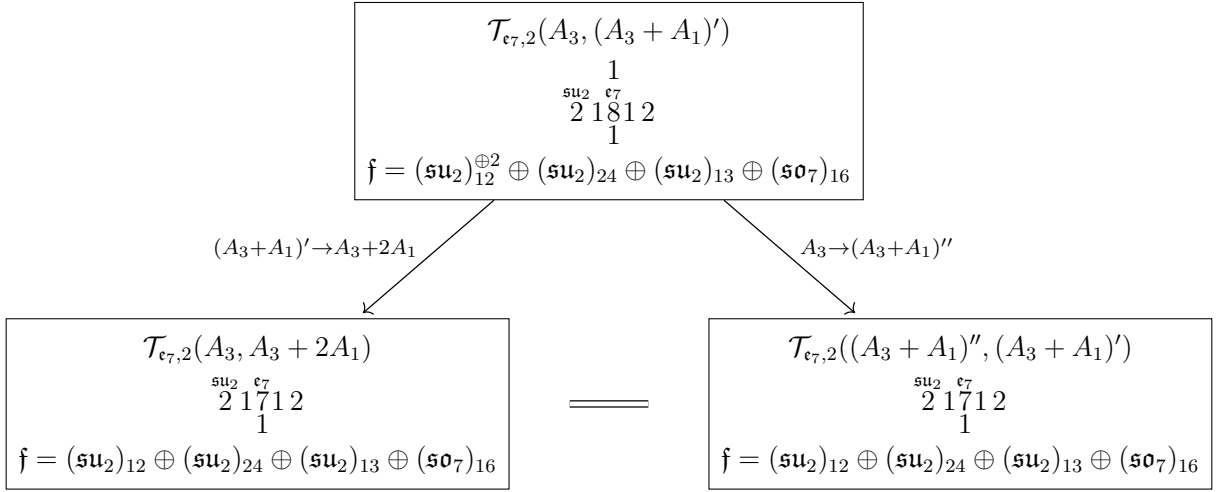

When $N=2$, we can see that the two theories in equations \eqref{eqn:eg1} and \eqref{eqn:eg2} have the same value of $\alpha$, in addition to the other properties that agree for all $N$.\footnote{It is important to note that $\alpha$, $\beta$, and $\rankfour$ are the same. In this example, they also have the same Higgs branch dimension and manifest flavor symmetries. However this is not necessary for the Higgsed conformal matter theories to have the same interacting sectors.} These are necessary conditions for the two SCFTs to be the same; now, we determine the tensor branch descriptions of the SCFTs when $N=2$, and we find that the Hasse diagram in Figure \ref{fig:hasse1} becomes the Hasse diagram in Figure \ref{fig:hasse2}.
It is clear that the tensor branch description of the parent theory possesses a $\mathbb{Z}_2$ automorphism that exchanges the two dangling $(-1)$-curves attached to the central $(-8)$-curve. Thus Higgsing by giving a nilpotent vev to either of the two $(\mathfrak{su}_2)_{12}$ moment maps yields the same tensor branch description and thus the same SCFTs, that is:
\begin{equation}
    \mathcal{T}_{\mathfrak{e}_7, 2}(A_3, A_3 + 2A_1) \quad = \quad \mathcal{T}_{\mathfrak{e}_7, 2}((A_3 + A_1)'', (A_3 + A_1)') \,.
\end{equation}

In the previous example, we have considered a solution to the question posed in this section where the Higgsed conformal matter theories are interacting SCFTs, with no free sector, and where the non-Abelian flavor algebra of the identical theories is simply the manifest flavor algebra associated to the nilpotent orbits. We now consider an example where each Higgsed conformal matter theory contains different numbers of free hypermultiplets and where there is flavor symmetry enhancement for both theories. Consider the theories, which we list together with their relevant physical properties, in Table \ref{tb:flavorenhancement}. We can see that when $N=2$, we have two SCFTs with the same values of $\alpha$, $\beta$, and $\rankfour$; these are then good candidates for theory with isomorphic interacting sectors. We determine that the tensor branch description for the interacting sector of both theories is
\begin{equation}
    \overset{\mathfrak{so}_9}{4}1\overset{\mathfrak{su}_3}{2} \,.
\end{equation}
The interacting SCFT associated to this tensor branch has
\begin{equation}\label{eqn:inter1}
  \operatorname{dim}\mathcal{H} = 41 \,, \qquad \mathfrak{f} = (\mathfrak{su}_2)_9 \oplus (\mathfrak{su}_6)_{18} \,.
\end{equation}
From the differences in the dimensions of the Higgs branches, we can see that $\mathcal{T}_{\mathfrak{e}_7, 2}(A_2, D_6(a_2))$ contains six free hypermultiplets, which rotate under an $(\mathfrak{su}_6)_2$ flavor algebra; similarly, $\mathcal{T}_{\mathfrak{e}_7, 2}(A_2 + A_1, A_5')$ contains two free hypermultiplets, which rotate under the fundamental of an $(\mathfrak{su}_2)_2$ flavor algebra. This is consistent with the non-Abelian flavor symmetry of the interacting sector in equation \eqref{eqn:inter1}. Thus, we learn that
\begin{equation}\label{eqn:dora}
    \begin{aligned}
      \mathcal{T}_{\mathfrak{e}_7, 2}(A_2, D_6(a_2)) \quad &= \quad \overset{\mathfrak{so}_9}{4}1\overset{\mathfrak{su}_3}{2} \quad + \quad \text{6 free hypers}\,, \\
      \mathcal{T}_{\mathfrak{e}_7, 2}(A_2 + A_1, A_5') \quad &= \quad \overset{\mathfrak{so}_9}{4}1\overset{\mathfrak{su}_3}{2} \quad + \quad \text{2 free hypers} \,.
    \end{aligned} 
\end{equation}

\begin{table}[H]
\centering
\footscriptsize
\renewcommand{\arraystretch}{2.0}
  \begin{threeparttable}$
  \begin{array}{ccc}
  \toprule
    \text{Theory} & \mathcal{T}_{\mathfrak{e}_7, N}(A_2, D_6(a_2)) & \mathcal{T}_{\mathfrak{e}_7, N}(A_2 + A_1, A_5') \\\midrule
    \begin{array}{@{}c@{}}
        \text{Tensor}\\[-8pt]
        \text{branch}
    \end{array} & 
    \begin{gathered}
    \overset{\mathfrak{so}_9}{4}1\overset{\mathfrak{g}_2}{3}\overset{\mathfrak{su}_2}{2}1\,\overbrace{
    \overset{\mathfrak{e}_7}{8}1\overset{\mathfrak{su}_2}{2}\overset{\mathfrak{so}_7}{3}\overset{\mathfrak{su}_2}{2}1\cdots\overset{\mathfrak{e}_7}{8}1\overset{\mathfrak{su}_2}{2}\overset{\mathfrak{so}_7}{3}\overset{\mathfrak{su}_2}{2}1
    }^{N-3}\,\overset{\mathfrak{e}_7}{8}1\overset{\mathfrak{su}_2}{2}\overset{\mathfrak{su}_4}{2} 
    \end{gathered} & 
    \begin{gathered}
    \overset{\mathfrak{so}_9}{4}1\overset{\mathfrak{so}_7}{3}\overset{\mathfrak{su}_2}{2}1\,\overbrace{
    \overset{\mathfrak{e}_7}{8}1\overset{\mathfrak{su}_2}{2}\overset{\mathfrak{so}_7}{3}\overset{\mathfrak{su}_2}{2}1\cdots\overset{\mathfrak{e}_7}{8}1\overset{\mathfrak{su}_2}{2}\overset{\mathfrak{so}_7}{3}\overset{\mathfrak{su}_2}{2}1
    }^{N-3}\,\overset{\mathfrak{e}_7}{8}1\overset{\mathfrak{su}_2}{2}\overset{\mathfrak{su}_3}{2} 
    \end{gathered} \\
    \alpha & 2304N^3 - 24958N + 39447 -13872/N & 2304N^3 - 23806N + 35607 -10800/N \\
    \beta & 663 - 382N & 663 - 382N \\
    \rankfour & 18N - 27 & 28N - 27 \\
    \operatorname{dim}\mathcal{H} & N + 45 & N + 41 \\
    \mathfrak{f}_\text{manifest} & (\mathfrak{su}_2)_9 \oplus (\mathfrak{su}_6)_{20} & (\mathfrak{su}_2)_9 \oplus (\mathfrak{su}_2)_{20} \oplus (\mathfrak{su}_4)_{18}\\\bottomrule
  \end{array}$
  \end{threeparttable}
  \caption{Some of the physical properties of the 6d $(1,0)$ SCFTs $\mathcal{T}_{\mathfrak{e}_7, N}(A_2, D_6(a_2))$ and $\mathcal{T}_{\mathfrak{e}_7, N}(A_2 + A_1, A_5')$.}
  \label{tb:flavorenhancement}
\end{table}

Similarly to the isomorphic interacting SCFTs that appear in Figure \ref{fig:hasse2}, we argue that there exists a parent theory, and that the isomorphism of the interacting sectors of the two SCFTs listed in Table \ref{tb:flavorenhancement} can similarly be understood as due to an enhanced symmetry in the parent theory that makes the equivalency of the two nilpotent Higgsings manifest. In this case, the parent theory is $\mathcal{T}_{\mathfrak{e}_7, N}(A_2, A_5')$, which has the tensor branch description
\begin{equation}
    \begin{gathered}
    \overset{\mathfrak{so}_9}{4}1\overset{\mathfrak{so}_7}{3}\overset{\mathfrak{su}_2}{2}1\,\overbrace{
    \overset{\mathfrak{e}_7}{8}1\overset{\mathfrak{su}_2}{2}\overset{\mathfrak{so}_7}{3}\overset{\mathfrak{su}_2}{2}1\cdots\overset{\mathfrak{e}_7}{8}1\overset{\mathfrak{su}_2}{2}\overset{\mathfrak{so}_7}{3}\overset{\mathfrak{su}_2}{2}1
    }^{N-3}\,\overset{\mathfrak{e}_7}{8}1\overset{\mathfrak{su}_2}{2}\overset{\mathfrak{su}_4}{2}  \,.
    \end{gathered}
\end{equation}
The non-Abelian flavor algebra of this SCFT is
\begin{equation}
    (\mathfrak{su}_2)_{9} \oplus (\mathfrak{su}_2)_{20} \oplus (\mathfrak{su}_6)_{20} \,.
\end{equation}
The Higgs branch renormalization group flows that lead to the two SCFTs in Table \ref{tb:flavorenhancement} are 
\begin{align}
    A_2 \rightarrow A_2 + A_1 \quad\text{and}\quad A_5' \rightarrow D_6(a_2)\,, 
\end{align}
which correspond to giving highest-root nilpotent vacuum expectation values to the moment maps of the $(\mathfrak{su}_6)_{20}$ and $(\mathfrak{su}_2)_{20}$ flavor algebras, respectively. When $N=2$ the tensor branch becomes
\begin{equation}
    \overset{\mathfrak{so}_9}{4}1\overset{\mathfrak{su}_4}{2} \,,
\end{equation}
and the two flavor symmetry factors with level $20$ recombine to the enhanced flavor algebra
\begin{equation}
    (\mathfrak{su}_2)_{20} \oplus (\mathfrak{su}_6)_{20} \,\, \rightarrow \,\, (\mathfrak{su}_8)_{20} \,.
\end{equation}
In the previous example, we observed that when $N=2$ there emerged a discrete $\mathbb{Z}_2$ symmetry of the tensor branch configuration, and this provided the physical justification for the isomorphism of the two theories in the Hasse diagram in Figure \ref{fig:hasse2}; in this case, instead of a emergent discrete symmetry we see that there is an enhanced continuous symmetry. When we see an enhanced flavor symmetry, we can Higgs by giving a vacuum expectation value to the highest-root moment-map of that enhanced flavor symmetry; this triggers a renormalization group flow to a new interacting SCFT. However, giving a VEV to the highest-root moment map of a subalgebra will lead to the same interacting SCFT, but with differing numbers of free hypermultiplets transforming under the unbroken part of the flavor symmetry. As such, it is clear that when there exists such a flavor symmetry enhancement, the two Higgs branch deformations corresponding to, in this case, 
\begin{align}
    A_2 \rightarrow A_2 + A_1 \quad\text{and}\quad A_5' \rightarrow D_6(a_2) \,,
\end{align}
lead to theories with the same interacting sector. The unbroken part of the flavor symmetry is, respectively, $\mathfrak{su}_2$ and $\mathfrak{su}_6$, which produces precisely the correct number of free hypermultiplets that we observed in equation \eqref{eqn:dora}.

\subsection{Isomorphic theories for \texorpdfstring{$\mathfrak{g} = \mathfrak{e}_7$}{g = E7}}\label{sec:isoe7}

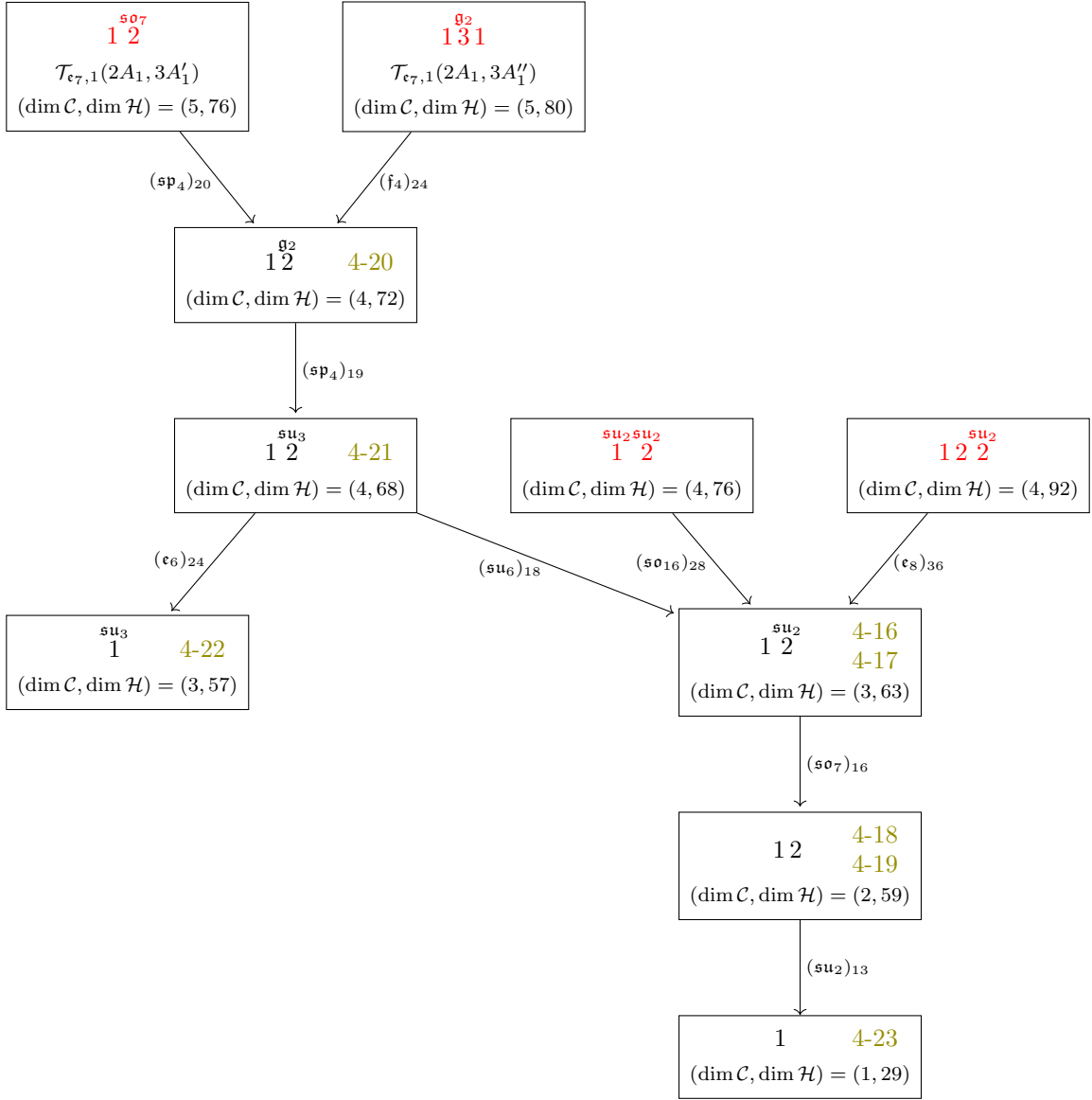
\begin{figure}[H]
    \centering
    \footnotesize
    \def\arraystretch{.9}
    \begin{tikzcd}[cells={nodes={draw=black}},column sep=-3em,row sep=3.8em,scale cd=1]
    % Row 0
    \begin{tabular}{@{}c@{}}
       {\color{red}$1\overset{\mathfrak{so}_7}{2}$} \\[2pt]
        \scriptsize$\mathcal{T}_{\mathfrak{e}_7, 1}(2A_1,3A_1')$\\[2pt]
        {\scriptsize$(\dim\mathcal{C},\dim\mathcal{H})=(5,76)$}
    \end{tabular}\arrow[shorten >=2pt]{rd}[swap,yshift=5pt]{(\mathfrak{sp}_4)_{20}} 
    & &
    \begin{tabular}{@{}c@{}}
       {\color{red} $1\overset{\mathfrak{g}_2}{3}1$} \\[2pt]
        \scriptsize$\mathcal{T}_{\mathfrak{e}_7, 1}(2A_1,3A_1'')$\\[2pt]
        {\scriptsize$(\dim\mathcal{C},\dim\mathcal{H})=(5,80)$}
    \end{tabular}\arrow[shorten >=2pt]{ld}[yshift=5pt]{(\mathfrak{f}_4)_{24}} & & & \\
     % Row 1
       & \begin{tabular}{@{}r@{}}
       $1\overset{\mathfrak{g}_2}{2}$\qquad {\color{olive}4-20\phantom{2}}\\[2pt]
        %\scriptsize$\mathcal{T}_{\mathfrak{e}_7, 1}(2A_1, 4A_1) - 1 = \mathcal{T}_{\mathfrak{e}_7, 1}(3A_1', 3A_1'') - 3$\\
        {\scriptsize$(\dim\mathcal{C},\dim\mathcal{H})=(4,72)$}
    \end{tabular}\arrow[shorten >=2pt]{d}{(\mathfrak{sp}_4)_{19}} & & & & \\ 
    % Row 2
       & \begin{tabular}{@{}r@{}}
        $1\overset{\mathfrak{su}_3}{2}$\quad\; {\color{olive}4-21\phantom{2}}\\[2pt]
        %\scriptsize$\mathcal{T}_{\mathfrak{e}_7, 1}(2A_1, A_2 + A_1) - 2 = \mathcal{T}_{\mathfrak{e}_7, 1}(A_2, 3A_1'') - 6$\\
        {\scriptsize$(\dim\mathcal{C},\dim\mathcal{H})=(4,68)$}
    \end{tabular}\arrow[shorten >=2pt]{ld}[swap,yshift=-4pt]{(\mathfrak{e}_6)_{24}}\arrow[shorten >=2pt]{rrrd}[swap,yshift=4pt]{(\mathfrak{su}_6)_{18}} & & \begin{tabular}{@{}c@{}}
        {\color{red}$\overset{\mathfrak{su}_2}{1}\overset{\mathfrak{su}_2}{2}$}\\[2pt]
        {\scriptsize$(\dim\mathcal{C},\dim\mathcal{H})=(4,76)$}
    \end{tabular}\arrow[shorten >=2pt]{rd}[swap,yshift=5pt]{(\mathfrak{so}_{16})_{28}} & & \begin{tabular}{@{}c@{}}
        {\color{red}$1\,2\overset{\mathfrak{su}_2}{2}$}\\[2pt]
        {\scriptsize$(\dim\mathcal{C},\dim\mathcal{H})=(4,92)$}
    \end{tabular}\arrow[shorten >=2pt]{ld}[yshift=5pt]{(\mathfrak{e}_8)_{36}}\\
    % Row 3
     \begin{tabular}{@{}r@{}}
        $\overset{\mathfrak{su}_3}{1}$\qquad {\color{olive}4-22\phantom{2}}\\[2pt]
        %\scriptsize$\mathcal{T}_{\mathfrak{e}_7, 1}(3A_1', A_2 + A_1) - 7 = \mathcal{T}_{\mathfrak{e}_7, 1}(A_2, 4A_1) - 9$\\
        {\scriptsize$(\dim\mathcal{C},\dim\mathcal{H})=(3,57)$}
     \end{tabular} & & & & \begin{tabular}{@{}r@{}}
        $1\overset{\mathfrak{su}_2}{2}$\qquad \begin{tabular}{@{}r@{}} {\color{olive}4-16\phantom{2}}\\
        {\color{olive}4-17\phantom{2}}
        \end{tabular}\\[2pt]
	    %\scriptsize$\mathcal{T}_{\mathfrak{e}_7, 1}(2A_1, A_2 + 2A_1) - 4 = \mathcal{T}_{\mathfrak{e}_7, 1}(2A_2, A_1) - 12 = \mathcal{T}_{\mathfrak{e}_7, 1}((A_3 + A_1)'', 0) - 28$\\
        {\scriptsize$(\dim\mathcal{C},\dim\mathcal{H})=(3,63)$}
       \end{tabular}\arrow[shorten >=2pt]{d}{(\mathfrak{so}_7)_{16}} & \\
    % Row 4
	& & & & \begin{tabular}{@{}r@{}}
        $1\,2$\qquad \begin{tabular}{@{}r@{}} {\color{olive}4-18\phantom{2}}\\
        {\color{olive}4-19\phantom{2}}
        \end{tabular}\\[2pt]
	    %\scriptsize$\mathcal{T}_{\mathfrak{e}_7, 1}(2A_1, A_2 + 3A_1) - 7 = \mathcal{T}_{\mathfrak{e}_7, 1}(2A_2 + A_1, A_1) - 13 = \mathcal{T}_{\mathfrak{e}_7, 1}(A_3 + 2A_1, 0) - 28$\\
        {\scriptsize$(\dim\mathcal{C},\dim\mathcal{H})=(2,59)$}
       \end{tabular}\arrow{d}{(\mathfrak{su}_2)_{13}} & \\
    % Row 5
	& & & & \begin{tabular}{@{}r@{}}
        $1$\qquad\; {\color{olive}4-23\phantom{2}}\\[2pt]
	    % \scriptsize$\mathcal{T}_{\mathfrak{e}_7, 1}(3A_1', A_2 + 3A_1) - 31 = \mathcal{T}_{\mathfrak{e}_7, 1}(A_3 + A_2, 0) - 56$\\
        {\scriptsize$(\dim\mathcal{C},\dim\mathcal{H})=(1,29)$}
       \end{tabular} & \\
\end{tikzcd}
\caption{The Hasse diagram of 6d $(1,0)$ SCFTs arising from Higgsed rank one $(\mathfrak{e}_7, \mathfrak{e}_7)$ conformal matter which occur via Higgsing by two distinct pairs of nilpotent orbits. We write the tensor branch description of the interacting sector. We have not depicted any non-minimal Higgs branch RG flows. Tensor branch configurations which are colored red are parent theories. The olive-colored ``x-y'' refers to Table x, entry y, where the pair of nilpotent orbits and its flavor symmetry are listed. This Hasse diagram is a subdiagram of Hasse diagram of nilpotent Higgsings of minimal $(\mathfrak{e}_7, \mathfrak{e}_7)$ conformal matter that appears in \cite{Baume:2021qho}. Note that this case contains some \emph{triples} of pairs corresponding to isomorphic theories.}\label{fig:e7N1}
\end{figure}

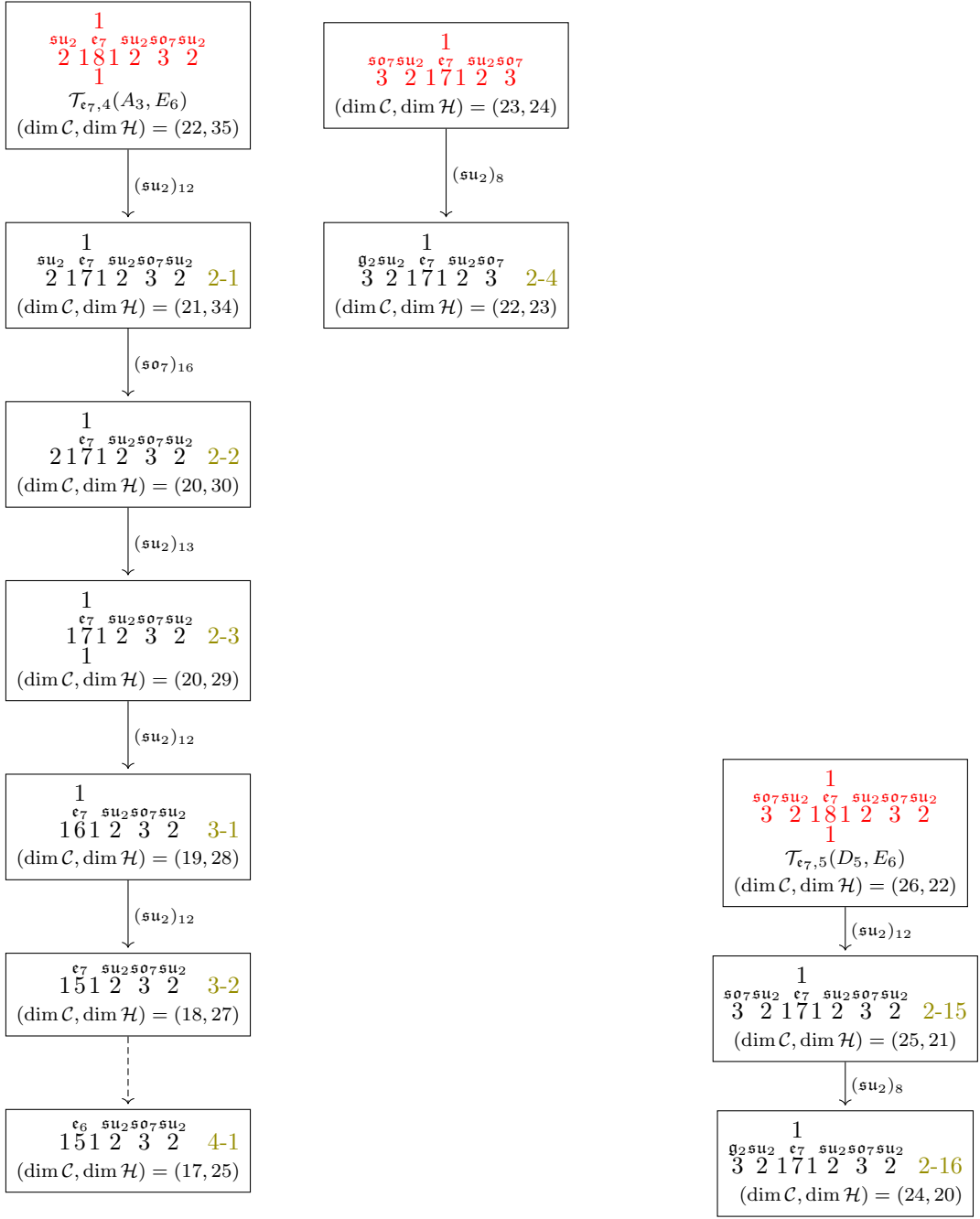
\begin{figure}[H]
    \centering
    \footnotesize
    \def\arraystretch{.7}
    \begin{subfigure}[b]{0.6\textwidth}
        \centering
    \begin{tikzcd}[cells={nodes={draw=black}},column sep=3em,row sep=3.0em,scale cd=1.0]
    % Row 0
    \begin{tabular}{@{}c@{}}
	     {\color{red}$\overset{\mathfrak{su}_2}{2}1\underset{\displaystyle 1}{\overset{\displaystyle 1}{\overset{\mathfrak{e}_7}{8}}}1\overset{\mathfrak{su}_2}{2}\overset{\mathfrak{so}_7}{3}\overset{\mathfrak{su}_2}{2}$}\\[8pt]
	     \scriptsize$\mathcal{T}_{\mathfrak{e}_7, 4}(A_3, E_6)$\\[2pt]
        {\scriptsize$(\dim\mathcal{C},\dim\mathcal{H})=(22,35)$}
    \end{tabular}\arrow[shorten >=2pt]{d}{\color{black}(\mathfrak{su}_2)_{12}}
%    \arrow[shorten >=2pt]{d}{(\mathfrak{so}_7)_{16}} 
    &
    \begin{tabular}{@{}c@{}}
	    {\color{red}$\overset{\mathfrak{so}_7}{3}\overset{\mathfrak{su}_2}{2}1\overset{\displaystyle 1}{\overset{\mathfrak{e}_7}{7}}1\overset{\mathfrak{su}_2}{2}\overset{\mathfrak{so}_7}{3}$}\\[2pt]
	%    \scriptsize$\mathcal{T}_{\mathfrak{e}_7, 4}(D_5, D_6(a_1))$\\[2pt]
        {\scriptsize$(\dim\mathcal{C},\dim\mathcal{H})=(23,24)$}
    \end{tabular}\arrow[shorten >=2pt]{d}{\color{black}(\mathfrak{su}_2)_{8}} \\
     % Row 1
      \begin{tabular}{@{}r@{}}
	     $\overset{\mathfrak{su}_2}{2}1\overset{\displaystyle 1}{\overset{\mathfrak{e}_7}{7}}1\overset{\mathfrak{su}_2}{2}\overset{\mathfrak{so}_7}{3}\overset{\mathfrak{su}_2}{2}$\;\;{\color{olive}2-1}\\[2pt]
	     % \scriptsize$\mathcal{T}_{\mathfrak{e}_7, 4}((A_3 + A_1)'', E_6) = \mathcal{T}_{\mathfrak{e}_7, 4}(A_3, E_7(a_2))$\\[2pt]
        {\scriptsize$(\dim\mathcal{C},\dim\mathcal{H})=(21,34)$}
    \end{tabular}\arrow[shorten >=2pt]{d}{(\mathfrak{so}_7)_{16}} & 
    % \begin{tabular}{@{}c@{}}
	   %   {\color{red}$2\,1\underset{\displaystyle 1}{\overset{\displaystyle 1}{\overset{\mathfrak{e}_7}{8}}}1\overset{\mathfrak{su}_2}{2}\overset{\mathfrak{so}_7}{3}\overset{\mathfrak{su}_2}{2}$}\\[8pt]
	   %   \scriptsize$\mathcal{T}_{\mathfrak{e}_7, 4}((A_3+A_1)', E_6)$\\[2pt]
    %     {\scriptsize$(\dim\mathcal{C},\dim\mathcal{H})=(fill,fill)$}
    % \end{tabular}\arrow[shorten >=2pt, color=red]{ld}{\color{black}(\mathfrak{su}_2)_{12}}\arrow[shorten >=2pt]{d}{(\mathfrak{su}_2)_{13}} 
    \begin{tabular}{@{}r@{}}
	    $\overset{\mathfrak{g}_2}{3}\overset{\mathfrak{su}_2}{2}1\overset{\displaystyle 1}{\overset{\mathfrak{e}_7}{7}}1\overset{\mathfrak{su}_2}{2}\overset{\mathfrak{so}_7}{3}$\;\;\;{\color{olive}2-4}\\[2pt]
	     % \scriptsize$\mathcal{T}_{\mathfrak{e}_7, 4}(D_5, E_7(a_4)) = \mathcal{T}_{\mathfrak{e}_7, 4}(D_6(a_1), D_5 + A_1)$\\[2pt]
        {\scriptsize$(\dim\mathcal{C},\dim\mathcal{H})=(22,23)$}
    \end{tabular} \\
    % Row 2
      \begin{tabular}{@{}r@{}}
	     $2\,1\overset{\displaystyle 1}{\overset{\mathfrak{e}_7}{7}}1\overset{\mathfrak{su}_2}{2}\overset{\mathfrak{so}_7}{3}\overset{\mathfrak{su}_2}{2}$\;\;{\color{olive}2-2}\\[2pt]
	     % \scriptsize$\mathcal{T}_{\mathfrak{e}_7, 4}(A_3 + 2A_1, E_6) = \mathcal{T}_{\mathfrak{e}_7, 4}((A_3+A_1)', E_7(a_2))$\\[2pt]
        {\scriptsize$(\dim\mathcal{C},\dim\mathcal{H})=(20,30)$}
    \end{tabular}\arrow[shorten >=2pt]{d}{(\mathfrak{su}_2)_{13}} & 
    % \begin{tabular}{@{}c@{}}
	   %   $\underset{\displaystyle 1\,\,1}{\overset{\displaystyle 1\,\,1}{\overset{\mathfrak{e}_7}{8}}}1\overset{\mathfrak{su}_2}{2}\overset{\mathfrak{so}_7}{3}\overset{\mathfrak{su}_2}{2}$\\[8pt]
	   %   \scriptsize$\mathcal{T}_{\mathfrak{e}_7, 4}(D_4(a_1), E_6)$
    % \end{tabular}\arrow[shorten >=2pt, color=red]{ld}{\color{black}(\mathfrak{su}_2)_{12}} 
     \\
    % Row 3
     \begin{tabular}{@{}r@{}}
	     $1\underset{\displaystyle 1}{\overset{\displaystyle 1}{\overset{\mathfrak{e}_7}{7}}}1\overset{\mathfrak{su}_2}{2}\overset{\mathfrak{so}_7}{3}\overset{\mathfrak{su}_2}{2}$\;\;{\color{olive}2-3}\\[8pt]
	     % \scriptsize$\mathcal{T}_{\mathfrak{e}_7, 4}(D_4(a_1) + A_1, E_6) = \mathcal{T}_{\mathfrak{e}_7, 4}(D_4(a_1), E_7(a_2))$
              {\scriptsize$(\dim\mathcal{C},\dim\mathcal{H})=(20,29)$}
    \end{tabular}\arrow[shorten >=2pt]{d}{\color{black}(\mathfrak{su}_2)_{12}} & \\
    % Row 4
     \begin{tabular}{@{}r@{}}
	     $1\overset{\displaystyle 1}{\overset{\mathfrak{e}_7}{6}}1\overset{\mathfrak{su}_2}{2}\overset{\mathfrak{so}_7}{3}\overset{\mathfrak{su}_2}{2}$\;\;\;{\color{olive}3-1}\\[2pt]
	     % \scriptsize$\mathcal{T}_{\mathfrak{e}_7, 4}(A_3 + A_2, E_6) = \mathcal{T}_{\mathfrak{e}_7, 4}(D_4(a_1) + A_1, E_7(a_2))$
              {\scriptsize$(\dim\mathcal{C},\dim\mathcal{H})=(19,28)$}
    \end{tabular}\arrow[shorten >=2pt]{d}{\color{black}(\mathfrak{su}_2)_{12}} & \\
    % Row 5
     \begin{tabular}{@{}r@{}}
	     $1\overset{\mathfrak{e}_7}{5}1\overset{\mathfrak{su}_2}{2}\overset{\mathfrak{so}_7}{3}\overset{\mathfrak{su}_2}{2}$\;\;\;{\color{olive}3-2}\\[2pt]
	     % \scriptsize$\mathcal{T}_{\mathfrak{e}_7, 4}(A_3 + A_2 + A_1, E_6) = \mathcal{T}_{\mathfrak{e}_7, 4}(A_3 + A_2, E_7(a_2))$
              {\scriptsize$(\dim\mathcal{C},\dim\mathcal{H})=(18,27)$}
     \end{tabular}
     % \arrow[shorten >=2pt]{d}{(\mathfrak{su}_2)_{224}}
     \arrow[dashed,shorten >=2pt]{d} & 
     % \begin{tabular}{@{}c@{}}
	    %  $1\overset{\displaystyle 1}{\overset{\mathfrak{e}_6}{6}}1\overset{\mathfrak{su}_2}{2}\overset{\mathfrak{so}_7}{3}\overset{\mathfrak{su}_2}{2}$\\[8pt]
	    %  \scriptsize$\mathcal{T}_{\mathfrak{e}_7, 4}(A_4, E_6)$
     % \end{tabular}\arrow[shorten >=2pt, color=red]{ld}{\color{black}(\mathfrak{su}_3)_{12}} 
     \\
    % Row 6
      \begin{tabular}{@{}r@{}}
	     $1\overset{\mathfrak{e}_6}{5}1\overset{\mathfrak{su}_2}{2}\overset{\mathfrak{so}_7}{3}\overset{\mathfrak{su}_2}{2}$\;\;\;{\color{olive}4-1}\\[2pt]
	     % \scriptsize$\mathcal{T}_{\mathfrak{e}_7, 4}(A_4 + A_1, E_6) = \mathcal{T}_{\mathfrak{e}_7, 4}(A_4, E_7(a_2)) - 1$
              {\scriptsize$(\dim\mathcal{C},\dim\mathcal{H})=(17,25)$}
     \end{tabular} & & \\
\end{tikzcd}
\caption{$N = 4$.}
\end{subfigure}
\begin{subfigure}[b]{0.3\textwidth}
        \centering
    \begin{tikzcd}[cells={nodes={draw=black}},column sep=3em,row sep=2.0em,scale cd=1.0]
    % Row 0
    \begin{tabular}{@{}c@{}}
	     {\color{red}$\overset{\mathfrak{so}_7}{3}\overset{\mathfrak{su}_2}{2}1\underset{\displaystyle 1}{\overset{\displaystyle 1}{\overset{\mathfrak{e}_7}{8}}}1\overset{\mathfrak{su}_2}{2}\overset{\mathfrak{so}_7}{3}\overset{\mathfrak{su}_2}{2}$}\\[8pt]
	     \scriptsize$\mathcal{T}_{\mathfrak{e}_7, 5}(D_5, E_6)$ \\[2pt]
      {\scriptsize$(\dim\mathcal{C},\dim\mathcal{H})=(26,22)$}
    \end{tabular}\arrow[shorten >=2pt]{d}{\color{black}(\mathfrak{su}_2)_{12}} \\
    % Row 1
     \begin{tabular}{@{}c@{}}
	     $\overset{\mathfrak{so}_7}{3}\overset{\mathfrak{su}_2}{2}1\overset{\displaystyle 1}{\overset{\mathfrak{e}_7}{7}}1\overset{\mathfrak{su}_2}{2}\overset{\mathfrak{so}_7}{3}\overset{\mathfrak{su}_2}{2}$\;\;{\color{olive}2-15}\\[2pt]
	     % \scriptsize$\mathcal{T}_{\mathfrak{e}_7, 5}(D_6(a_1), E_6) = \mathcal{T}_{\mathfrak{e}_7, 5}(D_5, E_7(a_2))$
      {\scriptsize$(\dim\mathcal{C},\dim\mathcal{H})=(25,21)$}
    \end{tabular}\arrow[shorten >=2pt]{d}{(\mathfrak{su}_2)_{8}}
    % \begin{tabular}{@{}c@{}}
	   %   $\overset{\mathfrak{g}_2}{3}\overset{\mathfrak{su}_2}{2}1\underset{\displaystyle 1}{\overset{\displaystyle 1}{\overset{\mathfrak{e}_7}{8}}}1\overset{\mathfrak{su}_2}{2}\overset{\mathfrak{so}_7}{3}\overset{\mathfrak{su}_2}{2}$\\[8pt]
	   %   \scriptsize$\mathcal{T}_{\mathfrak{e}_7, 5}(D_5 + A_1, E_6)$
    % \end{tabular}\arrow[shorten >=2pt, color=red]{ld}{\color{black}(\mathfrak{su}_2)_{12}} 
    \\
    % Row 2
     \begin{tabular}{@{}r@{}}
	     $\overset{\mathfrak{g}_2}{3}\overset{\mathfrak{su}_2}{2}1\overset{\displaystyle 1}{\overset{\mathfrak{e}_7}{7}}1\overset{\mathfrak{su}_2}{2}\overset{\mathfrak{so}_7}{3}\overset{\mathfrak{su}_2}{2}$\;\;{\color{olive}2-16} \\[2pt]
	     % \scriptsize$\mathcal{T}_{\mathfrak{e}_7, 5}(E_7(a_4), E_6) = \mathcal{T}_{\mathfrak{e}_7, 5}(D_5+A_1, E_7(a_2))$
      {\scriptsize$(\dim\mathcal{C},\dim\mathcal{H})=(24,20)$}
    \end{tabular}  \\
\end{tikzcd}
\caption{$N=5$.}
\end{subfigure}
	\caption{The Hasse diagrams of 6d $(1,0)$ SCFTs arising from Higgsed rank four and rank five $(\mathfrak{e}_7, \mathfrak{e}_7)$ conformal matter which occur via Higgsing by two distinct pairs of nilpotent orbits. See the caption of Figure \ref{fig:e7N1} for an explanation of the notation.}\label{fig:e7N4}
\end{figure}

We now possess a straightforward and algorithmic method to determine all answers to the question posted in this section, for a given $\mathfrak{g}$. For each theory $\mathcal{T}_{\mathfrak{g}, N}(O_L, O_R)$, it is well-known how to determine the tensor branch description for $N$ sufficiently large such that the nilpotent Higgsing on the left and right do not cross-correlate across the tensor branch. From this effective tensor branch description, we can compute $\rankfour$ of the SCFT and the anomaly polynomial. For any pair of theories $\mathcal{T}_{\mathfrak{g}, N}(O_L, O_R)$ and $\mathcal{T}_{\mathfrak{g}, N}(O_L', O_R')$ we then ask if there exists a value of $N$ such that $\alpha$, $\beta$, and the $\rankfour$ are the same. From the resulting list of putatively isomorphic theories, we determine the tensor branch descriptions of the interacting part of each pair, and if they are identical then the two SCFTs associated to each pair are identical. The results are conveniently summarized in a collection of Hasse diagrams describing Higgs branch renormalization group flows; for $\mathfrak{g}=\mathfrak{e}_7$ there exists pairs with isomorphic interacting sectors for $N=1, \cdots, 5$, and we depict these in Figures \ref{fig:e7N1}, \ref{fig:e7N2}, \ref{fig:e7N3}, and \ref{fig:e7N4}, respectively.

In the Hasse diagrams that we have drawn, we have generally shown 6d $(1,0)$ SCFTs which have two realizations by two different pairs of nilpotent orbits connected by arrows if there exists a (minimal) nilpotent Higgsing (that is, a Higgsing triggered by giving a vacuum expectation value to a highest-root moment map of a simple non-Abelian flavor symmetry factor) between two interacting SCFTs. We have depicted these nilpotent Higgsing by solid arrows labelled by the flavor algebra which is given a VEV. However, in the sets of theories that we consider, there also exist elementary slices in the Hasse diagram of nilpotent orbit closures, which are not associated to a nilpotent Higgsing. These are labeled by dashed arrows; they occur when we have a 6d tensor branch description which is of one of the following forms:
\begin{equation}\label{eqn:dashed1}
    \cdots 1 \overset{\mathfrak{e}_7}{5}1\cdots \,, \qquad \cdots 1 \overset{\mathfrak{e}_7}{4}1\cdots \,.
\end{equation}
Anomaly cancellation requires that the $\mathfrak{e}_7$ gauge algebra has $n_{\bm{56}} = 3$ or $n_{\bm{56}} = 4$  half-hypermultiplets, respectively, in the fundamental representation; thus there is an $\mathfrak{so}_{n_{\bm{56}}}$ flavor symmetry under which the half-hypermultiplets transform in the vector representation.\footnote{We note that this is different from the manifest flavor symmetry, but rather, an enhanced flavor symmetry for the case of $n_{\bm{56}} = 4$. This can be observed easily in the entry 13 of Table \ref{tb:diffmanifest}.} There is a Higgsing of these theories to SCFTs with tensor branch configurations
\begin{equation}\label{eqn:dashed2}
    \cdots 1 \overset{\mathfrak{e}_6}{5}1\cdots \,, \qquad \cdots 1 \overset{\mathfrak{e}_6}{4}1\cdots \,,
\end{equation}
respectively. This is not a nilpotent Higgsing of the $\mathfrak{so}_{n_{\bm{56}}}$ flavor factor. 
Understanding the Higgs branch renormalization group flows between conformal matter theories (in particular those which, like the ones discussed here, are not nilpotent Higgsings) is the subject of \cite{DKL}; we leave a fuller explanation to that work. 

For $\mathfrak{g} = \mathfrak{e}_7$, it is noteworthy that almost all pairs of theories
\begin{equation}
    \mathcal{T}_{\mathfrak{g}, N}(O_L, O_R) \quad \text{and} \quad \mathcal{T}_{\mathfrak{g}, N}(O_L', O_R') \,,
\end{equation}
with the same $\rankfour$ and the same 't Hooft anomaly coefficients $\alpha$ and $\beta$ have isomorphic interacting parts.\footnote{For obvious reasons, this statement requires that $\mathcal{T}_{\mathfrak{g}, N}(O_L, O_R)$ and $\mathcal{T}_{\mathfrak{g}, N}(O_L', O_R')$ have a non-trivial interacting sector.} In fact, there is precisely one counterexample. The theories
\begin{equation}
    \mathcal{T}_{\mathfrak{e}_7, 3}(D_4(a_1), A_5'') \quad \text{and} \quad \mathcal{T}_{\mathfrak{e}_7, 3}((A_3 + A_1)'', E_6(a_3)) \,,
\end{equation}
have the same $\alpha$, $\beta$, and $\rankfour$, however they correspond to SCFTs with tensor branch descriptions
\begin{equation}
    \begin{gathered}
        1\overset{\mathfrak{f}_4}{5}1 \overset{\mathfrak{g}_2}{3} \overset{\mathfrak{su}_2}{2}1 \underset{\displaystyle 1}{\overset{\displaystyle 1}{\overset{\mathfrak{e}_7}{8}}}1
    \end{gathered}
    \qquad \text{and} \qquad 
    \begin{gathered}
        \overset{\mathfrak{so}_8}{4}1 \overset{\mathfrak{so}_7}{3} \overset{\mathfrak{su}_2}{2}1 \overset{\mathfrak{e}_7}{7}1\overset{\mathfrak{su}_2}{2}
    \end{gathered} \,,
\end{equation}
respectively. This is merely a consequence of the fact that the (mixed) R-symmetry and gravitational anomalies do not provide sufficient data to distinguish any pair of 6d $(1,0)$ SCFTs, and it is thus necessary to determine the full tensor branch description: the tensor branch effective field theory is (modulo the aforementioned subtlety of \cite{Distler:2022yse}) a complete invariant of the SCFT, unlike the anomalies.

\begin{figure}[H]
    \centering
    \footnotesize
    \def\arraystretch{.7}
    \begin{tikzcd}[cells={nodes={draw=black}},column sep=-3em,row sep=2.0em,scale cd=.8]
    % Row 1
     &\begin{tabular}{@{}c@{}}
        {\color{red}$\overset{\mathfrak{su}_2}{2}1\overset{\displaystyle 1}{\underset{\displaystyle 1}{\overset{\mathfrak{e}_7}{8}}}1\,2$}\\[8pt]
	    \scriptsize$\mathcal{T}_{\mathfrak{e}_7, 2}(A_3,(A_3 + A_1)')$\\[2pt]
        {\scriptsize$(\dim\mathcal{C},\dim\mathcal{H})=(15,47)$}
    \end{tabular}\arrow[shorten >=2pt]{rd}[swap,yshift=5pt]{(\mathfrak{su}_2)_{12}} & & \begin{tabular}{@{}c@{}}
        {\color{red}$\overset{\mathfrak{su}_2}{2}1\underset{\displaystyle 1}{\overset{\mathfrak{e}_7}{7}}1\overset{\mathfrak{su}_2}{2}$}\\[8pt]
	    \scriptsize$\mathcal{T}_{\mathfrak{e}_7, 2}(A_3,(A_3 + A_1)'')$\\[2pt]
        {\scriptsize$(\dim\mathcal{C},\dim\mathcal{H})=(15,50)$}
    \end{tabular}\arrow[shorten >=2pt]{ld}[yshift=5pt]{(\mathfrak{so}_7)_{16}} & & & 
    \begin{tabular}{@{}c@{}}
        {\color{red} $\overset{\mathfrak{g}_2}{3}\overset{\mathfrak{su}_2}{2}2\,1$} \\[2pt]
        \scriptsize$\mathcal{T}_{\mathfrak{e}_7, 2}(0,A_6)$\\[2pt]
        {\scriptsize$(\dim\mathcal{C},\dim\mathcal{H})=(7,78)$}
    \end{tabular}\arrow{d}[yshift=0pt]{(\mathfrak{e}_8)_{36}} \\
    % Row 2
     & & \begin{tabular}{@{}r@{}}
        $\overset{\mathfrak{su}_2}{2}1\underset{\displaystyle 1}{\overset{\mathfrak{e}_7}{7}}1\,2$\quad\; {\color{olive}2-12\phantom{2}} \\[8pt]
        {\scriptsize$(\dim\mathcal{C},\dim\mathcal{H})=(14,46)$}
    \end{tabular}\arrow[shorten >=2pt]{d}{(\mathfrak{su}_2)_{13}} & & & & \begin{tabular}{@{}r@{}}
        $\overset{\mathfrak{g}_2}{3}\overset{\mathfrak{su}_2}{2}1$\quad\; {\color{olive}4-9\phantom{2}} \\[2pt]
        {\scriptsize$(\dim\mathcal{C},\dim\mathcal{H})=(6,49)$}
    \end{tabular}
    \\
    % Row 3
     & & \begin{tabular}{@{}r@{}}
        $\overset{\mathfrak{su}_2}{2}1\overset{\displaystyle 1}{\underset{\displaystyle 1}{\overset{\mathfrak{e}_7}{7}}}1$\quad\; {\color{olive}2-13\phantom{2}} \\[8pt]
        {\scriptsize$(\dim\mathcal{C},\dim\mathcal{H})=(14,45)$}
    \end{tabular}\arrow[shorten >=2pt]{ld}[swap,yshift=-3pt]{(\mathfrak{so}_7)_{16}}\arrow[shorten >=2pt]{rd}[yshift=-3pt]{(\mathfrak{su}_2)_{12}}
     \\
    % Row 4
    & \begin{tabular}{@{}r@{}}
        $2\,1\overset{\displaystyle 1}{\underset{\displaystyle 1}{\overset{\mathfrak{e}_7}{7}}}1$\quad\; {\color{olive}2-14\phantom{2}} \\[8pt]
        {\scriptsize$(\dim\mathcal{C},\dim\mathcal{H})=(13,41)$}
    \end{tabular}\arrow[shorten >=2pt]{d}{(\mathfrak{su}_2)_{12}} & & \begin{tabular}{@{}r@{}}
        $\overset{\mathfrak{su}_2}{2}1\underset{\displaystyle 1}{\overset{\mathfrak{e}_7}{6}}1$\;\, \quad {\color{olive}3-7\phantom{2}} \\[8pt]
        {\scriptsize$(\dim\mathcal{C},\dim\mathcal{H})=(13,44)$}
    \end{tabular}\arrow[shorten >=2pt]{lld}{(\mathfrak{so}_7)_{16}}\arrow[shorten >=2pt]{d}{(\mathfrak{su}_2)_{12}}
    \\
    % Row 5
    & \begin{tabular}{@{}r@{}}
        $2\,1{\underset{\displaystyle 1}{\overset{\mathfrak{e}_7}{6}}}1$\;\, \quad {\color{olive}3-8\phantom{2}} \\[8pt]
        {\scriptsize$(\dim\mathcal{C},\dim\mathcal{H})=(12,40)$}
    \end{tabular}\arrow[shorten >=2pt]{ld}[swap,yshift=-3pt]{(\mathfrak{su}_2)_{13}}\arrow[shorten >=2pt]{rd}[yshift=-3pt]{(\mathfrak{su}_2)_{12}} & & \begin{tabular}{@{}r@{}}
        $\overset{\mathfrak{su}_2}{2}1\overset{\mathfrak{e}_7}{5}1$\quad\; {\color{olive}3-10\phantom{2}} \\[2pt]
        {\scriptsize$(\dim\mathcal{C},\dim\mathcal{H})=(12,43)$}
    \end{tabular}\arrow[shorten >=2pt]{ld}[yshift=3pt]{(\mathfrak{so}_7)_{16}}
    % \arrow[shorten >=2pt]{rd}[yshift=-3pt]{(\mathfrak{su}_2)_{12}} 
    \arrow[dashed,shorten >=2pt]{rd}
    & & & 
    \begin{tabular}{@{}c@{}}
        {\color{red}$\overset{\mathfrak{so}_9}{4}1\overset{\mathfrak{so}_7}{3}1$} \\[2pt]
        \scriptsize$\mathcal{T}_{\mathfrak{e}_7, 2}(2A_1,A_5')$\\[2pt]
        {\scriptsize$(\dim\mathcal{C},\dim\mathcal{H})=(11,55)$}
    \end{tabular}
    \arrow{d}[yshift=0pt]{(\mathfrak{sp}_2)_{20}}
    \\
    % Row 6
    \begin{tabular}{@{}r@{}}
        $1\overset{\displaystyle 1}{\underset{\displaystyle 1}{\overset{\mathfrak{e}_7}{6}}}1$\quad\;\;\;\; {\color{olive}3-9\phantom{2}} \\[8pt]
        {\scriptsize$(\dim\mathcal{C},\dim\mathcal{H})=(12,39)$}
    \end{tabular}\arrow[shorten >=2pt]{d}{(\mathfrak{su}_2)_{12}} & & \begin{tabular}{@{}r@{}}
        $2\,1\overset{\mathfrak{e}_7}{5}1$\quad\; {\color{olive}3-11\phantom{2}} \\[2pt]
        {\scriptsize$(\dim\mathcal{C},\dim\mathcal{H})=(11,39)$}
    \end{tabular}\arrow[shorten >=2pt]{lld}{(\mathfrak{su}_2)_{13}} 
    % \arrow[shorten >=2pt]{rd}[yshift=-3pt]{(\mathfrak{su}_2)_{224}} 
    \arrow[dashed,shorten >=2pt]{rd}
    & & \begin{tabular}{@{}r@{}}
        $\overset{\mathfrak{su}_2}{2}1\overset{\mathfrak{e}_6}{5}1$\;\, \quad {\color{olive}4-4\phantom{2}} \\[2pt]
        {\scriptsize$(\dim\mathcal{C},\dim\mathcal{H})=(11,41)$}
    \end{tabular}\arrow[shorten >=2pt]{ld}[yshift=3pt]{(\mathfrak{so}_7)_{16}} & & 
    \begin{tabular}{@{}r@{}}
        $\overset{\mathfrak{so}_9}{4}1\overset{\mathfrak{g}_2}{3}1$\quad\; {\color{olive}3-14\phantom{2}} \\[2pt]
        {\scriptsize$(\dim\mathcal{C},\dim\mathcal{H})=(10,53)$}
    \end{tabular}\arrow[swap,shorten >=2pt]{ld}[yshift=-3pt]{(\mathfrak{su}_2)_{9}}\arrow[shorten >=2pt]{rd}[yshift=-3pt]{(\mathfrak{f}_4)_{24}}
    \\
    % Row 7-A
    \begin{tabular}{@{}r@{}}
        $1\underset{\displaystyle 1}{\overset{\mathfrak{e}_7}{5}}1$\quad\;\; {\color{olive}3-12\phantom{2}} \\[8pt]
        {\scriptsize$(\dim\mathcal{C},\dim\mathcal{H})=(11,38)$}
    \end{tabular}\arrow[shorten >=2pt]{d}{(\mathfrak{su}_2)_{12}}
    % \arrow[shorten >=2pt]{rrd}[yshift=-3pt]{(\mathfrak{su}_2)_{224}} 
    \arrow[dashed,shorten >=2pt]{rrd} & & & \begin{tabular}{@{}r@{}}
        $2\,1\overset{\mathfrak{e}_6}{5}1$\;\, \quad {\color{olive}4-5\phantom{2}} \\[2pt]
        {\scriptsize$(\dim\mathcal{C},\dim\mathcal{H})=(10,37)$}
    \end{tabular}\arrow[shorten >=2pt]{ld}[yshift=3pt]{(\mathfrak{su}_2)_{13}} & &   
    \begin{tabular}{@{}r@{}}
        $\overset{\mathfrak{so}_8}{4}1\overset{\mathfrak{g}_2}{3}1$\quad\; {\color{olive}3-15\phantom{2}} \\[2pt]
        {\scriptsize$(\dim\mathcal{C},\dim\mathcal{H})=(10,52)$}
    \end{tabular}\arrow{d}[yshift=0pt]{(\mathfrak{f}_4)_{24}}
    & & \begin{tabular}{@{}r@{}}
        $\overset{\mathfrak{so}_9}{4}1\overset{\mathfrak{g}_2}{2}$\quad\; {\color{olive}4-12\phantom{2}} \\[2pt]
        {\scriptsize$(\dim\mathcal{C},\dim\mathcal{H})=(9,45)$}
    \end{tabular}\arrow[shorten >=2pt]{lld}[yshift=3pt]{(\mathfrak{su}_2)_{9}}\arrow{d}[yshift=0pt]{(\mathfrak{sp}_4)_{19}} \\
    % Row 7-B
    \begin{tabular}{@{}r@{}}
        $1\overset{\mathfrak{e}_7}{4}1$\quad\; {\color{olive}3-13\phantom{2}} \\[2pt]
        {\scriptsize$(\dim\mathcal{C},\dim\mathcal{H})=(10,37)$}
    \end{tabular}
    % \arrow[shorten >=2pt]{rrd}[swap,yshift=3pt]{(\mathfrak{so}_4)_{112}} 
    \arrow[dashed,shorten >=2pt]{rrd} & & \begin{tabular}{@{}r@{}}
        $1\underset{\displaystyle 1}{\overset{\mathfrak{e}_6}{5}}1$\quad\;\;\;\; {\color{olive}4-6\phantom{2}} \\[8pt]
        {\scriptsize$(\dim\mathcal{C},\dim\mathcal{H})=(10,36)$}
    \end{tabular}\arrow[shorten >=2pt]{d}{(\mathfrak{su}_3)_{12}} & & &
    \begin{tabular}{@{}r@{}}
        $\overset{\mathfrak{so}_8}{4}1\overset{\mathfrak{g}_2}{2}$\quad\; {\color{olive}4-13\phantom{2}} \\[2pt]
        {\scriptsize$(\dim\mathcal{C},\dim\mathcal{H})=(9,44)$}
    \end{tabular}\arrow{d}[yshift=0pt]{(\mathfrak{sp}_4)_{19}}
    & & \begin{tabular}{@{}r@{}}
        $\overset{\mathfrak{so}_9}{4}1\overset{\mathfrak{su}_3}{2}$\quad\; {\color{olive}4-10\phantom{2}} \\[2pt]
        {\scriptsize$(\dim\mathcal{C},\dim\mathcal{H})=(9,41)$}
    \end{tabular}\arrow[shorten >=2pt]{lld}[yshift=3pt]{(\mathfrak{su}_2)_{9}}\arrow{d}[yshift=0pt]{(\mathfrak{su}_6)_{18}}
    \\
    % Row 8
     & & \begin{tabular}{@{}r@{}}
        $1\overset{\mathfrak{e}_6}{4}1$\quad\;\;\;\; {\color{olive}4-7\phantom{2}} \\[2pt]
        {\scriptsize$(\dim\mathcal{C},\dim\mathcal{H})=(9,34)$}
    \end{tabular}\arrow[shorten >=2pt]{d}{(\mathfrak{su}_3)_{12}} & & & 
    \begin{tabular}{@{}r@{}}
        $\overset{\mathfrak{so}_8}{4}1\overset{\mathfrak{su}_3}{2}$\quad\; {\color{olive}4-11\phantom{2}} \\[2pt]
        {\scriptsize$(\dim\mathcal{C},\dim\mathcal{H})=(9,40)$}
    \end{tabular}
    & & \begin{tabular}{@{}r@{}}
        $\overset{\mathfrak{so}_9}{4}1\overset{\mathfrak{su}_2}{2}$\quad\; {\color{olive}4-14\phantom{2}} \\[2pt]
        {\scriptsize$(\dim\mathcal{C},\dim\mathcal{H})=(8,36)$}
    \end{tabular}\arrow[shorten >=2pt]{ld}[yshift=3pt]{(\mathfrak{so}_7)_{16}}
    \\
    % Row 9
    & & \begin{tabular}{@{}r@{}}
        ${\overset{\mathfrak{e}_6}{3}}1$\quad\;\;\;\; {\color{olive}4-8\phantom{2}} \\[2pt]
        {\scriptsize$(\dim\mathcal{C},\dim\mathcal{H})=(8,32)$}
    \end{tabular} & & & &
    \begin{tabular}{@{}r@{}}
        $\overset{\mathfrak{so}_9}{4}1\,2$\quad\; {\color{olive}4-15\phantom{2}} \\[2pt]
        {\scriptsize$(\dim\mathcal{C},\dim\mathcal{H})=(7,32)$}
    \end{tabular}
    \\
    %\arrow[shorten >=3pt]{ld}[swap,yshift=-3pt]{a_3}\arrow[shorten >=3pt]{rd}[yshift=-3pt]{a_3}
    \end{tikzcd}
    \caption{The Hasse diagrams of 6d $(1,0)$ SCFTs arising from Higgsed rank two $(\mathfrak{e}_7, \mathfrak{e}_7)$ conformal matter which occur via Higgsing by two distinct pairs of nilpotent orbits. See the caption of Figure \ref{fig:e7N1} for an explanation of the notation.}\label{fig:e7N2}
\end{figure}

\begin{figure}[H]
    \centering
    \footnotesize
    \def\arraystretch{.7}
    \begin{tikzcd}[cells={nodes={draw=black}},column sep=-3em,row sep=2.9em,scale cd=.9]
     % Row 0
     & |[draw=none]|\hspace{10em} & \begin{tabular}{@{}c@{}}
	    {\color{red}$\overset{\mathfrak{su}_2}{2}1\underset{\displaystyle 1}{\overset{\displaystyle 1}{\overset{\mathfrak{e}_7}{8}}}1\overset{\mathfrak{su}_2}{2}\overset{\mathfrak{so}_7}{3}$}\\[8pt]
	    \scriptsize$\mathcal{T}_{\mathfrak{e}_7, 3}(A_3, D_5)$\\[2pt]
        {\scriptsize$(\dim\mathcal{C},\dim\mathcal{H})=(20,38)$}
      \end{tabular}
      % \arrow[shorten >=2pt]{ld}{(\mathfrak{su}_2)_{8}}
      \arrow[shorten >=2pt]{d}{\color{black}(\mathfrak{su}_2)_{12}} & |[draw=none]|\hspace{10em} &
     \begin{tabular}{@{}c@{}}
	     {\color{red}$\overset{\mathfrak{so}_9}{4}1\overset{\mathfrak{so}_7}{3}1\overset{\mathfrak{so}_8}{4}$}\\[2pt]
	      \scriptsize$\mathcal{T}_{\mathfrak{e}_7, 3}((A_5)', E_6(a_3))$ \\[2pt]
       {\scriptsize$(\dim\mathcal{C},\dim\mathcal{H})=(16,27)$}
     \end{tabular}\arrow[swap,shorten >=2pt]{rd}{(\mathfrak{sp}_2)_{20}}
     & &
     \begin{tabular}{@{}c@{}}
	     {\color{red}$\overset{\mathfrak{so}_9}{4}1\overset{\mathfrak{g}_2}{3}1\overset{\mathfrak{so}_9}{4}$}\\[2pt]
	      \scriptsize$\mathcal{T}_{\mathfrak{e}_7, 3}((A_5)', D_6(a_2))$ \\[2pt]
       {\scriptsize$(\dim\mathcal{C},\dim\mathcal{H})=(15,26)$}
     \end{tabular}
     \arrow[shorten >=2pt]{ld}{(\mathfrak{su}_2)_{9}}
      \\
     % Row 1 
     % \begin{tabular}{@{}c@{}}
     %    {\color{red}$\overset{\mathfrak{su}_2}{2}1\underset{\displaystyle 1}{\overset{\displaystyle 1}{\overset{\mathfrak{e}_7}{8}}}1\overset{\mathfrak{su}_2}{2}\overset{\mathfrak{g}_2}{3}$}\\[8pt]
     %    \scriptsize$\mathcal{T}_{\mathfrak{e}_7, 3}(A_3, D_5 + A_1)$\\[2pt]
     %    {\scriptsize$(\dim\mathcal{C},\dim\mathcal{H})=(fill,fill)$}
     % \end{tabular}\arrow[shorten >=2pt, color=red]{d}{\color{black}(\mathfrak{su}_2)_{12}} 
     & & \begin{tabular}{@{}r@{}}
	    $\overset{\mathfrak{su}_2}{2}1\overset{\displaystyle 1}{\overset{\mathfrak{e}_7}{7}}1\overset{\mathfrak{su}_2}{2}\overset{\mathfrak{so}_7}{3}$\; {\color{olive}2-5}\\[2pt]
	    %\scriptsize$\mathcal{T}_{\mathfrak{e}_7, 3}((A_3 + A_1)'', D_5) = \mathcal{T}_{\mathfrak{e}_7, 3}(A_3, D_6(a_1))$\\[2pt]
        {\scriptsize$(\dim\mathcal{C},\dim\mathcal{H})=(19,37)$}
         \end{tabular}\arrow[shorten >=2pt]{lld}{(\mathfrak{su}_2)_{8}}\arrow[shorten >=2pt]{d}{(\mathfrak{so}_7)_{16}} & & &
     \begin{tabular}{@{}r@{}}
	    $\overset{\mathfrak{so}_9}{4}1\overset{\mathfrak{g}_2}{3}1\overset{\mathfrak{so}_8}{4}$\; {\color{olive}2-11}\\[2pt]
	    %\scriptsize$\mathcal{T}_{\mathfrak{e}_7, 3}((A_5)', E_7(a_5)) - 1 = \mathcal{T}_{\mathfrak{e}_7, 3}(E_6(a_3), D_6(a_2)) - 1$\\[2pt]
        {\scriptsize$(\dim\mathcal{C},\dim\mathcal{H})=(15,25)$}
     \end{tabular} & \\
    % Row 2
    \begin{tabular}{@{}r@{}}
        $\overset{\mathfrak{su}_2}{2}1\overset{\displaystyle 1}{\overset{\mathfrak{e}_7}{7}}1\overset{\mathfrak{su}_2}{2}\overset{\mathfrak{g}_2}{3}$\quad {\color{olive}2-6}\\[2pt]
        %\scriptsize$\mathcal{T}_{\mathfrak{e}_7, 3}((A_3 + A_1)'', D_5 + A_1) = \mathcal{T}_{\mathfrak{e}_7, 3}(A_3, E_7(a_4))$\\[2pt]
        {\scriptsize$(\dim\mathcal{C},\dim\mathcal{H})=(18,36)$}
    \end{tabular}\arrow[shorten >=2pt]{d}{(\mathfrak{so}_7)_{16}}  & & \begin{tabular}{@{}r@{}}
	    $2\,1\overset{\displaystyle 1}{\overset{\mathfrak{e}_7}{7}}1\overset{\mathfrak{su}_2}{2}\overset{\mathfrak{so}_7}{3}$\;\;\; {\color{olive}2-7}\\[2pt]
	    %\scriptsize$\mathcal{T}_{\mathfrak{e}_7, 3}(A_3 + 2A_1, D_5) = \mathcal{T}_{\mathfrak{e}_7, 3}((A_3 + A_1)', D_6(a_1))$\\[2pt]
        {\scriptsize$(\dim\mathcal{C},\dim\mathcal{H})=(18,33)$}
    \end{tabular}\arrow[shorten >=2pt]{lld}{(\mathfrak{su}_2)_{8}}\arrow[shorten >=2pt]{d}{(\mathfrak{su}_2)_{13}} \\ 
    % Row 3
    \begin{tabular}{@{}r@{}}
        $2\,1\overset{\displaystyle 1}{\overset{\mathfrak{e}_7}{7}}1\overset{\mathfrak{su}_2}{2}\overset{\mathfrak{g}_2}{3}$\quad {\color{olive}2-8}\\[2pt]
        %\scriptsize$\mathcal{T}_{\mathfrak{e}_7, 3}(A_3 + 2A_1, D_5 + A_1) = \mathcal{T}_{\mathfrak{e}_7, 3}((A_3 + A_1)', E_7(a_4))$\\[2pt]
        {\scriptsize$(\dim\mathcal{C},\dim\mathcal{H})=(17,32)$}
    \end{tabular}\arrow[shorten >=2pt]{d}{(\mathfrak{su}_2)_{13}} & & \begin{tabular}{@{}r@{}}
	    $1\underset{\displaystyle 1}{\overset{\displaystyle 1}{\overset{\mathfrak{e}_7}{7}}}1\overset{\mathfrak{su}_2}{2}\overset{\mathfrak{so}_7}{3}$\quad\; {\color{olive}2-9}\\[8pt]
	    %\scriptsize$\mathcal{T}_{\mathfrak{e}_7, 3}(D_4(a_1) + A_1, D_5) = \mathcal{T}_{\mathfrak{e}_7, 3}(D_4(a_1), D_6(a_1))$\\[2pt]
        {\scriptsize$(\dim\mathcal{C},\dim\mathcal{H})=(18,32)$}
    \end{tabular}\arrow[shorten >=2pt]{d}{(\mathfrak{su}_2)_{12}}\arrow[shorten >=2pt]{lld}{(\mathfrak{su}_2)_{8}} \\
    % Row 4
    \begin{tabular}{@{}r@{}}
        $1\underset{\displaystyle 1}{\overset{\displaystyle 1}{\overset{\mathfrak{e}_7}{7}}}1\overset{\mathfrak{su}_2}{2}\overset{\mathfrak{g}_2}{3}$\; {\color{olive}2-10}\\[8pt]
        %\scriptsize$\mathcal{T}_{\mathfrak{e}_7, 3}(D_4(a_1) + A_1, D_5 + A_1) = \mathcal{T}_{\mathfrak{e}_7, 3}(D_4(a_1), E_7(a_4))$\\[2pt]
        {\scriptsize$(\dim\mathcal{C},\dim\mathcal{H})=(17,31)$}
    \end{tabular}\arrow[shorten >=2pt]{d}{(\mathfrak{su}_2)_{12}} &  & \begin{tabular}{@{}r@{}}
	    $1\overset{\displaystyle 1}{\overset{\mathfrak{e}_7}{6}}1\overset{\mathfrak{su}_2}{2}\overset{\mathfrak{so}_7}{3}$\quad\; {\color{olive}3-3}\\[2pt]
	    %\scriptsize$\mathcal{T}_{\mathfrak{e}_7, 3}(A_3 + A_2, D_5) = \mathcal{T}_{\mathfrak{e}_7, 3}(D_4(a_1) + A_1, D_6(a_1))$\\[2pt]
        {\scriptsize$(\dim\mathcal{C},\dim\mathcal{H})=(17,31)$}
    \end{tabular}\arrow[shorten >=2pt]{d}{(\mathfrak{su}_2)_{12}}\arrow[shorten >=2pt]{lld}{(\mathfrak{su}_2)_{8}} \\
    % Row 5
    \begin{tabular}{@{}r@{}}
        $1\overset{\displaystyle 1}{\overset{\mathfrak{e}_7}{6}}1\overset{\mathfrak{su}_2}{2}\overset{\mathfrak{g}_2}{3}$\;\;\; {\color{olive}3-4}\\[2pt]
        %\scriptsize$\mathcal{T}_{\mathfrak{e}_7, 3}(A_3 + A_2, D_5 + A_1) = \mathcal{T}_{\mathfrak{e}_7, 3}(D_4(a_1) + A_1, E_7(a_4))$\\[2pt]
        {\scriptsize$(\dim\mathcal{C},\dim\mathcal{H})=(16,30)$}
    \end{tabular}\arrow[shorten >=2pt]{d}{(\mathfrak{su}_2)_{12}}  & & \begin{tabular}{@{}r@{}}
	    $1\overset{\mathfrak{e}_7}{5}1\overset{\mathfrak{su}_2}{2}\overset{\mathfrak{so}_7}{3}$\quad\; {\color{olive}3-5}\\[2pt]
	    %\scriptsize$\mathcal{T}_{\mathfrak{e}_7, 3}(A_3 + A_2 + A_1, D_5) = \mathcal{T}_{\mathfrak{e}_7, 3}(A_3 + A_2, D_6(a_1))$\\[2pt]
        {\scriptsize$(\dim\mathcal{C},\dim\mathcal{H})=(16,30)$}
      \end{tabular}
      % \arrow[shorten >=2pt]{d}{(\mathfrak{su}_2)_{224}}
      \arrow[dashed,shorten >=2pt]{d}
      \arrow[shorten >=2pt]{lld}{(\mathfrak{su}_2)_{8}} \\
    % Row 6
    \begin{tabular}{@{}r@{}}
        $1\overset{\mathfrak{e}_7}{5}1\overset{\mathfrak{su}_2}{2}\overset{\mathfrak{g}_2}{3}$\;\;\; {\color{olive}3-6}\\[2pt]
        %\scriptsize$\mathcal{T}_{\mathfrak{e}_7, 3}(A_3 + A_2 + A_1, D_5 + A_1) = \mathcal{T}_{\mathfrak{e}_7, 3}(A_3 + A_2, E_7(a_4))$\\[2pt]
        {\scriptsize$(\dim\mathcal{C},\dim\mathcal{H})=(15,29)$}
      \end{tabular}
      % \arrow[shorten >=2pt]{d}{(\mathfrak{su}_2)_{224}} 
      \arrow[dashed,shorten >=2pt]{d} & & 
      \begin{tabular}{@{}r@{}}
	    $1\overset{\mathfrak{e}_6}{5}1\overset{\mathfrak{su}_2}{2}\overset{\mathfrak{so}_7}{3}$\quad\; {\color{olive}4-2} \\[2pt]
	    %\scriptsize$\mathcal{T}_{\mathfrak{e}_7, 3}(A_4 + A_1, D_5) = \mathcal{T}_{\mathfrak{e}_7, 3}(A_4, D_6(a_1)) - 1$\\[2pt]
        {\scriptsize$(\dim\mathcal{C},\dim\mathcal{H})=(15,28)$}
      \end{tabular}\arrow[shorten >=2pt]{lld}{(\mathfrak{su}_2)_{8}} \\
    % Row 7
    \begin{tabular}{@{}r@{}}
        $1\overset{\mathfrak{e}_6}{5}1\overset{\mathfrak{su}_2}{2}\overset{\mathfrak{g}_2}{3}$\;\;\; {\color{olive}4-3}\\[2pt]
        %\scriptsize$\mathcal{T}_{\mathfrak{e}_7, 3}(A_4 + A_1, D_5 + A_1) = \mathcal{T}_{\mathfrak{e}_7, 3}(A_4, E_7(a_4)) - 1$\\[2pt]
        {\scriptsize$(\dim\mathcal{C},\dim\mathcal{H})=(14,27)$}
      \end{tabular} & & \\
\end{tikzcd}
	\caption{The Hasse diagrams of 6d $(1,0)$ SCFTs arising from Higgsed rank three $(\mathfrak{e}_7, \mathfrak{e}_7)$ conformal matter which occur via Higgsing by two distinct pairs of nilpotent orbits. See the caption of Figure \ref{fig:e7N1} for an explanation of the notation.}\label{fig:e7N3}
\end{figure}

\section{Isomorphic 4d \texorpdfstring{$\mathcal{N}=2$}{N = 2 } SCFTs of class \texorpdfstring{$\mathcal{S}$}{S} with \texorpdfstring{$\mathfrak{g} = \mathfrak{e}_7$}{g = e7}}\label{sec:4dbis}

We can now return to the three-punctured spheres that were discussed in Section \ref{sec:4d}. Following Section \ref{sec:6d}, we know how to determine when 6d $(1,0)$ SCFTs associated to the data 
\begin{equation}
    (\mathfrak{g}, N, O_L, O_R) \,,
\end{equation}
are isomorphic. The 4d $\mathcal{N}=2$ SCFTs resulting from the toroidal compactification of these isomorphic 6d SCFTs are thus evidently themselves isomorphic. Each of the 4d SCFTs has a dual description in terms of class $\mathcal{S}$, and we can take the degeneration limit depicted in equation \eqref{eqn:chainofP1}.\footnote{In equation \eqref{eqn:chainofP1}, we drew the degeneration limit for $\mathfrak{g} = \mathfrak{e}_7$, however the generalization to arbitrary $\mathfrak{g}$ is obvious.} Finally, moving to the codimension one boundary of the conformal manifold we can see that the three-punctured spheres on the right in equation \eqref{eqn:chainofP1} decouple, and we have established the isomorphism. It is important to note that the third puncture, appearing in both three-punctured sphere, must belong to the quiver tail in the degeneration limit; for $\mathfrak{g} = \mathfrak{e}_7$, the possible choices of regular third puncture are $O = E_7(a_1), D_6, (A_5)'', (3A_1)''$. Further isomorphisms of three-punctured spheres can be obtained via nilpotent Higgsings of the non-Abelian flavor symmetry associated to the third puncture. 

\subsection{Pairs with the same manifest symmetries}\label{sec:same}

We begin by considering three-punctured spheres where the flavor symmetry is simply the manifest flavor symmetry induced by the choice of the three punctures. The discovering of such examples was a part of the analysis in \cite{Distler:2022nsn}; in particular, the putatively isomorphic pair satisfies equation \eqref{eqn:higgspair} together with equation \eqref{eqn:noenhanced}. The isomorphic pairs of three-punctured spheres satisfying this condition, and for which the isomorphism can be proven directly from six dimensions, are listed in Table \ref{tb:isopairs}. 

The last two entries in Table \ref{tb:isopairs} are a little special, as the collision of the two chosen punctures leads to an irregular fixture on the right. The quiver tails for theory A in row 15 looks like
\begin{align}
\begin{aligned}
    \scalebox{.73}{
    \begin{tikzpicture}
    \draw[radius=40pt,fill=lightred] circle;
    \draw[radius=2pt,fill=white]  (-.7,.9) circle node[right=2pt] {$E_7(a_1)$};
    \draw[radius=2pt,fill=white]  (-.7,-.9) circle node[right=2pt] {$E_7(a_1)$};
    \draw[radius=2pt,fill=white]  (1,0) circle node[left=2pt] (D6su2) {$\bigl(D_6,SU(2)\bigr)$};
    \node at (0,-2) {$\tfrac{1}{2}(2,1,1)$};
    \draw[radius=40pt,fill=lightred] (4,0) circle;
    \draw[radius=2pt,fill=white]  (4,1) circle node[below=2pt] {$E_7(a_1)$};
    \draw[radius=2pt,fill=white]  (3,0) circle node[right=2pt] (D6) {$D_6$};
    \draw[radius=2pt,fill=white]  (5,0) circle node[below left=5pt and -7pt] (A5ppG2){$\bigl({(A_5)}'',G_2\bigr)$};
    \path (1.1,0) edge node[above] {$SU(2)$}  (2.9,0);
    \node at (4,-2) {$\tfrac{1}{2}(2,7,1)$};
    \draw[radius=40pt,fill=lightblue] (8,0) circle;
    \draw[radius=2pt,fill=white]  (8,1) circle node[below=2pt] {$E_7(a_1)$};
    \draw[radius=2pt,fill=white]  (7,0) circle node[right=2pt] (A5pp) {${(A_5)}''$};
    \draw[radius=2pt,fill=white]  (9,0) circle node[below left=5pt and -7pt] (A5ppG2){$\bigl({(3A_1)}'',F_4\bigr)$};
    \path (5.1,0) edge node[above] {$G_2$}  (6.9,0);
    \node at (8,-2) {$[{(E_8)}_{12}\;\text{SCFT}]$};
    \draw[radius=40pt,fill=lightblue] (12,0) circle;
    \draw[radius=2pt,fill=white]  (12,1) circle node[below=2pt] {$E_7(a_1)$};
    \draw[radius=2pt,fill=white]  (11,0) circle node[below right=2pt and -5pt] (A1A1A1pp) {${(3A_1)}''$};
    \draw[radius=2pt,fill=white]  (13,0) circle node[left=2pt] {$A_1$};
    \path (9.1,0) edge node[above] {$F_4$}  (10.9,0);
    \node at (12.3,-2) {$[{(F_4)}_{24}\times {Spin(13)}_{28}\;\text{SCFT}]$};
    \draw[radius=40pt,fill=lightmauve] (17,0) circle;
    \draw[radius=2pt,fill=white]  (16,0) circle node[below right=1pt and -12pt] {$\bigl(A_1,Spin(12)\bigr)$};
    \draw[radius=2pt,fill=white]  (17.5,.9) circle node[left=2pt] {$D_5$};
    \draw[radius=2pt,fill=white]  (17.5,-.9) circle node[left=2pt] {$E_7(a_2)$};
    \path (13.1,0) edge node[above] {$Spin(12)$}  (15.9,0);
    \node at (17,-2) {$(12)+[{(E_7)}_{8}\;\text{SCFT}]$};
    \end{tikzpicture}} \,,
\end{aligned}
\end{align}
whereas the quiver tail for theory B in row 16 is
\begin{align}
\begin{aligned}
    \scalebox{.73}{
    \begin{tikzpicture}
    \draw[radius=40pt,fill=lightred] circle;
    \draw[radius=2pt,fill=white]  (-.7,.9) circle node[right=2pt] {$E_7(a_1)$};
    \draw[radius=2pt,fill=white]  (-.7,-.9) circle node[right=2pt] {$E_7(a_1)$};
    \draw[radius=2pt,fill=white]  (1,0) circle node[left=2pt] (D6su2) {$\bigl(D_6,SU(2)\bigr)$};
    \node at (0,-2) {$\tfrac{1}{2}(2,1,1)$};
    \draw[radius=40pt,fill=lightred] (4,0) circle;
    \draw[radius=2pt,fill=white]  (4,1) circle node[below=2pt] {$E_7(a_1)$};
    \draw[radius=2pt,fill=white]  (3,0) circle node[right=2pt] (D6) {$D_6$};
    \draw[radius=2pt,fill=white]  (5,0) circle node[below left=5pt and -7pt] (A5ppG2){$\bigl({(A_5)}'',G_2\bigr)$};
    \path (1.1,0) edge node[above] {$SU(2)$}  (2.9,0);
    \node at (4,-2) {$\tfrac{1}{2}(2,7,1)$};
    \draw[radius=40pt,fill=lightblue] (8,0) circle;
    \draw[radius=2pt,fill=white]  (8,1) circle node[below=2pt] {$E_7(a_1)$};
    \draw[radius=2pt,fill=white]  (7,0) circle node[right=2pt] (A5pp) {${(A_5)}''$};
    \draw[radius=2pt,fill=white]  (9,0) circle node[below left=5pt and -7pt] (A5ppG2){$\bigl({(3A_1)}'',F_4\bigr)$};
    \path (5.1,0) edge node[above] {$G_2$}  (6.9,0);
    \node at (8,-2) {$[{(E_8)}_{12}\;\text{SCFT}]$};
    \draw[radius=40pt,fill=lightblue] (12,0) circle;
    \draw[radius=2pt,fill=white]  (12,1) circle node[below=2pt] {$E_7(a_1)$};
    \draw[radius=2pt,fill=white]  (11,0) circle node[below right=2pt and -5pt] (A1A1A1pp) {${(3A_1)}''$};
    \draw[radius=2pt,fill=white]  (13,0) circle node[left=2pt] {$A_1$};
    \path (9.1,0) edge node[above] {$F_4$}  (10.9,0);
    \node at (12.3,-2) {$[{(F_4)}_{24}\times {Spin(13)}_{28}\;\text{SCFT}]$};
    \draw[radius=40pt,fill=lightred] (17,0) circle;
    \draw[radius=2pt,fill=white]  (16,0) circle node[below right=1pt and -12pt] {$\bigl(A_1,Spin(12)\bigr)$};
    \draw[radius=2pt,fill=white]  (17.5,.9) circle node[left=2pt] {$E_7(a_4)$};
    \draw[radius=2pt,fill=white]  (17.5,-.9) circle node[left=2pt] {$E_6$};
    \path (13.1,0) edge node[above] {$Spin(12)$}  (15.9,0);
    \node at (17,-2) {$(12)+(32)$};
    \end{tikzpicture}} \,.
\end{aligned}
\end{align}
The degeneration for the other element of each pair involves modifying only the two rightmost punctures.

For entries 1--11 in Table \ref{tb:isopairs}, we can do a chain of nilpotent Higgsings of the third puncture to establish other isomorphic pairs of theories. For entries 1--4 in Table \ref{tb:isopairs}, we can start with $(3A_1)''$ and Higgs it according to the following:
\begin{align}
\begin{aligned}
    \begin{tikzpicture}
        \node at (0,0) (3A1pp) {$\color{midgreen}(3A_1)''$};
        \node[right=1cm of 3A1pp] (A1A1A1A1) {$\color{midgreen}4A_1$};
        \node[right=1.5cm of A1A1A1A1] (A2A1) {$\color{midgreen}A_2+A_1$};
        \node[right=1.5cm of A2A1] (A2A1A1) {$\color{midgreen}{A_2+2A_1}$};
        \node[above right= .75cm and .75cm of A2A1A1] (A2A2) {$\color{purple}2A_2$};
        \node[below right= .75cm and .5cm of A2A1A1] (A2A1A1A1) {$\color{midgreen}A_2+3A_1$};
        \node[right=2.5cm of A2A1A1] (A2A2A1) {${\color{midgreen}2A_2+A_1}$};
        \path[->] 
        (3A1pp) edge node[above] {$(\mathfrak{f}_4)_{24}$} (A1A1A1A1)
        (A1A1A1A1) edge node[above] {$(\mathfrak{sp}_3)_{19}$} (A2A1)
        (A2A1) edge node[above] {$(\mathfrak{su}_4)_{18}$} (A2A1A1)
        (A2A1A1) edge node[above left=-.125cm and -.125cm] {$(\mathfrak{su}_2)_{28}$} (A2A2)
        (A2A1A1) edge node[below left=-.125cm and -.125cm] {$(\mathfrak{su}_2)_{16}$} (A2A1A1A1)
        (A2A2) edge node[above right=-.125cm and -.125cm] {$(\mathfrak{g}_2)_{16}$} (A2A2A1)
        (A2A1A1A1) edge node[below right=-.125cm and -.125cm] {$(\mathfrak{g}_2)_{28}$} (A2A2A1)
        ;
    \end{tikzpicture}
\label{familyHasse}
\end{aligned} \,.
\end{align}

The Higgsing to $\color{purple}2A_2$ is special. For these theories, the manifest $(\mathfrak{su}_2)_{16}\oplus (\mathfrak{su}_2)_{28}\oplus (\mathfrak{su}_2)_{84}$ symmetry of the $A_2+2A_1$ puncture is enhanced to $(\mathfrak{su}_2)_{16}\oplus (\mathfrak{su}_2)_{28}\oplus (\mathfrak{su}_2)_{28}\oplus (\mathfrak{su}_2)_{56}$. Of the two $(\mathfrak{su}_2)_{28}$ factors present in the SCFT, which one is the ``manifest'' one (whose Higgsing leads to the $2A_2$ puncture) differs between Theory A and Theory B. Hence, at that one step in the chain, the nilpotent Higgsing of the manifest $(\mathfrak{su}_2)_{28}$, which leads to $2A_2$, yields non-isomorphic SCFTs in Theories A and B.

The Higgsings for entries 5--11
in Table \ref{tb:isopairs} are simpler, yielding three additional pairs of isomorphic theories per entry. These Higgsings were discussed around equation \eqref{eqn:swiper}, and they are
\begin{equation}\label{smallerHiggsings}
\begin{tikzpicture}
\node[color=midgreen] (A5pp) at (0,0) {$(A_5)''$};
\node[color=midgreen] (A5A1) at (2,0) {$A_5+A_1$};
\node[color=midgreen] (D6a2) at (5,0) {$D_6(a_2)$};
\node[color=midgreen] (E7a5) at (7.5,0) {$E_7(a_5)$};
\path[thick, ->,draw,color=olive] 
(A5pp) edge[color=black] node[color=black,below] {$\scriptstyle (\mathfrak{g}_2)_{12}$} (A5A1)
(A5A1) edge[color=blue] node[color=black,below] {$\scriptstyle \mathfrak{su}(2)_{26}$} (D6a2)
(D6a2) edge[color=black] node[color=black,below] {$\scriptstyle \mathfrak{su}(2)_{9}$} (E7a5)
(0.3,0.3) arc[color=black,radius=4.2, start angle=120, end angle=60] node[color=black,above right=.5cm and -2.5cm] {$\scriptstyle (\mathfrak{g}_2)_{12}$} (D6a2)
;
\end{tikzpicture}
 \,.
\end{equation}
Nilpotent Higgsing of the $(\mathfrak{su}_2)_7$ flavor factor associated to the $D_6$ punctures in entries 12--14 yields bad theories, as would further Higgsings of $2A_2+A_1$ in equation \eqref{familyHasse}.

\begin{table}[H]
  \centering
  \footnotesize
  \renewcommand{\arraystretch}{1.25}
  \begin{threeparttable}
      \begin{tabular}{@{}ccc@{}c@{}c@{}c@{}}
        \toprule
        \#&Theory A&Theory B&$N_{\text{min}}$& 6d SCFT & Flavor symmetry \\\midrule
        1&$\begin{matrix}A_3\\ E_7(a_2)\end{matrix}\quad (3A_1)''$&$\begin{matrix}(A_3+A_1)''\\ E_6\end{matrix}\quad (3A_1)''$& $4$ & $\begin{gathered} \overset{\mathfrak{su}_2}{2}\overset{\mathfrak{so}_7}{3}\overset{\mathfrak{su}_2}{2}1\underset{\displaystyle 1}{\overset{\mathfrak{e}_7}{7}}1\overset{\mathfrak{su}_2}{2} \end{gathered}$ & $(\mathfrak{su}_2)_{12} \oplus (\mathfrak{so}_7)_{16}$ \\\hline

        2&$\begin{matrix}E_6\\ A_3+2A_1\end{matrix}\quad (3A_1)''$&$\begin{matrix}E_7(a_2)\\ (A_3+A_1)'\end{matrix}\quad (3A_1)''$& $4$ & $\begin{gathered} \overset{\mathfrak{su}_2}{2}\overset{\mathfrak{so}_7}{3}\overset{\mathfrak{su}_2}{2}1\underset{\displaystyle 1}{\overset{\mathfrak{e}_7}{7}}1\,2 \end{gathered}$ & $(\mathfrak{su}_2)_{12} \oplus (\mathfrak{su}_2)_{24} \oplus (\mathfrak{su}_2)_{13}$ \\\hline

        3&$\begin{matrix}D_4(a_1)\\ E_7(a_2)\end{matrix}\quad (3A_1)''$&$\begin{matrix}D_4(a_1)+A_1\\ E_6\end{matrix}\quad (3A_1)''$& $4$ & $\begin{gathered} \overset{\mathfrak{su}_2}{2}\overset{\mathfrak{so}_7}{3}\overset{\mathfrak{su}_2}{2}1\overset{\displaystyle 1}{\underset{\displaystyle 1}{\overset{\mathfrak{e}_7}{7}}}1 \end{gathered}$ & $(\mathfrak{su}_2)_{12}^{\oplus 3}$ \\\hline

        4&$\begin{matrix}D_5\\ E_7(a_4)\end{matrix}\quad (3A_1)''$&$\begin{matrix}D_6(a_1)\\ D_5+A_1\end{matrix}\quad (3A_1)''$& $4$ & $\begin{gathered} \overset{\mathfrak{so}_7}{3}\overset{\mathfrak{su}_2}{2}1\underset{\displaystyle 1}{\overset{\mathfrak{e}_7}{7}}1\overset{\mathfrak{su}_2}{2} \overset{\mathfrak{g}_2}{3} \end{gathered}$ & $(\mathfrak{su}_2)_{12} \oplus (\mathfrak{su}_2)_{8}$ \\\hline

        5&$\begin{matrix}A_3\\ D_6(a_1)\end{matrix}\quad (A_5)''$&$\begin{matrix}(A_3+A_1)''\\ D_5\end{matrix}\quad (A_5)''$& $3$ & $\begin{gathered} \overset{\mathfrak{so}_7}{3}\overset{\mathfrak{su}_2}{2}1\underset{\displaystyle 1}{\overset{\mathfrak{e}_7}{7}}1\overset{\mathfrak{su}_2}{2} \end{gathered}$ & $(\mathfrak{su}_2)_{12} \oplus (\mathfrak{su}_2)_{8} \oplus (\mathfrak{so}_7)_{16}$ \\\hline

        6&$\begin{matrix}A_3\\ E_7(a_4)\end{matrix}\quad (A_5)''$&$\begin{matrix}(A_3+A_1)''\\ D_5+A_1\end{matrix}\quad (A_5)''$& $3$ & $\begin{gathered} \overset{\mathfrak{su}_2}{2}1\underset{\displaystyle 1}{\overset{\mathfrak{e}_7}{7}}1\overset{\mathfrak{su}_2}{2} \overset{\mathfrak{g}_2}{3} \end{gathered}$ & $(\mathfrak{su}_2)_{12} \oplus (\mathfrak{so}_7)_{16}$ \\\hline

        7&$\begin{matrix}(A_3+A_1)'\\ D_6(a_1)\end{matrix}\quad (A_5)''$&$\begin{matrix}A_3+2A_1\\ D_5\end{matrix}\quad (A_5)''$& $3$ & $\begin{gathered} \overset{\mathfrak{so}_7}{3}\overset{\mathfrak{su}_2}{2}1\underset{\displaystyle 1}{\overset{\mathfrak{e}_7}{7}}1\,2 \end{gathered}$ & $\begin{array}{l} 
        (\mathfrak{su}_2)_{12} \oplus (\mathfrak{su}_2)_{24}\\[-3pt]
        \qquad\oplus (\mathfrak{su}_2)_{8} \oplus (\mathfrak{su}_2)_{13} 
        \end{array}$\\\hline

        8&$\begin{matrix}(A_3+A_1)'\\ E_7(a_4)\end{matrix}\quad (A_5)''$&$\begin{matrix}A_3+2A_1\\ D_5+A_1\end{matrix}\quad (A_5)''$& $3$ & $\begin{gathered} \overset{\mathfrak{g}_2}{3}\overset{\mathfrak{su}_2}{2}1\underset{\displaystyle 1}{\overset{\mathfrak{e}_7}{7}}1\,2 \end{gathered}$ & $(\mathfrak{su}_2)_{12} \oplus (\mathfrak{su}_2)_{24} \oplus (\mathfrak{su}_2)_{13}$ \\\hline

        9&$\begin{matrix}D_4(a_1)\\ D_6(a_1)\end{matrix}\quad (A_5)''$&$\begin{matrix}D_4(a_1)+A_1\\ D_5\end{matrix}\quad (A_5)''$& $3$ & $\begin{gathered} \overset{\mathfrak{so}_7}{3}\overset{\mathfrak{su}_2}{2}1\overset{\displaystyle 1}{\underset{\displaystyle 1}{\overset{\mathfrak{e}_7}{7}}}1 \end{gathered}$ & $(\mathfrak{su}_2)_{12}^{\oplus 3} \oplus (\mathfrak{su}_2)_{8}$ \\\hline

        10&$\begin{matrix}D_4(a_1)\\ E_7(a_4)\end{matrix}\quad (A_5)''$&$\begin{matrix}D_4(a_1)+A_1\\ D_5+A_1\end{matrix}\quad (A_5)''$& $3$ & $\begin{gathered} \overset{\mathfrak{g}_2}{3}\overset{\mathfrak{su}_2}{2}1\overset{\displaystyle 1}{\underset{\displaystyle 1}{\overset{\mathfrak{e}_7}{7}}}1 \end{gathered}$ & $(\mathfrak{su}_2)_{12}^{\oplus 3}$ \\\hline

        11&$\begin{matrix}(A_5)'\\ E_7(a_5)\end{matrix}\quad (A_5)''$&$\begin{matrix}E_6(a_3)\\ D_6(a_2)\end{matrix}\quad (A_5)''$& $3$ & $\begin{gathered} \overset{\mathfrak{so}_8}{4}1\overset{\mathfrak{g}_2}{3}1\overset{\mathfrak{so}_9}{4} \end{gathered}$ & $(\mathfrak{su}_2)_{19} \oplus (\mathfrak{su}_2)_{9}$ \\\hline

        12&$\begin{matrix}A_3\\ A_3+2A_1\end{matrix}\quad D_6$&$\begin{matrix}(A_3+A_1)''\\ (A_3+A_1)'\end{matrix}\quad D_6$& $2$ & $\begin{gathered} 2\,1\underset{\displaystyle 1}{\overset{\mathfrak{e}_7}{7}}1\overset{\mathfrak{su}_2}{2} \end{gathered}$ & $\begin{array}{l} (\mathfrak{su}_2)_{12} \oplus (\mathfrak{su}_2)_{24}\\[-3pt]
        \qquad\oplus (\mathfrak{su}_2)_{13} \oplus (\mathfrak{so}_7)_{16}
        \end{array}$ \\\hline

        13&$\begin{matrix}A_3\\ D_4(a_1)+A_1\end{matrix}\quad D_6$&$\begin{matrix}(A_3+A_1)''\\ D_4(a_1)\end{matrix}\quad D_6$& $2$ & $\begin{gathered} 1\overset{\displaystyle 1}{\underset{\displaystyle 1}{\overset{\mathfrak{e}_7}{7}}}1\overset{\mathfrak{su}_2}{2} \end{gathered}$ & $(\mathfrak{su}_2)_{12}^{\oplus 3} \oplus (\mathfrak{so}_7)_{16}$ \\\hline

        14&$\begin{matrix}(A_3+A_1)'\\ D_4(a_1)+A_1\end{matrix}\quad D_6$&$\begin{matrix}A_3+2A_1\\ D_4(a_1)\end{matrix}\quad D_6$& $2$ & $\begin{gathered} 1\overset{\displaystyle 1}{\underset{\displaystyle 1}{\overset{\mathfrak{e}_7}{7}}}1\,2 \end{gathered}$ & $(\mathfrak{su}_2)_{12}^{\oplus 3} \oplus (\mathfrak{su}_2)_{24} \oplus (\mathfrak{su}_2)_{13}$  \\\hline

        15&$\begin{matrix}D_5\\ E_7(a_2)\end{matrix}\; \bigl(A_1,Spin(12)\bigr)$&$\begin{matrix}D_6(a_1)\\ E_6\end{matrix}\; \bigl(A_1,Spin(12)\bigr)$&5&$
        \begin{gathered} \overset{\mathfrak{so}_7}{3}\,
        \overset{\mathfrak{su}_2}{2}\,1\,\underset{\displaystyle 1}{\overset{\mathfrak{e}_7}{7}}\,1\,
        \overset{\mathfrak{su}_2}{2}\,
        \overset{\mathfrak{so}_7}{3}\,
        \overset{\mathfrak{su}_2}{2}\end{gathered}
        $ & $(\mathfrak{su}_2)_{12} \oplus (\mathfrak{su}_2)_{8}$  \\\hline

        16&$\begin{matrix}D_5+A_1\\ E_7(a_2)\end{matrix}\; \bigl(A_1,Spin(12)\bigr)$&$\begin{matrix}E_7(a_4)\\ E_6\end{matrix}\; \bigl(A_1,Spin(12)\bigr)$&5&$
        \begin{gathered} \overset{\mathfrak{g}_2}{3}\,
        \overset{\mathfrak{su}_2}{2}1\,
        \underset{\displaystyle 1}{\overset{\mathfrak{e}_7}{7}}\,1\,
        \overset{\mathfrak{su}_2}{2}\,
        \overset{\mathfrak{so}_7}{3}\,
        \overset{\mathfrak{su}_2}{2}\end{gathered}
        $ & $(\mathfrak{su}_2)_{12}$ \\
        \bottomrule
        \end{tabular}
  \end{threeparttable}
  \caption{Isomorphic pairs of interacting three-punctured spheres for class $\mathcal{S}$ of type $\mathfrak{e}_7$, where the flavor symmetry is the manifest flavor symmetry from the individual punctures.}\label{tb:isopairs}
\end{table}

\begin{table}[H]
  \centering
  \footnotesize
  \renewcommand{\arraystretch}{1.4}
  \begin{threeparttable}
  \begin{tabular}{cccccc}
  \toprule
  \#&Theory A&Theory B&$N_{\text{min}}$& 6d SCFT & Flavor symmetry \\\midrule
  1&${\renewcommand{\arraystretch}{1.3}\begin{matrix}D_4(a_1)+A_1\\ E_7(a_2)\end{matrix}}\,\,\, (3A_1)''$&${\renewcommand{\arraystretch}{1.3}\begin{matrix}A_3+A_2\\ E_6\end{matrix}}\,\,\, (3A_1)''$& $4$ & $\begin{gathered} 1\underset{\displaystyle 1}{\overset{\mathfrak{e}_7}{6}}1\overset{\mathfrak{su}_2}{2}\overset{\mathfrak{so}_7}{3}\overset{\mathfrak{su}_2}{2} \end{gathered}$ & $(\mathfrak{su}_2)_{12}^{\oplus 2}$ \\\hline
  2&${\renewcommand{\arraystretch}{1.3}\begin{matrix}E_6\\ A_3+A_2+A_1\end{matrix}}\,\,\, (3A_1)''$&${\renewcommand{\arraystretch}{1.3}\begin{matrix}E_7(a_2)\\ A_3+A_2\end{matrix}}\,\,\, (3A_1)''$& $4$ & $\begin{gathered} 1\overset{\mathfrak{e}_7}{5}1\overset{\mathfrak{su}_2}{2}\overset{\mathfrak{so}_7}{3}\overset{\mathfrak{su}_2}{2} \end{gathered}$ & $(\mathfrak{su}_2)_{12} \oplus (\mathfrak{su}_2)_{224}$ \\\hline
  3&${\renewcommand{\arraystretch}{1.3}\begin{matrix}D_5\\ A_3+A_2\end{matrix}}\,\,\, (A_5)''$&${\renewcommand{\arraystretch}{1.3}\begin{matrix}D_6(a_1)\\ D_4(a_1)+A_1\end{matrix}}\,\,\, (A_5)''$& $3$ & $\begin{gathered} 1\underset{\displaystyle 1}{\overset{\mathfrak{e}_7}{6}}1\overset{\mathfrak{su}_2}{2}\overset{\mathfrak{so}_7}{3} \end{gathered}$ & $(\mathfrak{su}_2)_{12}^{\oplus 2} \oplus (\mathfrak{su}_2)_{8}$ \\\hline
  4&${\renewcommand{\arraystretch}{1.3}\begin{matrix}D_4(a_1)+A_1\\ E_7(a_4)\end{matrix}}\,\,\, (A_5)''$&${\renewcommand{\arraystretch}{1.3}\begin{matrix}A_3+A_2\\ D_5+A_1\end{matrix}}\,\,\, (A_5)''$& $3$ & $\begin{gathered} 1\underset{\displaystyle 1}{\overset{\mathfrak{e}_7}{6}}1\overset{\mathfrak{su}_2}{2}\overset{\mathfrak{g}_2}{3} \end{gathered}$ & $(\mathfrak{su}_2)_{12}^{\oplus 2}$ \\\hline
  5&${\renewcommand{\arraystretch}{1.3}\begin{matrix}A_3+A_2\\ D_6(a_1)\end{matrix}}\,\,\, (A_5)''$&${\renewcommand{\arraystretch}{1.3}\begin{matrix}A_3+A_2+A_1\\ D_5\end{matrix}}\,\,\, (A_5)''$& $3$ & $\begin{gathered} 1\overset{\mathfrak{e}_7}{5}1\overset{\mathfrak{su}_2}{2}\overset{\mathfrak{so}_7}{3} \end{gathered}$ & $(\mathfrak{su}_2)_{12} \oplus (\mathfrak{su}_2)_{224} \oplus (\mathfrak{su}_2)_8$ \\\hline
  6&${\renewcommand{\arraystretch}{1.3}\begin{matrix}A_3+A_2\\ E_7(a_4)\end{matrix}}\,\,\, (A_5)''$&${\renewcommand{\arraystretch}{1.3}\begin{matrix}A_3+A_2+A_1\\ D_5+A_1\end{matrix}}\,\,\, (A_5)''$& $3$ & $\begin{gathered} 1\overset{\mathfrak{e}_7}{5}1\overset{\mathfrak{su}_2}{2}\overset{\mathfrak{g}_2}{3} \end{gathered}$ & $(\mathfrak{su}_2)_{12} \oplus (\mathfrak{su}_2)_{224}$ \\\hline
  7&${\renewcommand{\arraystretch}{1.3}\begin{matrix}A_3\\ A_3+A_2\end{matrix}}\,\,\, D_6$&${\renewcommand{\arraystretch}{1.3}\begin{matrix}(A_3+A_1)''\\ D_4(a_1)+A_1\end{matrix}}\,\,\, D_6$& $2$ & $\begin{gathered} 1\underset{\displaystyle 1}{\overset{\mathfrak{e}_7}{6}}1\overset{\mathfrak{su}_2}{2} \end{gathered}$ & $(\mathfrak{su}_2)_{12}^{\oplus 2} \oplus (\mathfrak{so}_7)_{16}$ \\\hline
  8&${\renewcommand{\arraystretch}{1.3}\begin{matrix}(A_3+A_1)'\\ A_3+A_2\end{matrix}}\,\,\, D_6$&${\renewcommand{\arraystretch}{1.3}\begin{matrix}A_3+2A_1\\ D_4(a_1)+A_1\end{matrix}}\,\,\, D_6$& $2$ & $\begin{gathered} 1\underset{\displaystyle 1}{\overset{\mathfrak{e}_7}{6}}1\,2 \end{gathered}$ & $(\mathfrak{su}_2)_{12}^{\oplus 2} \oplus (\mathfrak{su}_2)_{24}\oplus (\mathfrak{su}_2)_{13}$ \\\hline
  9&${\renewcommand{\arraystretch}{1.3}\begin{matrix}D_4(a_1)\\ A_3+A_2\end{matrix}}\,\,\, D_6$&${\renewcommand{\arraystretch}{1.3}\begin{matrix}D_4(a_1)+A_1\\ D_4(a_1)+A_1\end{matrix}}\,\,\, D_6$& $2$ & $\begin{gathered} 1\overset{\displaystyle 1}{\underset{\displaystyle 1}{\overset{\mathfrak{e}_7}{6}}}1 \end{gathered}$ & $(\mathfrak{su}_2)_{12}^{\oplus 4}$ \\\hline
  10&${\renewcommand{\arraystretch}{1.3}\begin{matrix}A_3+A_2\\ (A_3+A_1)''\end{matrix}}\,\,\, D_6$&${\renewcommand{\arraystretch}{1.3}\begin{matrix}A_3+A_2+A_1\\ A_3\end{matrix}}\,\,\, D_6$& $2$ & $\begin{gathered} 1\overset{\mathfrak{e}_7}{5}1\overset{\mathfrak{su}_2}{2} \end{gathered}$ & $(\mathfrak{su}_2)_{12} \oplus (\mathfrak{su}_2)_{224} \oplus (\mathfrak{so}_7)_{16}$ \\\hline
  11&${\renewcommand{\arraystretch}{1.3}\begin{matrix}A_3+A_2\\ A_3+2A_1\end{matrix}}\,\,\, D_6$&${\renewcommand{\arraystretch}{1.3}\begin{matrix}A_3+A_2+A_1\\ (A_3+A_1)'\end{matrix}}\,\,\, D_6$& $2$ & $\begin{gathered} 1\overset{\mathfrak{e}_7}{5}1\,2 \end{gathered}$ & $\begin{array}{l}
    (\mathfrak{su}_2)_{12} \oplus (\mathfrak{su}_2)_{24}\\[-3pt]
    \qquad \oplus (\mathfrak{su}_2)_{224} \oplus (\mathfrak{su}_2)_{13}
  \end{array}$ \\\hline
  12&${\renewcommand{\arraystretch}{1.3}\begin{matrix}A_3+A_2\\ D_4(a_1)+A_1\end{matrix}}\,\,\, D_6$&${\renewcommand{\arraystretch}{1.3}\begin{matrix}A_3+A_2+A_1\\ D_4(a_1)\end{matrix}}\,\,\, D_6$& $2$ & $\begin{gathered} 1\underset{\displaystyle 1}{\overset{\mathfrak{e}_7}{5}}1 \end{gathered}$ & $(\mathfrak{su}_2)_{12}^{\oplus 3} \oplus (\mathfrak{su}_2)_{224}$ \\\hline
  13&${\renewcommand{\arraystretch}{1.3}\begin{matrix}A_3+A_2\\ A_3+A_2\end{matrix}}\,\,\, D_6$&${\renewcommand{\arraystretch}{1.3}\begin{matrix}A_3+A_2+A_1\\ D_4(a_1)+A_1\end{matrix}}\,\,\, D_6$& $2$ & $\begin{gathered} 1\overset{\mathfrak{e}_7}{4}1 \end{gathered}$ & $(\mathfrak{su}_2)_{12}^{\oplus 2} \oplus (\mathfrak{so}_4)_{112}$  \\\hline
  14&${\renewcommand{\arraystretch}{1.3}\begin{matrix}2A_1\\ D_6(a_2)\end{matrix}}\,\,\, D_6$&${\renewcommand{\arraystretch}{1.3}\begin{matrix}(3A_1)''\\ (A_5)'\end{matrix}}\,\,\, D_6$& $2$ & $\begin{gathered} \overset{\mathfrak{so}_9}{4}1\overset{\mathfrak{g}_2}{3}1\\[-5pt]
  +\;\text{a free hyper} \end{gathered}$ & $(\mathfrak{f}_4)_{24} \oplus (\mathfrak{su}_2)_9 \oplus (\mathfrak{su}_2)_{19}$ \\\hline
  15&${\renewcommand{\arraystretch}{1.3}\begin{matrix}2A_1\\ E_7(a_5)\end{matrix}}\,\,\, D_6$&${\renewcommand{\arraystretch}{1.3}\begin{matrix}(3A_1)''\\ E_6(a_3)\end{matrix}}\,\,\, D_6$& $2$ & $\begin{gathered} \overset{\mathfrak{so}_8}{4}1\overset{\mathfrak{g}_2}{3}1\\[-5pt]
  +\;\text{a free hyper} \end{gathered}$ & $(\mathfrak{f}_4)_{24} \oplus (\mathfrak{su}_2)_{19}$ \\\bottomrule
  \end{tabular}
  \end{threeparttable}
  \caption{Isomorphic pairs of three-punctured spheres for class $\mathcal{S}$ of type $\mathfrak{e}_7$, where the flavor symmetry is not the manifest flavor symmetry from the individual punctures.}\label{tb:diffmanifest}
\end{table}

\subsection{Interacting pairs with different manifest symmetries}\label{sec:different}

In Section \ref{sec:4d}, we constructed candidate pairs of isomorphic theories by starting with a parent theory with punctures $O_1, O'_2$, with an $\mathfrak{f}_k\oplus \mathfrak{f}_k$ flavor symmetry. Then, by Higgsing of one or the other of the $\mathfrak{f}_k$ factors, as in equation \eqref{eqn:higgspair}, we obtained a pair of theories whose conventional invariants coincided. As in \cite{Distler:2022nsn}, we imposed the restriction in equation \eqref{eqn:noenhanced} that the \emph{manifest} symmetries of the two theories coincide. We can relax this assumption: two three-punctured spheres can be isomorphic as long as the enhanced flavor symmetries match, even if the manifest flavor symmetries are different.

In the $\mathfrak{e}_7$ case, this leads to a slew of new pairs which we can show are isomorphic SCFTs using the 6d $(1,0)$ uplift. We list the interacting three-punctured sphere which are isomorphic and do not have identical manifest flavor symmetries in Table \ref{tb:diffmanifest}. As before, the third puncture in each of these pairs can be Higgsed as in equations \eqref{familyHasse} and \eqref{smallerHiggsings} to yield additional isomorphic pairs of SCFTs.

\subsection{Fixtures with isomorphic interacting sectors}\label{sec:isointer}

Thus far, we have demanded that each pair of fixtures be isomorphic on-the-nose as 4d SCFTs; that is, each fixture is associated to an interacting SCFTs plus some number of free hypermultiplets which is the same across the pair. More generally, we could allow fixtures which include differing numbers of free hypermultiplets, but whose interacting sectors are isomorphic. To construct examples of this behavior, we rely on the fact that if we have an embedding $\mathfrak{h}_k \subset \mathfrak{g}_k$, then a nilpotent Higgsing of $\mathfrak{h}_k$ will result in the same theory as a nilpotent Higgsing of $\mathfrak{g}_k$, with the addition of some number of free hypermultiplets. This results in two mechanisms for constructing candidate pairs which correspond to theories with isomorphic interacting parts.

The first occurs when two punctures both contribute simple flavor factors at level $k$, say $\mathfrak{h}_k$ and $\mathfrak{g}_k$. If we choose a third puncture low enough down on the Hasse diagram, it is possible that the $\mathfrak{h}_k \oplus \mathfrak{g}_k$ enhances to a $\mathfrak{g}_k\oplus \mathfrak{g}_k$ and there is an outer automorphism symmetry that exchanges the two $\mathfrak{g}$ factors. Then we can do a nilpotent Higgsing of either the $\mathfrak{g}_k$ or $\mathfrak{h}_k$, where the latter takes us to the same theory as the former in addition to some free hypermultiplets. Using 6d constructions we can prove the existence of such automorphism symmetries and thus prove the interacting sectors are indeed isomorphic.

The second mechanism occurs when two punctures contribute an $\mathfrak{f}_k$ and an $\mathfrak{h}_k$ that is enhanced to a $\mathfrak{g}_k$ global symmetry. Then the interacting part of the two fixtures obtained by Higgsing either the $\mathfrak{f}_k$ or $\mathfrak{h}_k$ should be the theory obtained by Higgsing the $\mathfrak{g}_k$, meaning they are isomorphic. In this case we do not require a 6d explanation as the isomorphism can be directly seen from the nilpotent Higgsing, though, of course, the 6d uplifts if they exist will also be isomorphic.

An example of the first mechanism is given by the following three-punctured sphere in the class $\mathcal{S}$ theory of type $\mathfrak{e}_8$: 
\begin{align}\label{eqn:joyforever}
\begin{aligned}
    \begin{tikzpicture}
    	\draw[radius=40pt,fill=lightblue] circle;
        \draw[radius=2pt,fill=white]  (-.5,.9) circle node[right=2pt] {$E_8(a_1)$};
        \draw[radius=2pt,fill=white]  (-.5,-1) circle node[above right=2pt and -15pt] {$A_3+A_1$};
        \draw[radius=2pt,fill=white]  (1,0) circle node[left=2pt] {$2A_2$};
        % \node at (0,-2) {$\mathfrak{f}_{\text{manifest}}=(\mathfrak{so}_7)_{24}\oplus (\mathfrak{g}_2)_{24}^2 \oplus (\mathfrak{su}_2)_{21}$};
    \end{tikzpicture} \,.
\end{aligned}
\end{align}
The manifest flavor symmetry is enhanced:
\begin{equation}
    \mathfrak{f}_{\text{manifest}}=(\mathfrak{so}_7)_{24}\oplus (\mathfrak{g}_2)_{24}^{\oplus 2} \oplus (\mathfrak{su}_2)_{21} \quad\rightarrow\quad \mathfrak{f}=(\mathfrak{so}_7)_{24}^{\oplus 3} \oplus (\mathfrak{su}_2)_{21} \,.
\end{equation}
Minimal nilpotent Higgsing of the one of the manifest $(\mathfrak{g}_2)_{24}$ factors is a local Higgsing that changes the puncture $2A_2 \rightarrow 2A_2 + A_1$, whereas minimal nilpotent Higgsing of the manifest $(\mathfrak{so}_7)_{24}$ factor changes the puncture $A_3 + A_1 \rightarrow A_3 + 2A_1$. These Higgsings lead to the following three-punctured spheres:
\begin{align}\label{eqn:potatoes}
\begin{aligned}
    \begin{tikzpicture}
        \draw[radius=40pt,fill=lightblue] circle;
        \draw[radius=2pt,fill=white]  (-.5,.9) circle node[right=2pt] {$E_8(a_1)$};
        \draw[radius=2pt,fill=white]  (-.5,-1) circle node[above right=2ptand -18pt] {$A_3+2A_1$};
        \draw[radius=2pt,fill=white]  (1,0) circle node[left=2pt] {$2A_2$};
        \draw[radius=40pt,fill=lightmauve] (4,0) circle;
        \draw[radius=2pt,fill=white]  (3.5,.9) circle node[right=2pt] {$E_8(a_1)$};
        \draw[radius=2pt,fill=white]  (3.5,-1) circle node[above right=2pt and -18pt] {$A_3+A_1$};
        \draw[radius=2pt,fill=white]  (5,0) circle node[left=2pt] {$2A_2+A_1 $};
    \end{tikzpicture} \,.
\end{aligned}
\end{align}
These are clearly different theories -- the theory on the right has a free hypermultiplet whereas the theory on the left does not. The parent theory, given by the three-punctured sphere in equation \eqref{eqn:joyforever}, has an uplift to a 6d $(1,0)$ curve configuration (via the usual quiver tail procedure), which is
\begin{align}
\begin{gathered} 
	1{\underset{\displaystyle 1}{\overset{\mathfrak{so}_9}{4}}}1 \,.
\end{gathered}
\end{align}
Evidently this curve configuration possesses an $S_3$ outer-automorphism group that permutes the three $(-1)$-curves, and thus the three $(\mathfrak{so}_7)_{24}$ flavor symmetry factors. Performing a minimal nilpotent Higgsing by any of the three $(\mathfrak{so}_7)_{24}$ factors, or by the subalgebra $(\mathfrak{g}_2)_{24}\subset (\mathfrak{so}_7)_{24}$, leads to 6d SCFTs with the same interacting sector. In conclusion, we can see that the interacting parts of the two 4d theories in equation \eqref{eqn:potatoes} are isomorphic. In fact, as we can see from \cite{Baume:2021qho}, the uplift of the interacting sector of both three-punctured spheres has the curve configuration
\begin{align}
\begin{gathered} 
	1{{\overset{\mathfrak{so}_9}{3}}}1 \,.
\end{gathered}
\end{align}

\begin{table}[H]
  \centering
  \footscriptsize
  \renewcommand{\arraystretch}{1.12}
  \begin{threeparttable}
      \begin{tabular}{@{}ccccccc@{}c@{}}
        \toprule
        \#&Theory A&Theory B&$n_A$&$n_B$&$N_{\text{min}}$& 6d SCFT & Flavor symmetry \\\midrule
        1&$\begin{matrix}A_4\\ E_7(a_2)\end{matrix}\quad (3A_1)''$&$\begin{matrix}A_4+A_1\\ E_6\end{matrix}\quad (3A_1)''$&1&0&4& $1\overset{\mathfrak{e}_6}{5}1\overset{\mathfrak{su}_2}{2}\overset{\mathfrak{so}_7}{3}\overset{\mathfrak{su}_2}{2}$ & $(\mathfrak{su}_3)_{12}$\\\hline

        2&$\begin{matrix}A_4\\ D_6(a_1)\end{matrix}\quad (A_5)''$&$\begin{matrix}A_4+A_1\\ D_5\end{matrix}\quad (A_5)''$&1&0&3& $1\overset{\mathfrak{e}_6}{5}1\overset{\mathfrak{su}_2}{2}\overset{\mathfrak{so}_7}{3}$ & $(\mathfrak{su}_3)_{12} \oplus (\mathfrak{su}_{2})_{8}$\\\hline
        
	3&$\begin{matrix}A_4\\ E_7(a_4)\end{matrix}\quad (A_5)''$&$\begin{matrix}A_4+A_1\\ D_5+A_1\end{matrix}\quad (A_5)''$&1&0&3& $1\overset{\mathfrak{e}_6}{5}1\overset{\mathfrak{su}_2}{2}\overset{\mathfrak{g}_2}{3}$ & $(\mathfrak{su}_3)_{12}$ \\\hline
	
	4&$\begin{matrix}A_4\\ (A_3+A_1)''\end{matrix}\quad D_6$&$\begin{matrix}A_4+A_1\\A_3\end{matrix}\quad D_6$&1&0&2& $1\overset{\mathfrak{e}_6}{5}1\overset{\mathfrak{su}_2}{2}$ & $(\mathfrak{su}_3)_{12} \oplus (\mathfrak{so}_{7})_{16}$ \\\hline
	
	5&$\begin{matrix}A_4\\ A_3+2A_1\end{matrix}\quad D_6$&$\begin{matrix}A_4+A_1\\(A_3+A_1)'\end{matrix}\quad D_6$&1&0&2& $1\overset{\mathfrak{e}_6}{5}1\,2$ & $(\mathfrak{su}_3)_{12} \oplus (\mathfrak{su}_3)_{24} \oplus (\mathfrak{su}_{2})_{13}$ \\\hline

	6&$\begin{matrix}A_4\\ D_4(a_1)+A_1\end{matrix}\quad D_6$&$\begin{matrix}A_4+A_1\\D_4(a_1)\end{matrix}\quad D_6$&1&0&2& $1\underset{\displaystyle 1}{\overset{\mathfrak{e}_6}{5}}1$ & $(\mathfrak{su}_3)_{12}^{\oplus 3}$ \\\hline
	
	7&$\begin{matrix}A_4\\ A_3+A_2\end{matrix}\quad D_6$&$\begin{matrix}A_4+A_1\\ D_4(a_1)+A_1\end{matrix}\quad D_6$& 2 & 1 &2& $1\overset{\mathfrak{e}_6}{4}1$ & $(\mathfrak{su}_3)_{12}^{\oplus 2} \oplus (\mathfrak{su}_{2})_{54}$ \\\hline
	
	8&$\begin{matrix}A_4\\ A_3+A_2+A_1\end{matrix}\quad D_6$&$\begin{matrix}A_4+A_1\\ A_3+A_2\end{matrix}\quad D_6$& 3 & 2 &2& $1\overset{\mathfrak{e}_6}{3}$ & $(\mathfrak{su}_3)_{12} \oplus (\mathfrak{su}_{3})_{54}$ \\\hline
	
	9&$\begin{matrix}0\\ E_7(a_4)\end{matrix}\quad D_6$&$\begin{matrix}A_1\\ A_6\end{matrix}\quad D_6$& 28 & 12 &2& $1\overset{\mathfrak{su}_2}{2}\overset{\mathfrak{g}_2}{3}$ & $(\mathfrak{e}_7)_{24}$ \\\hline
	
	10&$\begin{matrix}A_2\\ D_6(a_2)\end{matrix}\quad D_6$&$\begin{matrix}A_2+A_1\\ (A_5)'\end{matrix}\quad D_6$& 6 & 2 &2& $\overset{\mathfrak{so}_9}{4}1\overset{\mathfrak{su}_3}{2}$ & $(\mathfrak{su}_2)_{9} \oplus (\mathfrak{su}_{6})_{18}$ \\\hline
	
	11&$\begin{matrix}A_2\\ E_7(a_5)\end{matrix}\quad D_6$&$\begin{matrix}A_2+A_1\\ E_6(a_3)\end{matrix}\quad D_6$& 6 &2 &2& $\overset{\mathfrak{so}_8}{4}1\overset{\mathfrak{su}_3}{2}$ & $(\mathfrak{su}_{6})_{18}$ \\\hline
	
	12&$\begin{matrix}(3A_1)'\\ D_6(a_2)\end{matrix}\quad D_6$&$\begin{matrix}4A_1\\ (A_5)'\end{matrix}\quad D_6$&3&1&2& $\overset{\mathfrak{so}_9}{4}1\overset{\mathfrak{g}_2}{2}$ & $(\mathfrak{su}_2)_{9} \oplus (\mathfrak{sp}_{4})_{19}$ \\\hline
	
	13&$\begin{matrix}(3A_1)'\\ E_7(a_5)\end{matrix}\quad D_6$&$\begin{matrix}4A_1\\ E_6(a_3)\end{matrix}\quad D_6$&3&1&2& $\overset{\mathfrak{so}_8}{4}1\overset{\mathfrak{g}_2}{2}$ & $(\mathfrak{sp}_{4})_{19}$\\\hline
	
	14&$\begin{matrix}A_2 + 2A_1\\ A_5'\end{matrix}\quad D_6$&$\begin{matrix}2A_2\\ D_5(a_1)+A_1\end{matrix}\quad D_6$&4&3&2& $\overset{\mathfrak{so}_9}{4}1\overset{\mathfrak{su}_2}{2}$ & $\begin{aligned}(\mathfrak{so}_7)_{16} &\oplus (\mathfrak{su}_2)_{9} \\[-3pt] &\oplus (\mathfrak{su}_2)_{24} \oplus (\mathfrak{su}_2)_{48}\end{aligned}$\\\hline
	
	15&$\begin{matrix}A_2 + 3A_1\\ A_5'\end{matrix}\quad D_6$&$\begin{matrix}2A_2 + A_1\\ D_5(a_1)+A_1\end{matrix}\quad D_6$&7&4&2& $\overset{\mathfrak{so}_9}{4}1\,2$ & $(\mathfrak{so}_7)_{24} \oplus (\mathfrak{su}_2)_{9} \oplus (\mathfrak{su}_2)_{13}$\\\hline
	
	16&$\begin{matrix}0\\ (A_3+A_1)''\end{matrix}\quad E_7(a_1)$&$\begin{matrix}A_1\\ 2A_2\end{matrix}\quad E_7(a_1)$&28&12&1& $\begin{gathered} 1{\overset{\mathfrak{su}_2}{2}} \end{gathered}$ & $(\mathfrak{e}_7)_{24} \oplus (\mathfrak{so}_{7})_{16}$ \\\hline
	
	17&$\begin{matrix}A_1\\ 2A_2\end{matrix}\quad E_7(a_1)$&$\begin{matrix}2A_1\\ A_2+2A_1\end{matrix}\quad E_7(a_1)$&12&4&1& $\begin{gathered} 1{\overset{\mathfrak{su}_2}{2}} \end{gathered}$ & $(\mathfrak{e}_7)_{24} \oplus (\mathfrak{so}_{7})_{16}$ \\\hline
	
	18&$\begin{matrix}A_1\\ 2A_2+A_1\end{matrix}\quad E_7(a_1)$&$\begin{matrix}2A_1\\ A_2+3A_1\end{matrix}\quad E_7(a_1)$&13&7&1& $\begin{gathered} 1\, 2 \end{gathered}$ & $(\mathfrak{e}_8)_{24} \oplus (\mathfrak{su}_{2})_{13}$ \\\hline
	
	19&$\begin{matrix}0\\ A_3 + 2A_1\end{matrix}\quad E_7(a_1)$&$\begin{matrix}A_1\\ 2A_2+A_1\end{matrix}\quad E_7(a_1)$&28&13&1& $\begin{gathered} 1\, 2 \end{gathered}$ & $(\mathfrak{e}_8)_{24} \oplus (\mathfrak{su}_{2})_{13}$ \\\hline
	
	20&$\begin{matrix}2A_1\\ 4A_1\end{matrix}\quad E_7(a_1)$&$\begin{matrix}(3A_1)''\\ (3A_1)'\end{matrix}\quad E_7(a_1)$&1&3&1& $\begin{gathered} 1{\overset{\mathfrak{g}_2}{2}} \end{gathered}$ & $(\mathfrak{f}_4)_{24} \oplus (\mathfrak{sp}_{4})_{19}$ \\\hline
	
	21&$\begin{matrix}A_2\\ (3A_1)''\end{matrix}\quad E_7(a_1)$&$\begin{matrix}A_2+A_1\\ 2A_1\end{matrix}\quad E_7(a_1)$&6&2&1& $\begin{gathered} 1{{\overset{\mathfrak{su}_3}{2}}} \end{gathered}$ & $(\mathfrak{e}_6)_{24} \oplus (\mathfrak{su}_{6})_{18}$\\\hline
	
	22&$\begin{matrix}A_2\\ 4A_1\end{matrix}\quad E_7(a_1)$&$\begin{matrix}A_2+A_1\\ (3A_1)'\end{matrix}\quad E_7(a_1)$&9&7&1& $\begin{gathered} {{\overset{\mathfrak{su}_3}{1}}} \end{gathered}$ & $(\mathfrak{su}_{12})_{18}$ \\\hline
	
	23&$\begin{matrix}0\\ A_3+A_2\end{matrix}\quad E_7(a_1)$&$\begin{matrix}(3A_1)'\\ A_2+3A_1\end{matrix}\quad E_7(a_1)$&56&31&1& $\begin{gathered} 1 \end{gathered}$ & $(\mathfrak{e}_8)_{12}$ \\
        \bottomrule
        \end{tabular}
  \end{threeparttable}
  \caption{Isomorphic SCFTs arising from class $\mathcal{S}$ of type $\mathfrak{e}_7$ on three-punctured spheres with differing numbers of free hypermultiplets. $n_{A/B}$ is the number for free hypermultiplets in Theory A/B.}
  \label{tb:differingfree}
\end{table}

We now turn to an example of the second mechanism for generating isomorphic pairs with different numbers of free hypermultiplets. Consider class $\mathcal{S}$ of type $\mathfrak{e}_7$ and the following three-punctured sphere:
\begin{align}
\begin{aligned}
    \begin{tikzpicture}
        \draw[radius=40pt,fill=lightblue] circle;
        \draw[radius=2pt,fill=white]  (-.5,.9) circle node[right=2pt] {$E_7(a_1)$};
        \draw[radius=2pt,fill=white]  (-.5,-1) circle node[right=2pt] {$2A_1$};
        \draw[radius=2pt,fill=white]  (1,0) circle node[left=2pt] {$A_2$};
        % \node at (0,-2) {$\mathfrak{f}_{\text{manifest}}=(\mathfrak{so}_9)_{24}\oplus (\mathfrak{su}_2)_{20}\oplus (\mathfrak{su}_6)_{20}$};
    \end{tikzpicture} \,.
\end{aligned}
\end{align} 
The manifest flavor symmetry 
\begin{equation}
    \mathfrak{f}_{\text{manifest}}=(\mathfrak{so}_9)_{24}\oplus (\mathfrak{su}_2)_{20}\oplus (\mathfrak{su}_6)_{20} \,,
\end{equation}
is enhanced to 
\begin{equation}
    \mathfrak{f} = (\mathfrak{so}_{10})_{24}\oplus (\mathfrak{su}_8)_{20} \,.
\end{equation}
The manifest $(\mathfrak{su}_2)_{20}$ and $(\mathfrak{su}_6)_{20}$ have combined into a single simple $(\mathfrak{su}_8)_{20}$ factor. A minimal nilpotent Higgsing of the $(\mathfrak{su}_2)_{20}$ corresponds to partially closing the $2A_1$ puncture, $2A_1 \rightarrow (3A_1)''$; similarly, the minimal nilpotent Higgsing of the $(\mathfrak{su}_6)_{20}$ corresponds to $A_2 \rightarrow A_2 + A_1$. However, we know that the minimal nilpotent Higgsing of any subalgebra leads to the same interacting SCFT as the minimal nilpotent Higgsing of the full flavor factor, plus some number of free hypermultiplets fixed by the embedding. Indeed it is easy to check that the interacting sectors of the fixtures
\begin{align}
\begin{aligned}
    \begin{tikzpicture}
        \draw[radius=40pt,fill=lightmauve] circle;
        \draw[radius=2pt,fill=white]  (-.5,.9) circle node[right=2pt] {$E_7(a_1)$};
        \draw[radius=2pt,fill=white]  (-.5,-1) circle node[above right=2ptand -18pt] {$(3A_1)''$};
        \draw[radius=2pt,fill=white]  (1,0) circle node[left=2pt] {$A_2$};
        \draw[radius=40pt,fill=lightmauve] (4,0) circle;
        \draw[radius=2pt,fill=white]  (3.5,.9) circle node[right=2pt] {$E_7(a_1)$};
        \draw[radius=2pt,fill=white]  (3.5,-1) circle node[above right=2pt and -18pt] {$2A_1$};
        \draw[radius=2pt,fill=white]  (5,0) circle node[left=2pt] {$A_2+A_1$};
    \end{tikzpicture} \,,
\end{aligned}
\end{align}
have the correct conventional invariants, and their Schur indices (after removing 6 free hypers from the fixture on the left and 2 free hypers from the fixture on the right) agree to at least order $\tau^4$. Their interacting parts are isomorphic and this isomorphism of the interacting sectors of these two three-punctures spheres follows only from the enhanced flavor symmetry, without the need to uplift to six dimensions. However, one can read directly from \cite{Baume:2021qho} that the interacting parts of both three-punctured spheres are obtained from the torus-compactification of the 6d $(1,0)$ SCFT associated to the curve configuration
\begin{align}
\begin{gathered} 
	1{{\overset{\mathfrak{su}_3}{3}}} \,.
\end{gathered}
\end{align}

In Table \ref{tb:differingfree}, we list all the isomorphic pairs with differing numbers of free hypermultiplets obtained by the aforementioned methods for the class $\mathcal{S}$ theory of type $\mathfrak{e}_7$. We can see that the pairs with isomorphic interacting sectors in Table \ref{tb:differingfree}, together with those in Tables \ref{tb:isopairs} and \ref{tb:diffmanifest}, reproduce all of the isomorphic pairs of 6d $(1,0)$ SCFTs $\mathcal{T}_{\mathfrak{e}_7, N}(O_L, O_R)$, as summarized in Figures \ref{fig:e7N1}, \ref{fig:e7N2}, \ref{fig:e7N3}, and \ref{fig:e7N4}.

Interestingly, when considering theories where only the interacting sectors are required to coincide, we find that there are isomorphic \emph{triples} of three-punctured spheres.
As is evident from the 6d curve configurations in Table \ref{tb:differingfree}, we see that the following three-punctured spheres:
\begin{align}
\begin{aligned}
    \begin{tikzpicture}
        \draw[radius=40pt,fill=lightmauve] circle;
        \draw[radius=2pt,fill=white]  (-.5,.9) circle node[right=2pt] {$A_1$};
        \draw[radius=2pt,fill=white]  (-.5,-1) circle node[above right=2pt and -18pt] {$2A_2+A_1$};
        \draw[radius=2pt,fill=white]  (1,0) circle node[left=2pt] {$E_7(a_1)$};
        \node at (2,0) {$\sim$};
        \draw[radius=40pt,fill=lightmauve] (4,0) circle;
        \draw[radius=2pt,fill=white]  (3.5,.9) circle node[right=2pt] {$2A_1$};
        \draw[radius=2pt,fill=white]  (3.5,-1) circle node[ above right=2pt and -15pt] {$A_2+3A_1$};
        \draw[radius=2pt,fill=white]  (5,0) circle node[left=2pt] {$E_7(a_1)$};
        \node at (6,0) {$\sim$};
        \draw[radius=40pt,fill=lightmauve] (8,0) circle;
        \draw[radius=2pt,fill=white]  (7.5,.9) circle node[right=2pt] {$0$};
        \draw[radius=2pt,fill=white]  (7.5,-1) circle node[above right=2pt and -18pt] {$A_3+2A_1$};
        \draw[radius=2pt,fill=white]  (9,0) circle node[left=2pt] {$E_7(a_1)$};
    \end{tikzpicture} \,,
\end{aligned}
\end{align}
are different realizations of the rank-2 $E_8$ Minahan--Nemeschansky theory with $13$, $7$ and $28$ free hypermultiplets, respectively. Similarly,
\begin{align}
\begin{aligned}
    \begin{tikzpicture}
        \draw[radius=40pt,fill=lightmauve] circle;
        \draw[radius=2pt,fill=white]  (-.5,.9) circle node[right=2pt] {$0$};
        \draw[radius=2pt,fill=white]  (-.5,-1) circle node[above right=2pt and -18pt] {$(A_3+A_1)''$};
        \draw[radius=2pt,fill=white]  (1,0) circle node[left=2pt] {$E_7(a_1)$};
        \node at (2,0) {$\sim$};
        \draw[radius=40pt,fill=lightmauve] (4,0) circle;
        \draw[radius=2pt,fill=white]  (3.5,.9) circle node[right=2pt] {$A_1$};
        \draw[radius=2pt,fill=white]  (3.5,-1) circle node[ right=2pt] {$2A_2$};
        \draw[radius=2pt,fill=white]  (5,0) circle node[left=2pt] {$E_7(a_1)$};
        \node at (6,0) {$\sim$};
        \draw[radius=40pt,fill=lightmauve] (8,0) circle;
        \draw[radius=2pt,fill=white]  (7.5,.9) circle node[right=2pt] {$2A_1$};
        \draw[radius=2pt,fill=white]  (7.5,-1) circle node[above right=2pt and -18pt] {$A_2+2A_1$};
        \draw[radius=2pt,fill=white]  (9,0) circle node[left=2pt] {$E_7(a_1)$};
    \end{tikzpicture} \,,
\end{aligned}
\end{align}
are different realizations of the rank-3 $(E_7)_{24}\times Spin(7)_{16}$ SCFT with $28$, $12$ and $4$ free hypermultiplets, respectively.

\section{Examples where \texorpdfstring{\boldmath{$\mathfrak{g} \neq \mathfrak{e}_7$}}{g != e7}}\label{sec:other}

Class $\mathcal{S}$ theories of type $\mathfrak{e}_7$ have provided a wealth of examples of candidates for pairs of isomorphic SCFTs. We have explored the $\mathfrak{e}_7$ theories as they provide a representative sample of theories illustrating our six-dimensional methods for determining isomorphisms. However, class $\mathcal{S}$ theories of different ADE-types evince the same behavior. Using the procedures laid out in this paper, it is straightforward to determine candidate papers for any other $\mathfrak{g}$, and furthermore to verify they correspond to isomorphic 4d $\mathcal{N}=2$ SCFTs from the 6d $(1,0)$ uplift, as discussed in Section \ref{sec:6d}. In this section, we provide a small number of examples of isomorphic pairs when $\mathfrak{g}$ is a classical Lie algebra.\footnote{Class $\mathcal{S}$ theories of type $\mathfrak{g}$ can also contain punctures that are twisted by an outer-automorphism of $\mathfrak{g}$. Some such theories can alternatively be constructed from 6d $(1,0)$ compactified on a torus, now with the inclusion of a Stiefel--Whitney twist \cite{Ohmori:2018ona,Giacomelli:2020jel,Heckman:2022suy}; therefore the methods of this paper demonstrating the isomorphisms of 6d $(1,0)$ SCFTs can also prove isomorphisms between class $\mathcal{S}$ theories with twisted punctures.}

\subsection{An \texorpdfstring{$\mathfrak{so}_{12}$}{so(12) } example}\label{sec:so12}

One nice class of examples can be found in the $\mathfrak{g} = \mathfrak{so}_{12}$ theory. When the 6d $(1,0)$ conformal matter theory, Higgsed by nilpotent orbits $O_a$ and $O_b$, is compactified on the torus, the class $\mathcal{S}$ dual theory has a degeneration limit where it looks like
\begin{align}
\begin{aligned}
    \scalebox{.60}{
    \begin{tikzpicture}
        \draw[radius=40pt,fill=lightred] circle;
        \draw[radius=2pt,fill=white]  (-.5,.9) circle node[right=2pt] {$[9,3]$};
        \draw[radius=2pt,fill=white]  (-.5,-.9) circle node[right=2pt] {$[9,3]$};
        \draw[radius=2pt,fill=white]  (1,0) circle node[left=0pt] (7221) {$\scriptstyle ([7,2^2,1],SU(2))$};
        \node at (0,-2) {$\tfrac{1}{2}(2,1,1)$};
        \draw[radius=40pt,fill=lightred] (4,0) circle;
        \draw[radius=2pt,fill=white]  (4,1) circle node[below=2pt] {$[9,3]$};
        \draw[radius=2pt,fill=white]  (3,0) circle node[right=2pt] (D6) {$[7,2^2,1]$};
        \draw[radius=2pt,fill=white]  (5,0) circle node[below left=4pt and -10pt] (51111111G2){$\bigl([5,1^7],G_2\bigr)$};
        \path (1.1,0) edge node[above] {$SU(2)$}  (2.9,0);
        \node at (4,-2) {$\tfrac{1}{2}(2,7,1)$};
        \draw[radius=40pt,fill=lightblue] (8,0) circle;
        \draw[radius=2pt,fill=white]  (8,1) circle node[below=2pt] {$[9,3]$};
        \draw[radius=2pt,fill=white]  (7,0) circle node[right=2pt] (51111111) {$[5,1^7]$};
        \draw[radius=2pt,fill=white]  (9,0) circle node[below left=4pt and -10pt] (3111111111so9){$\scriptstyle ([3,1^9],SO(9))$};
        \path (5.1,0) edge node[above] {$G_2$}  (6.9,0);
        \node at (8,-2) {$[{(E_8)}_{12}\;\text{SCFT}]$};
        \draw[radius=40pt,fill=lightblue] (12,0) circle;
        \draw[radius=2pt,fill=white]  (12,1) circle node[below=2pt] {$[9,3]$};
        \draw[radius=2pt,fill=white]  (11,0) circle node[ right=2pt] (A1A1A1pp) {$[3,1^9]$};
        \draw[radius=2pt,fill=white]  (13,0) circle node[below left=4pt and -10pt] {$\scriptstyle ([1^{12}], SO(11))$};
        \path (9.1,0) edge node[above] {$SO(9)$}  (10.9,0);
        \node at (11.6,-2) {$[Spin(20)_{16}\;\text{SCFT}]$};
        \draw[radius=40pt,fill=lightblue] (16,0) circle;
        \draw[radius=2pt,fill=white]  (16,1) circle node[below=2pt] {$[9,3]$};
        \draw[radius=2pt,fill=white]  (15,0) circle node[right=2pt] (A1A1A1pp) {$[1^{12}]$};
        \draw[radius=2pt,fill=white]  (17,0) circle node[below left=2pt and -5pt] {$[1^{12}]$};
        \path (13.1,0) edge node[above] {$SO(11)$}  (14.9,0);
        \node at (16,-2) {$[Spin(12)_{20}^2\;\text{SCFT}]$};
        \node at (19,0) (dots) {$\dots$};
        \path (17.1,0) edge node[above right=0pt and -15pt] {$SO12)$}  (dots);
        \draw[radius=40pt,fill=lightblue] (22,0) circle;
        \draw[radius=2pt,fill=white]  (21,0) circle node[right=2pt] {$[1^{12}]$};
        \draw[radius=2pt,fill=white]  (22.5,.9) circle node[left=2pt] {$O_a$};
        \draw[radius=2pt,fill=white]  (22.5,-.9) circle node[left=2pt] {$O_b$};
        \path (dots) edge node[above left=0pt and -15pt] {$SO(12)$}  (20.9,0);
    \end{tikzpicture}}
\end{aligned} \,. \label{eqn:D6chainofP1}
\end{align}

If we follow the prescription of equation \eqref{eqn:higgspair}, we can take $(O_1,O_2)=([4^2,1^4], [5,3,2^2])$ and $(O'_1,O'_2)=({\color{red}[4^2,2^2]},[5,3,1^4])$ or $({\color{blue}[4^2,2^2]},[5,3,1^4])$. The puncture $[4^2,1^4]$ has an $(\mathfrak{su}_2)_{12}\oplus (\mathfrak{su}_2)_8^2$ flavor symmetry, and nilpotent Higgsing of one or the other of the $(\mathfrak{su}_2)_8$ factors leads to either $\color{red}[4^2,2^2]$ \emph{or} $\color{blue}[4^2,2^2]$. The resulting theories are related by the outer-automorphism of $\mathfrak{so}_{12}$ which exchanges (globally) {\color{red}red} and {\color{blue}blue}. This is an isomorphism of SCFTs (for any $N$). When $N=4$, the 6d curve configuration is
\begin{equation}\label{firstD6example}
\begin{gathered} \underset{[(\mathfrak{su}_2)_8\oplus(\mathfrak{su}_2)_8]}{\overset{\vphantom{\mathfrak{p}_1}\mathfrak{so}_8}{3}}\overset{\mathfrak{sp}_1}{1}
\underset{[(\mathfrak{su}_2)_{11}]}{\overset{\vphantom{\mathfrak{p}_1}\mathfrak{so}_{11}}{4}}\overset{\mathfrak{sp}_1}{1}
\underset{[(\mathfrak{su}_2)_{8}]}{\overset{\vphantom{\mathfrak{p}_1}\mathfrak{so}_7}{3}} \,.
\end{gathered}
\end{equation}
Peeling off the quiver tail establishes the isomorphism
\begin{align}\label{D6ex1}
\begin{aligned}
    \begin{matrix}
    \begin{tikzpicture}
        \draw[radius=40pt,fill=lightmauve] (0,0) circle;
        \draw[radius=2pt,fill=white]  (-1,0) circle node[right=2pt] {$[3,1^9]$};
        \draw[radius=2pt,fill=white]  (.5,.9) circle node[left=0pt] {$[4^2,1^4]$};
        \draw[radius=2pt,fill=white]  (.5,-.9) circle node[above left=0pt and -10pt] {$[5,3,2^2]$};
    \end{tikzpicture}
    \end{matrix}
    \quad\simeq\quad
    \begin{matrix}
    \begin{tikzpicture}
        \draw[radius=40pt,fill=lightmauve] (0,0) circle;
        \draw[radius=2pt,fill=white]  (-1,0) circle node[right=2pt] {$[3,1^9]$};
        \draw[radius=2pt,fill=white]  (.5,.9) circle node[left=0pt] {${\color{red}[4^2,2^2]}$};
        \draw[radius=2pt,fill=white]  (.5,-.9) circle node[above left=0pt and -10pt] {$[5,3,1^4]$};
    \end{tikzpicture}
    \end{matrix}
     \quad\simeq\quad
    \begin{matrix}
    \begin{tikzpicture}
        \draw[radius=40pt,fill=lightmauve] (0,0) circle;
        \draw[radius=2pt,fill=white]  (-1,0) circle node[right=2pt] {$[3,1^9]$};
        \draw[radius=2pt,fill=white]  (.5,.9) circle node[left=0pt] {${\color{blue}[4^2,2^2]}$};
        \draw[radius=2pt,fill=white]  (.5,-.9) circle node[above left=0pt and -10pt] {$[5,3,1^4]$};
    \end{tikzpicture} \,,
    \end{matrix}
\end{aligned}
\end{align}
which is a rank-8 interacting SCFT with flavor symmetry
\begin{equation}
\mathfrak{f}= (\mathfrak{su}_2)_8^{3}\oplus (\mathfrak{su}_2)_{11} \,,
\end{equation}
and one free hypermultiplet, transforming as $\tfrac{1}{2}(\bm{1,1,1,2})$ under the four manifest $\mathfrak{su}_2$ factors.

It should be emphasized that while the fixtures in class $\mathcal{S}$ are isomorphic if, upon attaching the quiver tail, the resulting theories arise from the compactification of isomorphic theories in 6d, the statement is \emph{\textbf{not}} an \emph{if and only if}. Consider the degeneration limit
\begin{align}
\begin{aligned}
    \scalebox{.62}{
    \begin{tikzpicture}
        \draw[radius=40pt,fill=lightred] circle;
        \draw[radius=2pt,fill=white]  (-.5,.9) circle node[right=2pt] {$[9,3]$};
        \draw[radius=2pt,fill=white]  (-.5,-.9) circle node[right=2pt] {$[9,3]$};
        \draw[radius=2pt,fill=white]  (1,0) circle node[left=0pt] (7221) {$\scriptstyle ([7,2^2,1],SU(2))$};
        \node at (0,-2) {$\tfrac{1}{2}(2,1,1)$};
        \draw[radius=40pt,fill=lightred] (4,0) circle;
        \draw[radius=2pt,fill=white]  (4,1) circle node[below=2pt] {$[9,3]$};
        \draw[radius=2pt,fill=white]  (3,0) circle node[right=2pt] (D6) {$[7,2^2,1]$};
        \draw[radius=2pt,fill=white]  (5,0) circle node[below left=4pt and -10pt] (51111111G2){$\bigl([5,1^7],G_2\bigr)$};
        \path (1.1,0) edge node[above] {$SU(2)$}  (2.9,0);
        \node at (4,-2) {$\tfrac{1}{2}(2,7,1)$};
        \draw[radius=40pt,fill=lightblue] (8,0) circle;
        \draw[radius=2pt,fill=white]  (8,1) circle node[below=2pt] {$[9,3]$};
        \draw[radius=2pt,fill=white]  (7,0) circle node[right=2pt] (51111111) {$[5,1^7]$};
        \draw[radius=2pt,fill=white]  (9,0) circle node[below left=4pt and -10pt] (3111111111so9){$\scriptstyle ([3,1^9],Spin(9))$};
        \path (5.1,0) edge node[above] {$G_2$}  (6.9,0);
        \node at (8,-2) {$[{(E_8)}_{12}\;\text{SCFT}]$};
        \draw[radius=40pt,fill=lightblue] (12.5,0) circle;
       \draw[radius=2pt,fill=white]  (12.5,1) circle node[below=2pt] {$[9,3]$};
       \draw[radius=2pt,fill=white]  (11.5,0) circle node[ right=2pt] (A1A1A1pp) {$[3,1^9]$};
       \draw[radius=2pt,fill=white]  (13.5,0) circle node[below left=4pt and -10pt] {$\scriptstyle ([1^{12}], Spin(11))$};
        \path (9.1,0) edge node[above] {$Spin(9)$}  (11.4,0);
        \node at (12,-2) {$[Spin(20)_{16}\;\text{SCFT}]$};
        \draw[radius=40pt,fill=lightmauve] (17,0) circle;
    \draw[radius=2pt,fill=white]  (17,1) circle node[below=2pt] {$[9,3]$};
    \draw[radius=2pt,fill=white]  (16,0) circle node[right=2pt] (A1A1A1pp) {$[1^{12}]$};
    \draw[radius=2pt,fill=white]  (18,0) circle node[below left=4pt and -5pt] {$[2^2,1^8]$};
        \path (13.6,0) edge node[above] {$Spin(11)$}  (15.9,0);
        \node at (17,-2.25) {$\begin{gathered}
    [Spin(20)_{16}\;\text{SCFT}]\\
        + 1(1)+1(11)
        \end{gathered}$};
    \draw[radius=40pt,fill=lightred] (21.5,0) circle;  
     \path (18.1,0) edge   node[above] {$Spin(7)$}
     (20.4,0);     
       \draw[radius=2pt,fill=white]  (20.5,0) circle node[below right=2pt and -12pt] {$\scriptstyle ([2^2,1^8],Spin(7))$};
       \draw[radius=2pt,fill=white]  (22,.9) circle node[left=2pt] {$O_a$};
       \draw[radius=2pt,fill=white]  (22,-.9) circle node[left=2pt] {$O_b$};
       \node at (21.5,-2) {$1(8)$};
    \end{tikzpicture}}
\end{aligned}. \label{eqn:D6nonisom1}
\end{align}
For $(O_a,O_b)=({\color{red}[4^2,2^2])},[9,3])$, $({\color{blue}[4^2,2^2])},[9,3])$ or $([4^2,3,1]),[9,1^3])$, these 8-punctured spheres have all the same conventional invariants. But they are not isomorphic SCFTs. Indeed, the 6d curve configuration for $(O_a,O_b)=({\color{red}[4^2,2^2])},[9,3])$ and $({\color{blue}[4^2,2^2])},[9,3])$ is 
\begin{equation}
\begin{gathered} \overset{\vphantom{\mathfrak{p}_1}\mathfrak{so}_7}{3}\,
\overset{\mathfrak{sp}_1}{1}\,
\overset{\vphantom{\mathfrak{p}_1}\mathfrak{so}_{11}}{4}\,
\overset{\mathfrak{sp}_1}{1}\,
\overset{\vphantom{\mathfrak{p}_1}\mathfrak{so}_9}{3}\,
\overset{\mathfrak{sp}_1}{1}\,
\overset{\vphantom{\mathfrak{p}_1}\mathfrak{g}_2}{3}\,
\overset{\mathfrak{sp}_1}{1} \,,
\end{gathered} 
\end{equation}
whereas, for $(O_a,O_b)=([4^2,3,1]),[9,1^3])$, the curve configuration is 
\begin{equation}
\begin{gathered} \overset{\vphantom{\mathfrak{p}_1}\mathfrak{g}_2}{3}\,
\overset{\mathfrak{sp}_1}{1}\,
\overset{\vphantom{\mathfrak{p}_1}\mathfrak{so}_{11}}{4}\,
\overset{\mathfrak{sp}_1}{1}\,
\overset{\vphantom{\mathfrak{p}_1}\mathfrak{so}_9}{3}\,
\overset{\mathfrak{sp}_1}{1}\,
\overset{\vphantom{\mathfrak{p}_1}\mathfrak{so}_7}{3}\,
\overset{\mathfrak{sp}_1}{1} \,.
\end{gathered} 
\end{equation}
Nevertheless, if we peel off the quiver tail, the fixture on the far right is 8 free hypermultiplets for all three choices.

The point is that there are three distinct embeddings of the $Spin(7)$ gauge group in the $Spin(8)$ flavor symmetry of the $[2^2,1^8]$ puncture. Two of them are exchanged under the $\mathfrak{so}(12)$ outer-automorphism that exchanges ${\color{red}[4^2,2^2]}\leftrightarrow{\color{blue}[4^2,2^2]}$, which is, of course, a symmetry of the SCFT. But the third embedding (in which the vector of $Spin(8)$ decomposes as $(\bm{7}+\bm{1})$) is distinct\footnote{In fact, the $Spin(7)$ is embedded in the centralizer of $Spin(11)\subset Spin(20)$. Even in that context the two embeddings are not conjugate to each other.}, leading to a distinct SCFT in equation \eqref{eqn:D6nonisom1} for the pair $(O_a,O_b)=([4^2,3,1]),[9,1^3])$. 

\subsection{A family of examples in \texorpdfstring{$\mathfrak{su}_{n}$}{su(n)}}

In class $\mathcal{S}$ theories of type $\mathfrak{su}_n$, it is easy to see that the condition in equation \eqref{eqn:noenhanced} is never satisfied, and thus there can only be pairs of isomorphic SCFTs if there is enhanced global symmetry, which may arise from the presence of a free sector. In this section, we present one example of an isomorphic pair for class $\mathcal{S}$ theories of type $\mathfrak{su}_n$, for each $n \geq 6$, and explain how the isomorphism can be verified from the 6d $(1,0)$ uplift. 

We consider class $\mathcal{S}$ of type $\mathfrak{su}_{n+6}$, where the nilpotent orbits describing the punctures are in one-to-one correspondence with integer partitions of $n+6$. Consider then the following pair of three-punctured spheres:
\begin{align}
\begin{aligned}
    \begin{matrix}
    \begin{tikzpicture}
        \draw[radius=40pt,fill=lightmauve] (0,0) circle;
        \draw[radius=2pt,fill=white]  (-1.1,0) circle node[right=2pt] {$[n+1,1^5]$};
        \draw[radius=2pt,fill=white]  (.7,.8) circle node[left=2pt] {$[3,1^{n+3}]$};
        \draw[radius=2pt,fill=white]  (.8,-.8) circle node[left=2pt] {$[3^2,1^n]$};
    \end{tikzpicture}
    \end{matrix}
    \quad\simeq\quad
    \begin{matrix}
    \begin{tikzpicture}
        \draw[radius=40pt,fill=lightmauve] (0,0) circle;
        \draw[radius=2pt,fill=white]  (-1.1,0) circle node[right=2pt] {$[n+1,1^5]$};
        \draw[radius=2pt,fill=white]  (.8,.7) circle node[left=2pt] {$[2^2,1^{n+2}]$};
        \draw[radius=2pt,fill=white]  (.8,-.7) circle node[left=2pt] {$[4,2,1^n]$};
    \end{tikzpicture} \,.
    \end{matrix} 
\end{aligned}\label{eqn:supair}
\end{align}
The manifest flavor symmetry, which can be read off directly from the integer partitions, for the theory associated to the the fixture on the left in equation \eqref{eqn:supair} is
\begin{align}\label{eqn:suAmani}
\begin{aligned}
    \mathfrak{f}^{\text{manifest}}= (\mathfrak{su}_{n+3})_{2n+8}\oplus
    (\mathfrak{su}_2)_{2n+12}\oplus
    (\mathfrak{su}_n)_{2n+4}\oplus
    (\mathfrak{su}_5)_{12}\oplus
    \mathfrak{u}_1^{\oplus 3} \,,
\end{aligned}
\end{align}
whereas the fixture on the right has manifest flavor symmetry
\begin{align}\label{eqn:suBmani}
\begin{aligned}
    \mathfrak{f}^{\text{manifest}}= (\mathfrak{su}_{n+2})_{2n+8}\oplus
    (\mathfrak{su}_2)_{2n+12}\oplus
    (\mathfrak{su}_n)_{2n+4}\oplus
    (\mathfrak{su}_5)_{12}\oplus
    \mathfrak{u}_1^{\oplus 4} \,.
\end{aligned}
\end{align}
For all values of $n$, these are mixed fixtures. When $n \geq 1$, there are two free hypermultiplets which transform in the $(\bm{1,2,1,1})$ representation of the manifest flavor symmetries appearing in both equations \eqref{eqn:suAmani} and \eqref{eqn:suBmani}; thus for both theories, in the interacting sector the $(\mathfrak{su}_2)_{2n+12}$ factor is replaced by an $(\mathfrak{su}_2)_{2n+10}$ factor. For $n \geq 2$, this is the only enhancement for the theory on the left, however the in the theory on the right the manifest $(\mathfrak{su}_{n+2})_{2n+8}$ combines with one of the $\mathfrak{u}_1$ factors to produce a $(\mathfrak{su}_{n+3})_{2n+8}$ factor. Thus, we can see that the enhanced flavor symmetries agree between the two interacting sectors:
\begin{align}
\begin{aligned}
    \mathfrak{f} = (\mathfrak{su}_{n+3})_{2n+8}\oplus
    (\mathfrak{su}_2)_{2n+10}\oplus
    (\mathfrak{su}_n)_{2n+4}\oplus
    (\mathfrak{su}_5)_{12}\oplus
    \mathfrak{u}_1^{\oplus 3} \,.
\end{aligned}
\end{align}
We can also see, for example when $n=3$, that the Schur indices of the interacting sectors of the fixtures match (up to the order we could compute them):
\begin{align}
\begin{aligned}
    I_{\text{Schur}}(\tau)=\ & 1 + 73 \tau^2 + 34 \tau^3 + 2823 \tau^4 + 2626 \tau^5 \\
    & + 77298 \tau^6 + 107048 \tau^7 + 1689006 \tau^8 + 3064288 \tau^9 +O(\tau^{10}) \,.
\end{aligned}
\end{align} 
For $n=1$, the flavor symmetry is further enhanced on both sides to
\begin{align}
    (\mathfrak{su}_4)_{10}\oplus(\mathfrak{su}_7)_{12}\oplus\mathfrak{u}_1^{\oplus 2} \,.
\end{align}
Finally, in the extremal case where $n=0$, there are twelve free hypermultiplets, and the interacting sector of both theories is the $(E_7)_8$ Minahan--Nemeschansky theory. The free hypers transform in the $(\bm{2,6})$ of the manifest $(\mathfrak{su}_2)_{12}\oplus (\mathfrak{su}_6)_{12}$. We emphasize once again that, for all values of $n \geq 0$, the conventional invariants and the Schur index (insofar as we can compute it) are consistent with the two fixtures in equation \eqref{eqn:supair} being isomorphic 4d SCFTs. 

The theories can be proven to be isomorphic from our 6d considerations. If we replace the $[n+1, 1^5]$ punctures that appear in both the fixtures in equation \eqref{eqn:supair} with $N$ copies of the simple puncture, each corresponding to the partition $[n+5, 1]$, then each of the resulting $N + 2$ punctured spheres arise as torus-compactifications of 6d $(1,0)$ SCFTs. In particular, the uplift of the fixture on the left in equation \eqref{eqn:supair} is rank $N$ $(\mathfrak{su}_{n+6}, \mathfrak{su}_{n+6})$ conformal matter, where one of the $\mathfrak{su}_{n+6}$ flavor symmetries is Higgsed by the nilpotent orbit associated to the partition $[3, 1^{n+3}]$ and the other is Higgsed by $[3^2, 1^n]$. It is well-known how to map from a pair of partitions to a tensor branch description, and in this case that description is
\begin{equation}\label{eqn:suA6}
    \underset{[(\mathfrak{su}_{n+3})_{2n+8}]}{\overset{\mathfrak{su}_{n+4}}{2}}\,
        \overset{\mathfrak{su}_{n+5}}{2}\,
        \underset{[1]}{\overset{\mathfrak{su}_{n+6}}{2}}\,
        \overset{\mathfrak{su}_{n+6}}{2}\,
        \dots
        \overset{\mathfrak{su}_{n+6}}{2}\,
        \overset{\mathfrak{su}_{n+6}}{2}\,
        \underset{[(\mathfrak{su}_{2})_{2n+12}]}{\overset{\mathfrak{su}_{n+6}}{2}}\,
        \overset{\mathfrak{su}_{n+4}}{2}\,
        \underset{[(\mathfrak{su}_{n})_{2n+4}]}{\overset{\mathfrak{su}_{n+2}}{2}} \,,
\end{equation}
where we have written the non-Abelian flavor factors directly in the quiver. Similarly, the 6d $(1,0)$ uplift of the fixture on the right in equation \eqref{eqn:supair} has the tensor branch description:
\begin{equation}\label{eqn:suB6}
    \underset{[(\mathfrak{su}_{n+2})_{2n+8}]}{\overset{\mathfrak{su}_{n+4}}{2}}\,
        \underset{[(\mathfrak{su}_{2})_{2n+12}]}{\overset{\mathfrak{su}_{n+6}}{2}}\,
        \overset{\mathfrak{su}_{n+6}}{2}\,
        \overset{\mathfrak{su}_{n+6}}{2}\,
        \dots
        \overset{\mathfrak{su}_{n+6}}{2}\,
        \underset{[1]}{\overset{\mathfrak{su}_{n+6}}{2}}\,
        \overset{\mathfrak{su}_{n+5}}{2}\,
        \underset{[1]}{\overset{\mathfrak{su}_{n+4}}{2}}\,
        \underset{[(\mathfrak{su}_{n})_{2n+4}]}{\overset{\mathfrak{su}_{n+2}}{2}} \,.
\end{equation}
In both cases, the total number of $(-2)$-curves is $N-1$. Each of these theories, which are clearly not isomorphic, arise via Higgs branch renormalization group flow from a ``parent'' theory, with tensor branch description
\begin{equation}
    \underset{[(\mathfrak{su}_{n+2})_{2n+8}]}{\overset{\mathfrak{su}_{n+4}}{2}}\,
    \underset{[(\mathfrak{su}_{2})_{2n+12}]}{\overset{\mathfrak{su}_{n+6}}{2}}\,
    \overset{\mathfrak{su}_{n+6}}{2}\,
    \overset{\mathfrak{su}_{n+6}}{2}\,
    \dots
    \overset{\mathfrak{su}_{n+6}}{2}\,
    \overset{\mathfrak{su}_{n+6}}{2}\,
    \underset{[(\mathfrak{su}_{2})_{2n+12}]}{\overset{\mathfrak{su}_{n+6}}{2}}\,
    \overset{\mathfrak{su}_{n+4}}{2}\,
    \underset{[(\mathfrak{su}_{n})_{2n+4}]}{\overset{\mathfrak{su}_{n+2}}{2}} \,.
\end{equation}
Nilpotent Higgsing of one or the other of the $(\mathfrak{su}_2)_{2n+12}$ factors leads to SCFTs with tensor branches given in equations \eqref{eqn:suA6} and \eqref{eqn:suB6}. When $N = 5$, the parent theory becomes
\begin{equation}
    \underset{[(\mathfrak{su}_{n+2})_{2n+8}]}{\overset{\mathfrak{su}_{n+4}}{2}}\,
    \overset{[(\mathfrak{su}_{2})_{2n+12}]}{\underset{[(\mathfrak{su}_{2})_{2n+12}]}{\overset{\mathfrak{su}_{n+6}}{2}}}\,
    \overset{\mathfrak{su}_{n+4}}{2}\,
    \underset{[(\mathfrak{su}_{n})_{2n+4}]}{\overset{\mathfrak{su}_{n+2}}{2}} \,,
\end{equation}
which has a $\mathbb{Z}_2$ outer-automorphism that exchanges the two $(\mathfrak{su}_2)_{2n+12}$ flavor symmetry factors. In fact, the flavor symmetry enhances and we find
\begin{equation}\label{eqn:samgyetang}
    (\mathfrak{su}_2)_{2n+12} \oplus (\mathfrak{su}_2)_{2n+12} \rightarrow (\mathfrak{su}_4)_{2n+12} \,.
\end{equation}
Performing a Higgs branch deformation by giving a vacuum expectation value to the highest root moment map of the $\mathfrak{su}_4$ or either of the $\mathfrak{su}_2$ subalgebras leads to the same interacting SCFT, where the latter two options contain an additional two free hypermultiplets. The two nilpotent Higgsings, which take either $[2^2, 1^{n+2}] \rightarrow [3, 1^{n+3}]$ or $[3^2, 1^n] \rightarrow [4, 2, 1^n]$, thus both lead to the same interacting 6d $(1,0)$ SCFT; its tensor branch description is
\begin{equation}\label{Acurveconfig}
    \underset{[(\mathfrak{su}_{n+3})_{2n+8}]}{\overset{\mathfrak{su}_{n+4}}{2}}\,
    \underset{[(\mathfrak{su}_{2})_{2n+12}]}{\overset{\mathfrak{su}_{n+5}}{2}}\,
    \underset{[1]}{\overset{\mathfrak{su}_{n+4}}{2}}\,
    \underset{[(\mathfrak{su}_{n})_{2n+4}]}{\overset{\mathfrak{su}_{n+2}}{2}} \,.
\end{equation}
In this way, we have verified that the two class $\mathcal{S}$ descriptions in equation \eqref{eqn:supair} are isomorphic as 4d $\mathcal{N}=2$ quantum field theories when the $[n+1, 1^5]$ punctures are replaced with five copies of the simple puncture, $[n+5,1]$. To prove the isomorphism for the three-punctured spheres, we go to a different degeneration limit, where the simple punctures form a chain of 3-punctured spheres, reproducing the standard quiver tail of \cite{Gaiotto:2009we} --- with gauge group $SU(2)\times SU(3)\times SU(4)\times SU(5)$ and bifundamental hypermultiplets (\emph{i.e.}, the representation $(\bm{2,1,1,1})\oplus(\bm{2,3,1,1})\oplus (\bm{1,3,4,1})\oplus (\bm{1,1,4,5})$). Sending the $SU(5)$ gauge coupling to zero establishes the isomorphism in equation \eqref{eqn:supair}.

Having established the isomorphism of the two theories where the third puncture is given by the partition $[n+1,1^5]$, we can then do a chain of nilpotent Higgsings of that third puncture to establish further isomorphic pairs
\begin{equation}
	{\color{midgreen}[n+1,1^5]}\longrightarrow {\color{midgreen}[n+1,2,1^3]}\longrightarrow {\color{midgreen}[n+1,2^2,1]}\longrightarrow {\color{midgreen}[n+1,3,1^2]}\longrightarrow {\color{midgreen}[n+1,3,2]} \,.
\end{equation}
Note that for $n=2$, the nilpotent Higgsings 
\begin{equation}
    {\color{midgreen}[3^2,1^2]}\longrightarrow{\color{red}[4,2,1^2]}\longrightarrow{\color{red}[4,2^2]}\longrightarrow{\color{red}[4,3,1]} \,,
\end{equation}
lead to bad theories, and so do not generate any additional isomorphic pairs.

\section{Oddballs}\label{sec:oddballs}

The mechanism described in equation \eqref{eqn:higgspair} for generating candidate pairs of isomorphic 4d SCFTs led to a 6d \emph{proof} when the third puncture in the fixture was chosen from the ``quiver tail'' formed by fusing together $N$ simple punctures (in the type $\mathfrak{e}_7$ theory, this was the sequence of punctures $\{(3A)'', (A_5)'', D_6, E_7(a_1)\}$ -- see equation \eqref{eqn:chainofP1}). We could then find additional isomorphic SCFTs by Higgsing down from this puncture.

This does not preclude the possibility of finding isomorphic pairs of SCFTs where the third puncture is not part of the quiver tail (or a nilpotent Higgsing thereof). For instance, consider the pair of interacting fixtures
\begin{align}\label{oddball1}
\begin{aligned}
    \begin{tikzpicture}
        \draw[radius=40pt,fill=lightblue] circle;
        \draw[radius=2pt,fill=white]  (-.5,.9) circle node[right=2pt] {$A_6$};
        \draw[radius=2pt,fill=white]  (-.5,-1) circle node[above right=2ptand -18pt] {$(A_3+A_1)''$};
        \draw[radius=2pt,fill=white]  (1,0) circle node[left=2pt] {${\color{red}{O}} $};
        \draw[radius=40pt,fill=lightblue] (4,0) circle;
        \draw[radius=2pt,fill=white]  (3.5,.9) circle node[right=2pt] {$E_7(a_4)$};
        \draw[radius=2pt,fill=white]  (3.5,-1) circle node[right=2pt] {$2A_2$};
        \draw[radius=2pt,fill=white]  (5,0) circle node[left=2pt] {${\color{red}{O}} $};
    \end{tikzpicture}  \,.
\end{aligned}
\end{align}
These fixtures appear to correspond to isomorphic SCFTs when ${\color{red}{O}}$ is chosen from the four punctures related by the following Hasse diagram
\begin{align}\label{D5higgsing}
\begin{aligned}
    \begin{tikzpicture}
        \node (D5) at (0,0) {${\color{midgreen}D_5}$};
        \node[above right=.5cm and .8cm of D5] (D6a1) {${\color{midgreen}D_6(a_1)}$};
        \node[below=1.75cm of D6a1] (D5A1) {${\color{midgreen}D_5+A_1}$};
        \node[right=3cm of D5] (E7a4) {${\color{midgreen}E_7(a_4)}$};
        \path[->] (D5) edge node[above left=-.125cm and -.125cm] {$(\mathfrak{su}_2)_{12}$} (D6a1)
        (D5) edge node[below left=-.125cm and -.125cm] {$(\mathfrak{su}_2)_{8}$} (D5A1)
        (D6a1) edge node[above right=-.125cm and -.125cm] {$(\mathfrak{su}_2)_{8}$} (E7a4)
        (D5A1) edge node[below right=-.125cm and -.125cm] {$(\mathfrak{su}_2)_{12}$} (E7a4);
    \end{tikzpicture} \,.
\end{aligned}
\end{align}
The Schur indices of all four pairs agree to at least $O(\tau^{10})$. E.g., for ${\color{red}O}=D_5$, the Schur indices of both SCFTs are
\begin{equation}
I_\text{Schur}=1+37\tau^2+853\tau^4 +15305\tau^6+233552\tau^8+
3168458\tau^{10}+O(\tau^{11}) \,.
\end{equation}
However, since none of the punctures in equation \eqref{D5higgsing} belong to the $E_7$ quiver tail, these SCFTs have no avatars as 6d $(1,0)$ SCFTs; thus, we cannot provide a proof that these 4d $\mathcal{N}=2$ SCFTs are isomorphic.

The theories in equation \eqref{oddball1} appear to be isomorphic on-the-nose. We can also find pairs of theories whose interacting parts appear to be isomorphic, but differ in the number of free hypermultiplets. For example,
%%%%%
%% This is oddball3
%%%%%
\begin{align}
\begin{aligned}
    \begin{tikzpicture}
        \draw[radius=40pt,fill=lightmauve] circle;
        \draw[radius=2pt,fill=white]  (-.5,.9) circle node[right=2pt] {$(A_5)''$};
        \draw[radius=2pt,fill=white]  (-.5,-1) circle node[above right=2ptand -18pt] {$D_6(a_1)$};
        \draw[radius=2pt,fill=white]  (1,0) circle node[left=2pt] {$A_4+A_2 $};
        \draw[radius=40pt,fill=lightblue] (4,0) circle;
        \draw[radius=2pt,fill=white]  (3.5,.9) circle node[below right=2ptand -15pt] {$A_5+A_1$};
        \draw[radius=2pt,fill=white]  (3.5,-1) circle node[above right=2pt and -18pt] {$D_5$};
        \draw[radius=2pt,fill=white]  (5,0) circle node[left=2pt] {$A_4+A_2 $};
        \node at (2,0) {$\simeq$};
        \node at (7,0){$+\; 2\;\text {free hypers}$};
    \end{tikzpicture} \,.
\end{aligned}
\end{align}
After removing two free hypermultiplets from the fixture on the left, the Schur indices are
\begin{align}
I_{\text{Schur}}= 1+23\tau^2+10\tau^3+344\tau^4+308\tau^5+4170\tau^6+5720\tau^7+O(\tau^{8}) \,.
\label{eqn:oddball3}
\end{align}
The interacting SCFT has flavor symmetry
\begin{equation}
\mathfrak{f}=(\mathfrak{su}_2)_{72}\oplus (\mathfrak{su}_2)_{26}\oplus (\mathfrak{su}_2)_8\oplus (\mathfrak{g}_2)_{12} \,.
\end{equation}
On the left, the $\mathfrak{su}(2)_8$ is associated to the $D_6(a_1)$ puncture; on the right it is associated to the $D_5$ puncture. If we do a nilpotent Higgsing of the $(\mathfrak{su}_2)_8$ on both sides, we arrive at
%%%
%% This is oddball2
%%%
\begin{align}
\begin{aligned}
    \begin{tikzpicture}
        \draw[radius=40pt,fill=lightmauve] circle;
        \draw[radius=2pt,fill=white]  (-.5,.9) circle node[right=2pt] {$(A_5)''$};
        \draw[radius=2pt,fill=white]  (-.5,-1) circle node[above right=2ptand -18pt] {$E_7(a_4)$};
        \draw[radius=2pt,fill=white]  (1,0) circle node[left=2pt] {$A_4+A_2 $};
        \draw[radius=40pt,fill=lightblue] (4,0) circle;
        \draw[radius=2pt,fill=white]  (3.5,.9) circle node[below right=2ptand -15pt] {$A_5+A_1$};
        \draw[radius=2pt,fill=white]  (3.5,-1) circle node[above right=2pt and -18pt] {$D_5+A_1$};
        \draw[radius=2pt,fill=white]  (5,0) circle node[left=2pt] {$A_4+A_2 $};
        \node at (2,0) {$\simeq$};
        \node at (7,0){$+\; 2\;\text {free hypers}$};
    \end{tikzpicture} \,.
\end{aligned}
\end{align}
After removing two free hypermultiplets from the fixture on the left, the Schur indices are
\begin{align}
I_{\text{Schur}}=1 + 20 \tau^2 + 14 \tau^3 + 272 \tau^4 + 380 \tau^5 + 3186 \tau^6 + 6338\tau^7 +O(\tau^{8}) \,,
\label{eqn:oddball2}
\end{align}
which, again, indicates that they correspond to isomorphic SCFTs.

As another example, we consider the following pair of fixtures
\begin{align}\label{oddball4}
\begin{aligned}
    \begin{tikzpicture}
        \draw[radius=40pt,fill=lightblue] circle;
        \draw[radius=2pt,fill=white]  (-.5,.9) circle node[right=2pt] {$2A_2$};
        \draw[radius=2pt,fill=white]  (-.5,-1) circle node[above right=2ptand -18pt] {$(A_3+A_1)'$};
        \draw[radius=2pt,fill=white]  (1,0) circle node[left=2pt] {$E_6$};
        \draw[radius=40pt,fill=lightmauve] (6.75,0) circle;
        \draw[radius=2pt,fill=white]  (6.25,.9) circle node[below right=2ptand -15pt] {$2A_2+A_1$};
        \draw[radius=2pt,fill=white]  (6.25,-1) circle node[above right=2pt and -18pt] {$A_3$};
        \draw[radius=2pt,fill=white]  (7.75,0) circle node[left=2pt] {$E_6$};
        \node at (4.75,0) {$\simeq$};
        \node at (2.9,0){$+\;1 \;\text {free hyper}$};
    \end{tikzpicture} \,.
\end{aligned}
\end{align}
After removing one free hypermultiplet from the fixture on the right, the Schur indices of both interacting SCFTs are
\begin{align}
    I_{\text{Schur}}=1+39\tau^2+42\tau^3+970\tau^4+2068\tau^5+20059\tau^6+O(\tau^7) \,.
    \label{eqn:oddball4}
\end{align}
The flavor symmetry of the interacting theory is
\begin{equation}
\mathfrak{f}=
 (\mathfrak{so}_7)_{16}\oplus(\mathfrak{su}_2)_{24}^2\oplus(\mathfrak{su}_2)_{12}^3 \oplus(\mathfrak{su}_2)_{13} \,.
\end{equation}
Of the three $(\mathfrak{su}_2)_{12}$ flavor algebra factors, two are manifest (associated to the $E_6$ puncture and to the $(A_3+A_1)'$ on the left or the $A_3$ on the right). The other arises as an enhancement of the $(\mathfrak{su}_2)_{36}\subset (\mathfrak{su}_2)_{24}\oplus(\mathfrak{su}_2)_{12}$ symmetry (associated to the $2A_2$ puncture on the left or the $2A_2+A_1$ puncture on the right). 

One of the $(\mathfrak{su}_2)_{12}$ factors (the one associated to $(A_3+A_1)'$ on the right or $A_3$ on the left) is the same on both sides of the isomorphism. But the role of the other two (the manifest one associated to the $E_6$ puncture and the enhanced one) is swapped between the two theories. This is exactly the same phenomenon we encountered in equation \eqref{familyHasse}, where the roles of the two $(\mathfrak{su}_2)_{28}$ (the manifest one and the enhanced one) associated to the $A_2+2A_1$ puncture were swapped between the two theories. There, when we Higgsed $A_2+2A_1\xrightarrow{(\mathfrak{su}_2)_{28}} 2A_2$, we obtained \emph{different} theories. Here, too, if we Higgs the $E_6\xrightarrow{(\mathfrak{su}_2)_{12}}E_7(a_2)$, we obtain non-isomorphic theories
\begin{align}
\begin{aligned}
    \begin{tikzpicture}
        \draw[radius=40pt,fill=lightblue] circle;
        \draw[radius=2pt,fill=white]  (-.5,.9) circle node[right=2pt] {$2A_2$};
        \draw[radius=2pt,fill=white]  (-.5,-1) circle node[above right=2ptand -18pt] {$(A_3+A_1)'$};
        \draw[radius=2pt,fill=white]  (1,0) circle node[left=2pt] {$E_7(a_2)$};
        \draw[radius=40pt,fill=lightmauve] (6.75,0) circle;
        \draw[radius=2pt,fill=white]  (6.25,.9) circle node[below right=2ptand -15pt] {$2A_2+A_1$};
        \draw[radius=2pt,fill=white]  (6.25,-1) circle node[above right=2pt and -18pt] {$A_3$};
        \draw[radius=2pt,fill=white]  (7.75,0) circle node[left=2pt] {$E_7(a_2)$};
        \node at (4.75,0) {$\not\simeq$};
        \node at (2.9,0){$+\; 1\;\text {free hyper}$};
    \end{tikzpicture} \,.
\end{aligned}
\end{align}
On the left, the $(\mathfrak{su}_2)_{24}^2$ is enhanced to $(\mathfrak{sp}_2)_{24}$, whereas on the right, it is unenhanced.

On the other hand, if we Higgs the
$(\mathfrak{su}_2)_{12}$ flavor symmetry of the fixtures in equation \eqref{oddball4} associated with the $(A_3+A_1)'$ puncture on the left and with the $A_3$ puncture on the right, we obtain theories with isomorphic interacting sectors:
\begin{align}
\begin{aligned}
    \begin{tikzpicture}
        \draw[radius=40pt,fill=lightblue] circle;
        \draw[radius=2pt,fill=white]  (-.5,.9) circle node[right=2pt] {$2A_2$};
        \draw[radius=2pt,fill=white]  (-.5,-1) circle node[above right=2ptand -18pt] {$A_3+2A_1$};
        \draw[radius=2pt,fill=white]  (1,0) circle node[left=2pt] {$E_6$};
        \draw[radius=40pt,fill=lightmauve] (6.75,0) circle;
        \draw[radius=2pt,fill=white]  (6.25,.9) circle node[below right=2ptand -15pt] {$2A_2+A_1$};
        \draw[radius=2pt,fill=white]  (6.25,-1) circle node[above right=2pt and -18pt] {$(A_3+A_1)''$};
        \draw[radius=2pt,fill=white]  (7.75,0) circle node[left=2pt] {$E_6$};
        \node at (4.75,0) {$\simeq$};
        \node at (2.9,0){$+\; 1\;\text {free hyper}$};
    \end{tikzpicture} \,.
\end{aligned}
\end{align}
The interacting sector has Schur index
\begin{align}
I_{\text{Schur}}=1+40\tau^2+58\tau^3+1048\tau^4+2848\tau^5+23541\tau^6+O(\tau^7) \,.
\label{eqn:oddball6}
\end{align}
This time, Higgsing the $E_6\xrightarrow{(\mathfrak{su}_2)_{12}}E_7(a_2)$ yields another pair of isomorphic theories:
\begin{align}
\begin{aligned}
    \begin{tikzpicture}
        \draw[radius=40pt,fill=lightmauve] circle;
        \draw[radius=2pt,fill=white]  (-.5,.9) circle node[right=2pt] {$2A_2$};
        \draw[radius=2pt,fill=white]  (-.5,-1) circle node[above right=2ptand -18pt] {$A_3+2A_1$};
        \draw[radius=2pt,fill=white]  (1,0) circle node[left=2pt] {$E_7(a_2)$};
        \draw[radius=40pt,fill=lightmauve] (6.75,0) circle;
        \draw[radius=2pt,fill=white]  (6.25,.9) circle node[below right=2ptand -15pt] {$2A_2+A_1$};
        \draw[radius=2pt,fill=white]  (6.25,-1) circle node[above right=2pt and -18pt] {$(A_3+A_1)''$};
        \draw[radius=2pt,fill=white]  (7.75,0) circle node[left=2pt] {$E_7(a_2)$};
        \node at (4.75,0) {$\simeq$};
        \node at (2.9,0){$+\; 1\;\text {free hyper}$};
    \end{tikzpicture} \,.
\end{aligned}
\end{align}
We need to remove one free hypermultiplet from the fixture on the left, and two free hypermultiplets from the fixture on the right to obtain isomorphic interacting SCFTs, which have Schur index 
\begin{align}
I_{\text{Schur}}=1+37 \tau ^2+78 \tau ^3+985 \tau ^4+3500 \tau ^5+O\left(\tau ^6\right)
\label{eqn:oddball7}
\end{align}

As we have highlighted, the ``oddball'' theories discussed in this section are pairs of class $\mathcal{S}$ theories which appear to be isomorphic, based on their conventional invariants and their Schur indices (to the extent that we were able to compute them). However, as they are unrelated to torus-compactifications of 6d $(1,0)$ SCFTs, we are unable to use the techniques from 6d to \emph{prove} that they are indeed isomorphic. Nevertheless, the insights from 6d point towards a possible direction for a direct 4d proof. The key insight was that the pair of isomorphic 4d SCFTs have a parent 4d SCFT in common. Turning on a VEV for certain operators in the parent theory triggers a Higgs branch renormalization group flow to one or the other of the ``child'' theories. Moreover, the parent SCFT has a $\mathbb{Z}_2$ symmetry which exchanges the two operators in question, and hence the RG flows that they trigger. This symmetry is manifest in the 6d $(1,0)$ uplift of the parent 4d SCFT. However, it might be possible to show that the symmetry is present directly in the 4d SCFT. For instance, the parent of the pair in equation \eqref{oddball1} is
\begin{align}\label{oddball1parent}
\begin{aligned}
    \begin{tikzpicture}
        \draw[radius=40pt,fill=lightblue] circle;
        \draw[radius=2pt,fill=white]  (-.5,.9) circle node[right=2pt] {$A_6$};
        \draw[radius=2pt,fill=white]  (-.5,-1) circle node[above right=2ptand -18pt] {$(A_3+A_1)''$};
        \draw[radius=2pt,fill=white]  (1,0) circle node[left=2pt] {$D_5$};
        \end{tikzpicture} \,,
\end{aligned}
\end{align}
which has flavor symmetry
\begin{equation}
\mathfrak{f}= (\mathfrak{g}_2)_{16}\oplus(\mathfrak{su}_2)_{36}^2\oplus(\mathfrak{su}_2)_8 \oplus(\mathfrak{su}_2)_{12} \,.
\end{equation}
This theory has a $\mathbb{Z}_2$ outer automorphism which exchanges the two $(\mathfrak{su}_2)_{36}$ factors. The two RG flows which lead to the pair of SCFTs in equation \eqref{oddball1} are triggered by turning on a VEV for the highest root moment map of one or the other of the $(\mathfrak{su}_2)_{36}$s. If we could show that this $\mathbb{Z}_2$ extends to a $\mathbb{Z}_2$ symmetry of the full SCFT in equation \eqref{oddball1parent}, we would establish the isomorphism in equation \eqref{oddball1}.

\section{Discussion}\label{sec:discussion}

The pairs of isomorphic class $\mathcal{S}$ theories we have found share the feature that they arise as (different) Higgsings of a ``parent'' SCFT. Upon uplifting to 6d, we found that the parent 6d (1,0) SCFT has a $\mathbb{Z}_2$ automorphism which exchanges the two Higgsings. It is striking that --- for the 6d $(1,0)$ SCFTs of $(\mathfrak{e},\mathfrak{e})$ conformal matter --- this automorphism has a \emph{geometrical} realization as an automorphism of the curve configuration on $B$. By contrast, for $(\mathfrak{su}_n,\mathfrak{su}_n)$ and $(\mathfrak{so}_{2n},\mathfrak{so}_{2n})$ conformal matter, the automorphism had a more subtle origin. This section is focused on explaining this behavior.

The curve configuration that gives rise to the intersecting part of the 6d $(1,0)$ SCFT is composed of non-Higgsable clusters (NHCs), which are given by
\begin{equation}
    \begin{gathered}
      \overset{\mathfrak{su}_3}{3} \,, \quad \overset{\mathfrak{so}_8}{4} \,, \quad \overset{\mathfrak{f}_4}{5} \,,\quad \overset{\mathfrak{e}_6}{6} \,, \quad \overset{\mathfrak{e}_7}{7} \,, \quad \overset{\mathfrak{e}_7}{8} \,,\quad \overset{\mathfrak{e}_{8}}{12} \,,\quad \overset{\mathfrak{su}_{2}}{2}\overset{\mathfrak{g}_{2}}{3} \,,\quad 
      2\overset{\mathfrak{su}_{2}}{2}\overset{\mathfrak{g}_{2}}{3} \,,\quad 
      \overset{\mathfrak{su}_{2}}{2}\overset{\mathfrak{so}_{7}}{3}\overset{\mathfrak{su}_{2}}{2} \,,\\
      \underbrace{2\cdots 2}_{N-1} \,,\quad 
      \underbrace{2\cdots 2}_{N-3}\overset{\displaystyle 2}{2}2 \,,\quad
      22\overset{\displaystyle 2}{2}22 \,,\quad 
      222\overset{\displaystyle 2}{2}22 \,,\quad 
      2222\overset{\displaystyle 2}{2}22 \,,
    \end{gathered}\label{eqn:nhc}
\end{equation}
where we have used the negative of the self-intersection number of the curves and the algebras $\mathfrak{g}$ associated to the singular fibers. Curve configurations are then constructed by connecting these non-Higgsable clusters via $(-1)$-curves, while requiring that the resulting curve configuration has a negative-definite intersection matrix. This restrict the number of $(-1)$-curves that can be attached to a $(-n)$-curve to be $\leq (n-1)$. 

In all of our examples drawn from the $(\mathfrak{e}_7, \mathfrak{e}_7)$ conformal matter theories, the curve configuration of the parent theory had a central $(-n)$-curve with 2 (or more) ``dangling'' $(-1)$-curves --- exchanged  by the $\mathbb{Z}_2$ automorphism --- in addition to the $(-1)$-curves which attach it to the rest of the diagram. The nilpotent Higgsings of the flavor symmetries associated to the dangling $(-1)$-curves are exchanged by the $\mathbb{Z}_2$ automorphism. For this to occur, we must have $n\geq5$. But, for
\begin{align}
    \mathfrak{g} = \mathfrak{su}_n, \mathfrak{so}_{2n}, \mathfrak{e}_6, \mathfrak{e}_7, \mathfrak{e}_8
\end{align} 
the maximally negative self-intersection curve can have self-intersection 
\begin{align}
    (-2), (-4), (-6), (-8), (-12) ,
\end{align}
respectively. Thus these \emph{geometrical}  $\mathbb{Z}_2$ automorphisms of the curve configuration only occur for the exceptional algebras.

For the classical algebras the isomorphisms are generated not by the automorphisms of the curve configuration, but rather as automorphisms of the flavor symmetry algebras decorating the central node. That is, they are implemented as automorphisms of the elliptic fiber over that exceptional curve on $B$. For example, in equation \eqref{eqn:samgyetang}, the central node of the parent theory had an $(\mathfrak{su}_2)_{2n+12}\oplus(\mathfrak{su}_2)_{2n+12}$ flavor symmetry which was enhanced to $(\mathfrak{su}_4)_{2n+12}$. The $\mathbb{Z}_2$ automorphism (which exchanges the two $(\mathfrak{su}_2)_{2n+12}$ factors) is an automorphism, not of the curve configuration on $B$, but of the elliptic fiber over the central $(-2)$-curve.

Similarly for the example in equation \eqref{firstD6example}, the parent theory has an 
\begin{align}
    \bigl((\mathfrak{su_2})_8\oplus(\mathfrak{su_2})_8\bigr)\oplus
(\mathfrak{su_2})_{11}\oplus\bigl((\mathfrak{su_2})_8\oplus(\mathfrak{su_2})_8\bigr) . 
\end{align}
There is a $\mathbb{Z}_2\times \mathbb{Z}_2$ automorphism of the flavor symmetry algebra. The first $\mathbb{Z}_2$ simultaneously swaps the two $(\mathfrak{su}_2)_8$s within each parenthesis. This $\mathbb{Z}_2$ corresponds to the choice of ambiguity of $\mathfrak{sp}^\pm$ for the theories with very-even punctures of type $\mathfrak{g}=\mathfrak{so}_{2n}$ \cite{Distler:2022yse}. The second $\mathbb{Z}_2$ exchanges the two parenthesized factors and, thereby, the two Higgsing that lead to the distinct $(\mathfrak{so}_{12},\mathfrak{so}_{12})$ conformal matter theories.

This paper made extensive use of two ingredients: the Higgs branch RG flows between SCFTs and the correspondence between a subclass of 6d (1,0) theories and a subclass of 4d theories of class $\mathcal{S}$. We have been very circumspect in the Higgsings we considered: focusing exclusively on nilpotent Higgsings. This rather limited the isomorphisms we could explore and expanding the class of Higgsing that we have under good control \cite{DKL} will expand the reach of our methods. Moreover, we found evidence, through the computation of Schur indices in Section \ref{sec:oddballs}, for isomorphisms in 4d which have no avatars in 6d $(1,0)$ SCFTs. Still, an important lesson emerged from the 6d $(1,0)$ analysis: the automorphisms of the UV CFT exchange naively distinct Higgs branch RG flows, thus leading to isomorphisms between the naively distinct IR CFTs.

\section*{Acknowledgements}
While we did not take up his bet, we would like to thank Sergei Gukov for encouraging us to actually prove something.
J.~D., M.~J.~K., and C.~L.~would like to thank the Simons Center Summer Workshop 2022 for the stimulating environment in which this work was initiated. Further progress was made during the SCGP Workshop on ``5d $\mathcal{N}=1$ SCFTs and Gauge Theories on Brane Webs.'' M.~J.~K.~and C.~L.~also wish to thank the Weinberg Institute of UT Austin for hospitality during the later stage of this project. C.~L.~further thanks Florent Baume for collaboration on the development of the software {\tt hexacon}. The work of J.~D.~and G.~E.~is supported in part by the National Science Foundation under Grant No.~PHY--2210562. M.~J.~K.~is supported by a Sherman Fairchild Postdoctoral Fellowship and the U.~S.~Department of Energy, Office of Science, Office of High Energy Physics, under Award Number DE-SC0011632. C.~L.~acknowledges support from DESY (Hamburg, Germany), a member of the Helmholtz Association HGF.

\bibliography{references}{}
\bibliographystyle{sortedbutpretty}

\end{document}